\documentclass[twocolumn]{aastex631}
\usepackage{graphicx}

\newcommand\aastex{AAS\TeX}

\received{\today}
\submitjournal{AAS}

\shorttitle{\aastex\ ULIRG 3D}  
\shortauthors{Imanishi et al.}
\begin{document}

\title{Spectrally and spatially resolved (sub)millimeter 
HCN-to-HCO$^{+}$ flux ratios in nearby ultraluminous infrared galaxies}

\correspondingauthor{Masatoshi Imanishi}
\email{masa.imanishi@nao.ac.jp, m.imanishi.astro@gmail.com}

\author[0000-0001-6186-8792]{Masatoshi Imanishi}
\affil{National Astronomical Observatory of Japan, National Institutes 
of Natural Sciences (NINS), 2-21-1 Osawa, Mitaka, Tokyo 181-8588, Japan}
\affil{Department of Astronomy, School of Science, Graduate
University for Advanced Studies (SOKENDAI), Mitaka, Tokyo 181-8588,
Japan} 
\affil{Toyo University, 5-28-20, Hakusan, Bunkyo-ku, Tokyo 112-8606, 
Japan}

\author[0000-0003-0563-067X]{Yuri Nishimura}
\affil{Institute of Pure and Applied Sciences, University of Tsukuba, 
1-1-1 Tennodai, Tsukuba, Ibaraki, 305-8577, Japan} 
\affil{Tomonaga Center for the History of the Universe, University of 
Tsukuba, 1-1-1 Tennodai, Tsukuba, Ibaraki, 305-8577, Japan} 
\affil{Tsukuba Institute for Advanced Research (TIAR), University of
Tsukuba, 1-1-1 Tennodai, Tsukuba, Ibaraki, 305-8577, Japan}

\author[0000-0002-9850-6290]{Shunsuke Baba}
\affil{Institute of Space and Astronautical Science, 
Japan Aerospace Exploration Agency, 
3-1-1 Yoshinodai, Chuo-ku, Sagamihara, Kanagawa 252-5210, Japan}

\author[0000-0002-6939-0372]{Kouichiro Nakanishi}
\affil{National Astronomical Observatory of Japan, National Institutes 
of Natural Sciences (NINS), 2-21-1 Osawa, Mitaka, Tokyo 181-8588, Japan}
\affil{Department of Astronomy, School of Science, Graduate
University for Advanced Studies (SOKENDAI), Mitaka, Tokyo 181-8588,
Japan} 

\author[0000-0001-9452-0813]{Takuma Izumi}
\affil{National Astronomical Observatory of Japan, National Institutes 
of Natural Sciences (NINS), 2-21-1 Osawa, Mitaka, Tokyo 181-8588, Japan}
\affil{Department of Astronomy, School of Science, Graduate
University for Advanced Studies (SOKENDAI), Mitaka, Tokyo 181-8588,
Japan} 

%\author[0000-0000-0000-0001]{et al.}
%\affil{}

%% Note that the \and command from previous versions of AASTeX is now
%% depreciated in this version as it is no longer necessary. AASTeX 
%% automatically takes care of all commas and "and"s between authors names.

%% AASTeX 6.1 has the new \collaboration and \nocollaboration commands to
%% provide the collaboration status of a group of authors. These commands 
%% can be used either before or after the list of corresponding authors. The
%% argument for \collaboration is the collaboration identifier. Authors are
%% encouraged to surround collaboration identifiers with ()s. The 
%% \nocollaboration command takes no argument and exists to indicate that
%% the nearby authors are not part of surrounding collaborations.

%% Mark off the abstract in the ``abstract'' environment. 
\begin{abstract}

We present the results of our investigations of spectrally and
spatially resolved (sub)millimeter HCN-to-HCO$^{+}$ flux ratios at
J=2--1, J=3--2, and/or J=4--3 in 18 nearby ($z <$ 0.15) ultraluminous
infrared galaxies (ULIRGs), using ALMA $\lesssim$0$\farcs$2
($\lesssim$500 pc) resolution data. 
The geometry of elevated HCN-to-HCO$^{+}$ flux ratios (with 
$>$3$\sigma$ detections for both molecular lines) in
position-position-velocity (PPV) space is visually classified into 
(i) spherical shell (spectrally and spatially distinct),
(ii) spectrally distinct and spatially compact, and
(iii) filled (spectrally filled and spatially compact).
These can naturally be explained by the elevation of the flux ratio due to 
(i) a spatially resolved outflow,
(ii) an AGN and/or a spatially unresolved outflow with blueshifted and
redshifted emission components, and
(iii) an AGN and/or a spatially confined outflow with not clearly
separated blueshifted and redshifted velocity components, respectively.
Signatures of elevated HCN-to-HCO$^{+}$ flux ratios originated
from (a) spatially resolved outflow and (b) AGN and/or spatially unresolved 
outflow are seen in seven and nine ULIRGs, respectively.
In the former spatially resolved outflow-origin case, 
modest-velocity components relative
to the maximum outflow velocity tend to be probed by spaxels with
elevated HCN-to-HCO$^{+}$ flux ratios.
The spectrally and spatially resolved HCN-to-HCO$^{+}$ flux ratios can
provide additional information on the physical origin of the elevated
flux ratios in nearby ULIRG nuclei, compared to previously
conducted spatially integrated and/or velocity-integrated analyses.

\end{abstract}

%% Keywords should appear after the \end{abstract} command. 
%% See the online documentation for the full list of available subject
%% keywords and the rules for their use.
%\keywords{galaxies: active --- galaxies: nuclei --- quasars: general ---
%galaxies: Seyfert --- galaxies: starburst --- submillimeter: galaxies}

%% From the front matter, we move on to the body of the paper.
%% Sections are demarcated by \section and \subsection, respectively.
%% Observe the use of the LaTeX \label
%% command after the \subsection to give a symbolic KEY to the
%% subsection for cross-referencing in a \ref command.
%% You can use LaTeX's \ref and \label commands to keep track of
%% cross-references to sections, equations, tables, and figures.
%% That way, if you change the order of any elements, LaTeX will
%% automatically renumber them.

%% We recommend that authors also use the natbib \citep
%% and \citet commands to identify citations.  The citations are
%% tied to the reference list via symbolic KEYs. The KEY corresponds
%% to the KEY in the \bibitem in the reference list below. 

\section{Introduction} 

According to the widely accepted cold dark matter-based galaxy
formation scenario, mergers of gas-rich galaxies hosting supermassive
black holes (SMBHs) at the centers are common throughout the history
of the universe \citep{hop08}.
In such galaxy mergers, not only are many stars formed rapidly, but
also mass accretion onto the existing SMBH can be enhanced, and thus
luminous active galactic nucleus (AGN) activity, energetically and
radiatively powered by a mass-accreting SMBH, can emerge.
Both rapid star formation (starburst) and AGN activity occur in
nuclear regions surrounded by large amounts of dust and gas
\citep{hop06}.
The bulk of the energetic UV--optical radiation from the starburst and
AGN is absorbed by the surrounding dust and re-emitted as infrared
dust thermal radiation.
Thus, such gas-rich galaxy mergers usually become very luminous in
the infrared and are often observed as ultraluminous infrared galaxies
(ULIRGs; infrared luminosity L$_{\rm IR}$ $\gtrsim$
10$^{12}$L$_{\odot}$) and/or luminous infrared galaxies (LIRGs;
L$_{\rm IR}$ = 10$^{11}$--10$^{12}$L$_{\odot}$) \citep{sam96}.

The primary energy source of AGN activity powered by a
mass-accreting SMBH is spatially very compact ($\lesssim$10 pc),
compared to nuclear starburst activity (several 10 pc -- a few kpc).
For this reason, AGN activity in (U)LIRG nuclei can be deeply embedded
in dust and gas, and becomes very difficult to detect observationally,
particularly if the AGN is obscured by a sufficient column density of
dust and dense gas along virtually all lines of sight.  
It is essential to detect such elusive buried AGNs and properly
estimate their energetic contributions to the bolometric luminosities
of (U)LIRGs, if we are to understand the true nature of the (U)LIRG
population, as well as the co-evolution of star formation and SMBH
mass growth in gas-rich galaxy mergers.

For this purpose, observations at wavelengths with low dust and gas
extinction are clearly needed.
Infrared 3--40 $\mu$m and hard X-ray (10--80 keV) spectroscopy have
been applied to many nearby ($z <$ 0.15) (U)LIRGs to elucidate the
putative deeply buried but intrinsically luminous AGNs in the nuclear
regions, and signatures of such AGNs have indeed been detected in a
certain fraction of (U)LIRG nuclei
\citep[e.g.,][]{gen98,rig99,tra01,ima06a,arm07,ima07a,ima08,nar08,vei09,nar09,nar10,ima10,ten15,ric17,ric21}.
However, (sub)millimeter spectroscopy at 0.35--3.5 mm can be an even
more powerful tool, because extinction effects are smaller by a factor
of $\gtrsim$20 than those at infrared 3--40 $\mu$m or hard X-rays
(10--80 keV) \citep{hil83}.
Even infrared-elusive and/or hard X-ray-elusive, extremely deeply
buried AGNs in (U)LIRG nuclei could be detected in the
(sub)millimeter \citep[e.g.,][]{aal15b,ima18,ima19,fal19,fal21}.
At (sub)millimeter wavelengths (0.35--3.5 mm), rotational (J)
transition lines of abundant molecules (e.g., CO, HCN, HCO$^{+}$,
HNC, H$_{2}$O, CS, etc.) are present.
A luminous AGN radiatively powered by a mass-accreting SMBH has a
different energy generation mechanism from a starburst, whose
luminosity originates from nuclear fusion inside stars.
Thus, the physical and chemical effects on the surrounding
dense molecular gas can differ between an AGN and a starburst, possibly
producing different (sub)millimeter molecular line flux ratios
depending on the primary energy source.

HCN and HCO$^{+}$ dense molecular lines have often been used to
distinguish between AGN and starburst activity, because it has been
argued that luminous AGNs tend to show higher (sub)millimeter
HCN-to-HCO$^{+}$ flux ratios than starburst-dominated galaxies
\citep[e.g.,][]{koh05,kri08,izu16,but22,isr23,but25}, most likely 
because of enhanced HCN abundance, relative to HCO$^{+}$, in the
vicinity of AGNs \citep[e.g.,][]{vol22,but22,but25}. 
This (sub)millimeter HCN-to-HCO$^{+}$ flux ratio method has been
applied to nearby ($z <$ 0.15) (U)LIRGs and it has been found that
(U)LIRGs with luminous AGN signatures at other wavelengths tend to
show higher ratios than starburst-dominated (U)LIRGs without AGN
signatures 
\citep[e.g.,][]{ima06b,ima07b,gra08,ima09,cos11,pri15,ima16b,ima18,ima19,ima23a},
although it is also argued that there are a non-negligible number of
counter-examples \citep[e.g.,][]{cos11,pri15,pri20}.
For nearby ULIRGs, most investigations of the (sub)millimeter
HCN-to-HCO$^{+}$ flux ratios were conducted in a spatially integrated
manner, particularly in the pre-ALMA era
\citep[e.g.,][]{ima06b,ima07b,gra08,ima09,cos11,ala15,pri15,sli17,ala18,pri20},
because nearby  ULIRGs are usually energetically dominated
by compact ($\lesssim$1 kpc) nuclear regions
\citep[e.g.,][]{soi00,dia10,ima11,bar17,per21}, which correspond to
only $\lesssim$1 arcsec at $z \gtrsim 0.05$.

With the advent of ALMA, HCN and HCO$^{+}$ line observations with
$\lesssim$0$\farcs$2 angular resolution are now routinely possible,
enabling investigations of the (sub)millimeter HCN-to-HCO$^{+}$ flux
ratios in a spatially resolved manner even for compact ($\lesssim$1
kpc) nearby ULIRG nuclei. 
Thus, it has become possible to obtain spatial variation information
on the (sub)millimeter HCN-to-HCO$^{+}$ flux ratios within 
nearby ULIRG nuclei \citep[e.g.,][]{ima19,ima23b}.
In addition to the improvement in angular (spatial) resolution,
\citet{nis24} have recently proposed a new method to investigate the
(sub)millimeter HCN-to-HCO$^{+}$ flux ratios in a spatially and
spectrally resolved manner simultaneously, the so-called
position-position-velocity (PPV) plot.
\citet{nis24} applied this PPV analysis to the 0$\farcs$2--0$\farcs$5
resolution millimeter HCN J=3--2 and HCO$^{+}$ J=3--2 line data of
very nearby (mostly $z \lesssim$ 0.03) LIRGs and several ULIRGs at 
$z =$ 0.04--0.07.
If a luminous AGN is responsible for the higher HCN-to-HCO$^{+}$ flux
ratios, such spaxels are expected to be spatially concentrated in very
compact regions in (U)LIRG nuclei.
%The distribution of spaxels with elevated HCN-to-HCO$^{+}$ flux ratios
%can be spectrally distinct or filled, depending on whether the
%molecular gas in the vicinity of, and largely affected by, a luminous
%AGN is rotation-dominated or random-motion-dominated (i.e., has a
%continuous velocity distribution).
If a spatially resolved molecular outflow with blueshifted and
redshifted components plays an important role in elevating the
HCN-to-HCO$^{+}$ flux ratios
\citep[e.g.,][]{aal12,izu13,lin16,pri17,bar18,ima20,sai22,nis24,ima25}, 
the geometry of the spaxels with elevated HCN-to-HCO$^{+}$ flux ratios
can appear as a ``spherical shell (spectrally and spatially
distinct)'' 
(the spaxels are separated in velocity and position space)
\citep{nis24}.   
\citet{nis24} found that the possible signatures of 
elevated HCN-to-HCO$^{+}$ flux ratios originated from spatially
resolved molecular outflow are frequently seen in ULIRGs, but are rare 
in LIRGs. 

The presence of molecular outflows in ULIRGs has been investigated
mostly through the detection of (1) blueshifted OH absorption features
in the far-infrared 60--120 $\mu$m
\citep[e.g.,][]{spo13,vei13,gon17}, and/or (2) broad emission wings in CO
J-transition lines at (sub)millimeter wavelengths
\citep[e.g.,][]{cic14,ima17,per18,flu19,lut20,lam22}.
However, molecular outflow geometry cannot be directly derived from
the former OH absorption studies.
In the latter CO emission line wing studies, outflow geometry can be
constrained if the spatial resolution is sufficiently high
\citep[e.g.,][]{per18,lam22}, but cannot if it is not high
\citep[e.g.,][]{cic14,flu19,lut20}.
The PPV plot of elevated HCN-to-HCO$^{+}$ flux ratios, obtained with
sufficiently high angular and velocity resolutions, may therefore help
better understand whether AGN effects and/or outflow activity are
responsible for elevating the ratios in compact ($\lesssim$1 kpc)
nearby ULIRG nuclei, compared to the widely investigated spatially
unresolved and/or velocity-integrated HCN-to-HCO$^{+}$ flux ratios.

In this paper, we conduct spectrally and spatially resolved PPV
investigations of elevated HCN-to-HCO$^{+}$ flux ratios in nearby 
($z <$ 0.15) ULIRGs with available
$\lesssim$0$\farcs$2-resolution HCN and HCO$^{+}$ line data at
J=2--1, J=3--2, and/or J=4--3, to better understand the physical
origin of the observed high HCN-to-HCO$^{+}$ flux ratios in ULIRG
nuclei.
We adopt the cosmological parameters $H_{0} = 71$ km s$^{-1}$
Mpc$^{-1}$, $\Omega_{\rm M} = 0.27$, and $\Omega_{\Lambda} = 0.73$ in
this paper.

\section{Targets and Data analysis} 

More than 30 nearby ($z <$ 0.15) ULIRGs have been observed in both the
(sub)millimeter HCN and HCO$^{+}$ lines at J=2--1, J=3--2, and/or
J=4--3 with $\lesssim$1 arcsec resolution, mostly using ALMA 
\citep[e.g.,][]{ima13,ima14,aal15a,ima16a,ima16b,ima17,pri17,ima18,ima19,fal21,sak21,ima23b,nis24}.
For ULIRGs at $z =$ 0.07--0.14, 1 arcsec corresponds to 1.3--2.4 kpc.
HCN and HCO$^{+}$ line data with $\sim$1 arcsec resolution are useful
only for investigating whether the HCN-to-HCO$^{+}$ flux ratios are
elevated or not over the entire compact ($\lesssim$1 kpc) nuclear
regions of nearby ULIRGs.
Substantially higher angular resolution ($\ll$1 arcsec) is needed to
investigate the spatial variation of the HCN-to-HCO$^{+}$ flux ratios
within the compact ULIRG nuclei.
We limit the target ULIRGs for our PPV study to those with available
$\lesssim$0$\farcs$2 resolution HCN and HCO$^{+}$ line data
\citep{ima19,ima23b}, to investigate the HCN-to-HCO$^{+}$
flux ratios in a spatially resolved manner at ULIRG nuclei.
Although our ULIRG sample is generally farther than the (U)LIRGs
studied by \citet{nis24}, the achieved angular resolution 
($\lesssim$0$\farcs$2) is smaller than that
($\sim$0$\farcs$3--0$\farcs$4) of \citet{nis24}.  
Velocity-integrated, spatially resolved maps of the HCN-to-HCO$^{+}$
flux ratios of our ULIRG sample are presented in previous
publications \citep{ima19,ima23b}. 

ALMA observations of our ULIRG sample were conducted in Cycles 5
and 7 \citep{ima19,ima23b}.
We reanalyzed the ALMA data by slightly modifying the method adopted by
\citet{ima19,ima23b}.
Specifically, in \citet{ima19,ima23b}, the clean procedure was
performed using the CASA task ``tclean'' \citep{CASA22}, with the same
number of channels for HCN and HCO$^{+}$.
The resulting velocity resolution slightly differs between HCN and
HCO$^{+}$ at the same J-transition because of their slightly different
rest frequencies.
We therefore applied ``tclean'' (Briggs weighting; robust = 0.5, gain =
0.1) with the same velocity width (20 km s$^{-1}$ was finally adopted;
see Section~3), the same velocity range of $\pm$1500 km s$^{-1}$
around each line, and the same pixel scale of 0$\farcs$02
pixel$^{-1}$, for both HCN and HCO$^{+}$.
Because the HCN and HCO$^{+}$ line data at the same J-transition
were taken simultaneously, their relative astrometry is regarded as
very accurate and is unaffected by possible absolute pointing
uncertainties in individual ALMA observations.
In this way, we can compare the HCN-to-HCO$^{+}$ flux ratios on a
spectrally and spatially resolved basis in a straightforward and
reliable manner.

For the ULIRGs with available $\lesssim$0$\farcs$2-resolution HCN and
HCO$^{+}$ data \citep{ima19,ima23b}, we created PPV plots of the
HCN-to-HCO$^{+}$ flux ratios for (1) all spaxels with $>$3$\sigma$
detections for both HCN and HCO$^{+}$, and (2) only spaxels with
elevated HCN emission relative to HCO$^{+}$ emission 
(with $>$3$\sigma$ detections for both lines).
For the latter, we select (a) the top 10\% of the HCN-to-HCO$^{+}$
flux ratios (i.e., ``relatively'' high), following \citet{nis24}.
However, this top 10\% criterion does not necessarily reflect truly
elevated HCN-to-HCO$^{+}$ flux ratios if the majority of the
$>$3$\sigma$-detected spaxels show ratios below or around unity.
We therefore investigate also the distribution of (b) spaxels with
HCN-to-HCO$^{+}$ flux ratios $\gtrsim$1.5 (i.e., ``absolutely'' high).
To investigate the spectral and spatial distributions of spaxels with
elevated HCN-to-HCO$^{+}$ flux ratios in a meaningful manner, ULIRGs
with faint HCN and/or HCO$^{+}$ emission need to be excluded.
We present results only for ULIRGs in which the number of spaxels in
the top 10\% of the HCN-to-HCO$^{+}$ flux ratios exceeds 25 in any of
the available J=2--1, J=3--2, and/or J=4--3 data.
In general, if the detection significance in the integrated-intensity
(moment 0) map is $\gtrsim$10$\sigma$ for both the HCN and HCO$^{+}$
emission lines \citep{ima19,ima23b}, this numerical requirement for
spaxels is usually met.

Table~\ref{tab:object} summarizes the final 18 ULIRGs used for our PPV
study of the elevated HCN-to-HCO$^{+}$ flux ratios.
%Our ULIRG sample is not statistically complete.
All ULIRGs are optically classified as non-Seyferts
(Table~\ref{tab:object}, column 9), and thus are not classical optically 
identified luminous AGNs that are surrounded by a torus-shaped
distribution of dust and gas, and have well-developed narrow line
regions along the torus axis above the torus scale height
\citep[e.g.,][]{ant85,vei87}. 
However, signatures of optically elusive, deeply buried
luminous AGNs have been found in a large fraction of the
ULIRGs, based on infrared and/or (sub)millimeter spectroscopy,
because of substantially reduced dust extinction effects,
compared to the optical. 
Information of such infrared- and (sub)millimeter-identified luminous
buried AGN signatures are summarized in columns 10 and 11 of
Table~\ref{tab:object}. 
Molecular outflow activity possibly caused by the central AGN feedback 
is expected to be detected strongly in such luminous buried AGNs
surrounded by large column densities of gas and dust in virtually all
sightlines.   
Hard X-ray (10--80 keV) spectroscopy can also be another powerful tool
to detect luminous buried AGNs 
\citep[e.g.,][]{ten15,ric17,ric21}. 
Among the 18 ULIRGs, \citet{ric21} presented hard X-ray observational
results for two $z <$ 0.1 ULIRGs, IRAS~12112$+$0305 and
IRAS~14348$-$1447, and reported no clear detection of luminous buried
AGNs. 
For the remaining 16 ULIRGs, we find no published hard 
X-ray observational studies searching for luminous buried AGNs.
Hard X-ray detection of buried AGNs in ULIRGs is difficult when the
X-ray absorbing hydrogen column density (N$_{\rm H}$) in front of the
X-ray continuum-emitting regions exceeds N$_{\rm H}$ $\gtrsim$ 10$^{25}$
cm$^{-2}$ (i.e., heavily Compton thick), and when ULIRGs lie at $z >$ 
0.1, because the observed X-ray fluxes are faint. 
In obscured AGNs, a large amount of dust-free gas inside the dust
sublimation radius around AGNs makes X-ray absorption N$_{\rm H}$ (by
gas and dust) much higher than that expected from dust extinction
(A$_{\rm V}$) for infrared and (sub)millimeter emission, and the
Galactic N$_{\rm H}$/A$_{\rm V}$ ratio of $\sim$2 $\times$ 10$^{21}$
cm$^{-2}$ mag$^{-1}$ \citep[e.g.,][]{alo97,bur16,ich19,miz22}. 
Even though luminous buried AGN signatures are found through infrared
and/or (sub)millimeter spectroscopy in a large fraction of the 18
ULIRGs (Table~\ref{tab:object}), hard X-ray detection of luminous buried
AGNs is not reported in the literature in any of the 18 ULIRGs at $z =$
0.07--0.14.  

Eight ULIRGs have available $\lesssim$0$\farcs$2-resolution HCN and
HCO$^{+}$ data for all of the J=2--1, J=3--2, and J=4--3 transitions.
The PPV plots at multiple J-transition lines for the same objects
enable us to investigate the consistency and/or possible differences in
the overall spectral and spatial properties of the elevated 
HCN-to-HCO$^{+}$ flux ratios.
For the remaining ten ULIRGs, the $\lesssim$0$\farcs$2-resolution HCN
and HCO$^{+}$ data are available only for one or two J-transition
lines among J=2--1, J=3--2, and J=4--3.
The achieved physical resolutions range from $\sim$100
pc to $\sim$500 pc (Table~\ref{tab:object}, columns 6--8).

%%%%%%%%%% Table 1 (object) %%%%%%%%%
\begin{deluxetable*}{lcccccccccc}[!hbt]
\tabletypesize{\scriptsize}
\rotate
\tablecaption{Basic Properties of the Studied Ultraluminous Infrared 
Galaxies \label{tab:object}}  
\tablewidth{0pt}
\tablehead{
\colhead{Object} & \colhead{Redshift} & 
\colhead{d$_{\rm L}$} & \colhead{Scale} & 
\colhead{log L$_{\rm IR}$} & 
\multicolumn{3}{c}{Beam size [$''$ $\times$ $''$] 
(pc $\times$ pc)} & 
\colhead{Optical} & \colhead{IR} & \colhead{(Sub)mm} \\ 
\colhead{} & \colhead{} & \colhead{[Mpc]} & \colhead{[kpc/$''$]}  
& \colhead{[L$_{\odot}$]} & \colhead{J=2--1} & \colhead{J=3--2} &
\colhead{J=4--3} & \colhead{Class} & \colhead{AGN} & \colhead{AGN} \\
\colhead{(1)} & \colhead{(2)} & \colhead{(3)} & \colhead{(4)} & 
\colhead{(5)} & \colhead{(6)} & \colhead{(7)} & \colhead{(8)} & 
\colhead{(9)} & \colhead{(10)} & \colhead{(11)}  
}
\startdata
IRAS~00091$-$0738 & 0.1180 & 543 & 2.1 & 12.3 & 0.19
$\times$ 0.15 (400 $\times$ 310) & 0.18 $\times$ 0.13 (390 $\times$ 280)
& 0.17 $\times$ 0.10 (360 $\times$ 220) & HII & Y$^{a,b,c}$ & Y$^{g,h}$ \\ 
IRAS~00188$-$0856 & 0.1285 & 596 & 2.3 & 12.4 & 0.22
$\times$ 0.17 (490 $\times$ 380) & 0.18 $\times$ 0.13 (420 $\times$ 290)
& 0.16 $\times$ 0.10 (350 $\times$ 230) & LINER & Y$^{a,b,c,d,e,f}$ & Y$^{g,h}$ \\ 
IRAS~00456$-$2904 & 0.1100 & 504 & 2.0 & 12.2 & 0.22
$\times$ 0.17 (450 $\times$ 330) & 0.16 $\times$ 0.12 (320 $\times$ 240)
& 0.19 $\times$ 0.14 (370 $\times$ 270) & HII & N$^{a,b,c}$ & Y$^{g,h}$ \\
IRAS~01166$-$0844 & 0.1172 & 539 & 2.1 & 12.1 & 0.21
$\times$ 0.17 (450 $\times$ 350) & 0.12 $\times$ 0.092 (250 $\times$
190) & 0.15 $\times$ 0.090 (310 $\times$ 190) & HII & Y$^{a,b,c,e}$ & Y$^{g,h}$ \\ 
IRAS~01569$-$2939 & 0.1402 & 655 & 2.4 & 12.3 & 0.21
$\times$ 0.15 (520 $\times$ 360) & 0.11 $\times$ 0.11 (270 $\times$ 270)
& 0.21 $\times$ 0.087 (520 $\times$ 210) & HII & Y$^{a,b}$ & N$^{g,h}$ \\ 
IRAS~16090$-$0139 & 0.1334 & 621 & 2.3 & 12.6 & 0.20
$\times$ 0.15 (460 $\times$ 350) & 0.17 $\times$ 0.15 (390 $\times$ 350)
& 0.18 $\times$ 0.10 (420 $\times$ 240) & LINER & Y$^{a,b,c,d,e}$ &
N$^{g,h}$ \\ 
IRAS~22206$-$2715 & 0.1320 & 614 & 2.3 & 12.2 & 0.18
$\times$ 0.15 (420 $\times$ 340) & 0.18 $\times$ 0.13 (420 $\times$ 310)
& 0.16 $\times$ 0.099 (360 $\times$ 230) & HII & N$^{a,b,c,e}$ & Y$^{h}$ \\ 
IRAS~22491$-$1808 & 0.0776  & 347 & 1.5 & 12.2 & 0.17
$\times$ 0.12 (240 $\times$ 180) & 0.23 $\times$ 0.13 (340 $\times$ 180)
& 0.16 $\times$ 0.10 (240 $\times$ 150) & HII & Y$^{c,f}$ &
Y$^{g,h,i,j}$ \\ \hline  
IRAS~01004$-$2237 & 0.1180 & 543 & 2.1 & 12.3  & \nodata & 0.17
$\times$ 0.12 (350 $\times$ 250) & \nodata & HII & Y$^{a,b,c,e,f}$ &
N$^{g}$ \\   
IRAS~01298$-$0744 & 0.1368 & 638 & 2.4 & 12.4 & \nodata & 0.15
$\times$ 0.12 (350 $\times$ 280) & \nodata & HII & Y$^{a,b,c}$ & N$^{g}$ \\
IRAS~09039$+$0503\tablenotemark{$\alpha$} & 0.1257 & 578 & 2.2 & 12.1 &
\nodata & 0.18 $\times$ 0.14 (410 $\times$ 310) & \nodata &
LINER & Y$^{c,d}$ & Y$^{g}$ \\ 
IRAS~10190$+$1322 & 0.0762 & 341 & 1.4 & 12.0 & \nodata & 0.17
$\times$ 0.15 (250 $\times$ 210) & \nodata & HII & N$^{a,b,c,d}$ & N$^{g}$ \\
IRAS~10378$+$1108 & 0.1365 & 636 & 2.4 & 12.3 & 0.17
$\times$ 0.15 (390 $\times$ 360) & 0.17 $\times$ 0.15 (400
$\times$ 350) & \nodata & LINER & Y$^{b,c,d,f}$ & N$^{g}$ \\ 
IRAS~11095$-$0238 NE\tablenotemark{$\beta$} & 0.1066 & 484 & 1.9 & 12.3 &
\nodata & 0.13 $\times$ 0.12 (250 $\times$ 230) & \nodata & LINER &
Y$^{a,b,e,f,g}$ & Y$^{g}$ \\ 
IRAS~12112$+$0305 NE\tablenotemark{$\beta$}  & 0.0730 & 326 & 1.4 & 12.3 & 
\nodata & \nodata & 0.11 $\times$ 0.071 (160 $\times$ 100) &
LINER & N$^{a,b,c,d,e}$ & Y$^{g,i,j}$ \\ 
IRAS~13509$+$0442 & 0.1365 & 636 & 2.4 & 12.3 & \nodata & 0.22
$\times$ 0.16 (520 $\times$ 390) & \nodata & HII & N$^{a,b,d,e}$ & N$^{g}$ \\
IRAS~14348$-$1447 SW\tablenotemark{$\beta$} & 0.0830 & 373 & 1.5 & 12.3 &
\nodata & 0.14 $\times$ 0.10 (220 $\times$ 160) & \nodata &
LINER & Y$^{e,f}$ & Y$^{g}$ \\ 
IRAS~23234$+$0946 & 0.1280 & 593 & 2.3 & 12.1 & \nodata & 0.15
$\times$ 0.13 (350 $\times$ 290) & \nodata & LINER & N$^{a,b,c,d}$ &
N$^{g}$ \\ \hline 
\enddata

\tablecomments{
Col.(1): Object name.
Col.(2): Redshift adopted from the ALMA dense molecular line data
\citep{ima16b,ima19}, which is slightly different from the optically
derived value \citep{kim98} in some cases.
Col.(3): Luminosity distance (d$_{\rm L}$) in Mpc.
Col.(4): Physical scale in kpc arcsec$^{-1}$.
Col.(5): Decimal logarithm of the infrared (8--1000 $\mu$m) luminosity
in units of solar luminosity (L$_{\odot}$), calculated as
L$_{\rm IR}$ = 2.1 $\times$ 10$^{39}$ $\times$ d$_{\rm L}$ ({\rm Mpc})$^{2}$
$\times$ (13.48 $f_{12}$ + 5.16 $f_{25}$ + 2.58 $f_{60}$ + $f_{100}$)
[erg s$^{-1}$] \citep{sam96}, where $f_{12}$, $f_{25}$, $f_{60}$, and
$f_{100}$ are IRAS fluxes at 12 $\mu$m, 25 $\mu$m, 60 $\mu$m, and 
100 $\mu$m, respectively, taken from \citet{kim98}.
Cols.(6)--(8): Synthesized beam sizes ([$''$ $\times$ $''$]) for
J=2--1, J=3--2, and J=4--3.
Data are taken from \citet{ima19} and \citet{ima23b}.
The sizes in pc $\times$ pc are also shown in parentheses. 
``$\cdots$'': no data.
Col.(9): Optical spectroscopic classification \citep{vei99}:
LINER or HII-region (HII). 
All of the observed ULIRGs are optically classified as non-Seyferts,
that is, they show no obvious optical AGN signatures.
Col.(10): Presence (``Y'') of optically elusive, 
but intrinsically luminous, infrared-identified buried AGNs.
We first refer to \citet{nar10} as the initial classification of
AGN-important ULIRGs (i.e., infrared-estimated AGN bolometric
contribution $\gtrsim$20\%). Additional references are also shown.
$^{a}$: \citet{nar10};  
$^{b}$: \citet{ima07a};  
$^{c}$: \citet{vei09};  
$^{d}$: \citet{ima06a};  
$^{e}$: \citet{ima10};  
$^{f}$: \citet{var25}.  
For \citet{vei09}, ULIRGs with infrared-estimated AGN bolometric
contributions $\gtrsim$20\% are classified as AGN-important.  
``N'': No clear luminous buried AGN signature.
Col.(11): Presence (``Y'') of optically elusive, 
but intrinsically luminous, (sub)millimeter-identified buried AGNs.
$^{g}$: \citet{ima19};  
$^{h}$: \citet{ima23b};  
$^{i}$: \citet{ima16b};  
$^{j}$: \citet{ima18}.  
We classify ULIRGs with nuclear HCN-to-HCO$^{+}$ flux ratios
$\gtrsim$1.5 as possible AGN signatures.
``N'': No clear luminous buried AGN signature.
}

\tablenotetext{\alpha}{The PPV plots of both the
  IRAS~09039$+$0503 SW and NE nuclei are displayed, but each
  plot is partially contaminated by line emission from the other nucleus.}

\tablenotetext{\beta}{Only the brighter primary nucleus is
investigated, because the secondary nucleus is too faint in
dense molecular line emission to allow a meaningful PPV analysis.}

\end{deluxetable*}
%%%%%%%%%%%%%%%%%%%%%%%%%%%%%%%%%%%

\section{Results}

\begin{figure*}[!hbt]
%\vspace*{0.6cm}
\begin{center}
%\hspace{-0.4cm}
\includegraphics[scale=0.45]{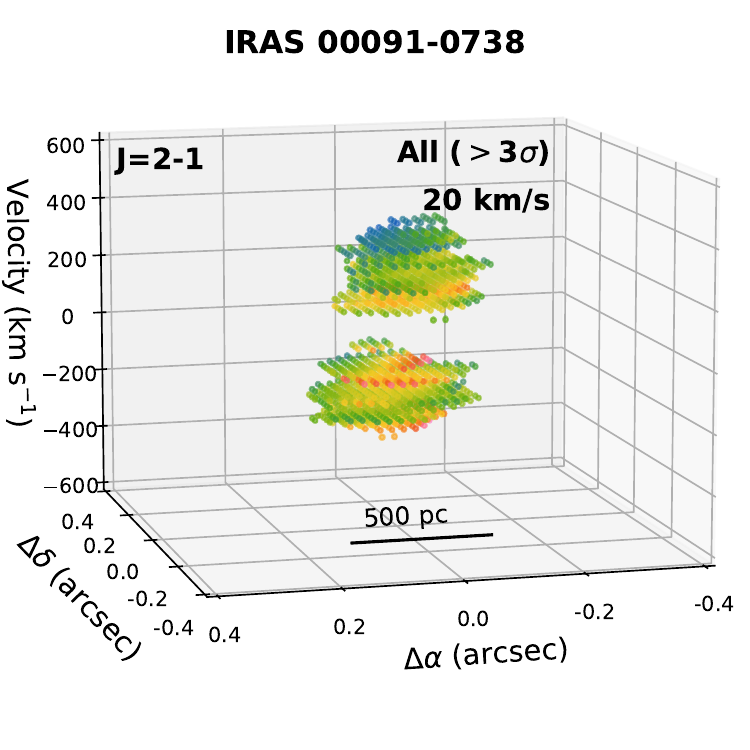} 
\includegraphics[scale=0.45]{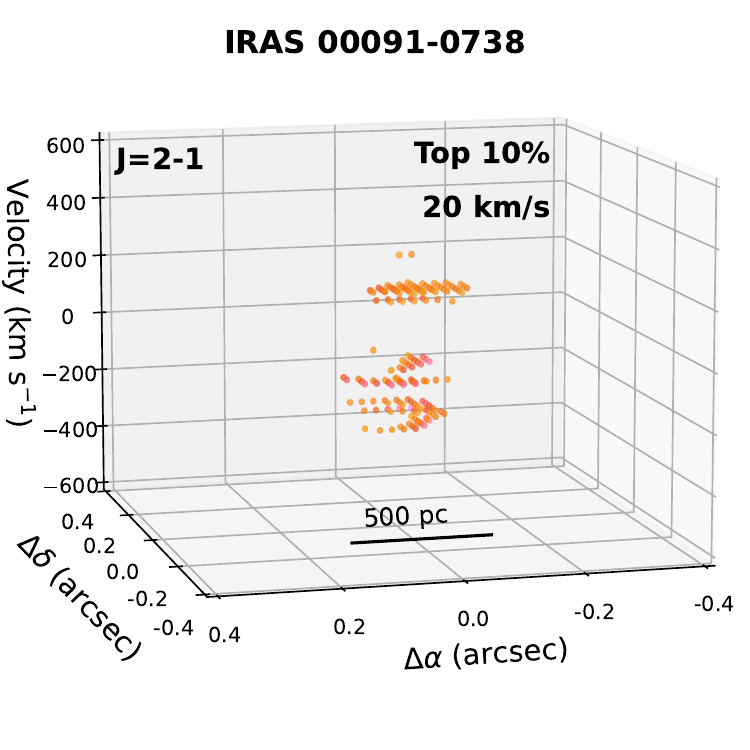} 
\includegraphics[scale=0.45]{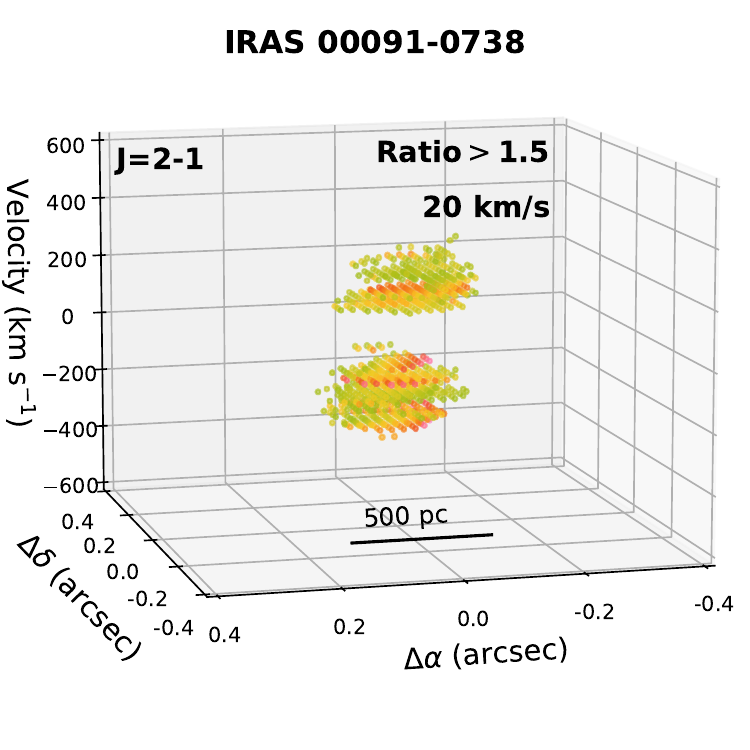} \\
\includegraphics[scale=0.45]{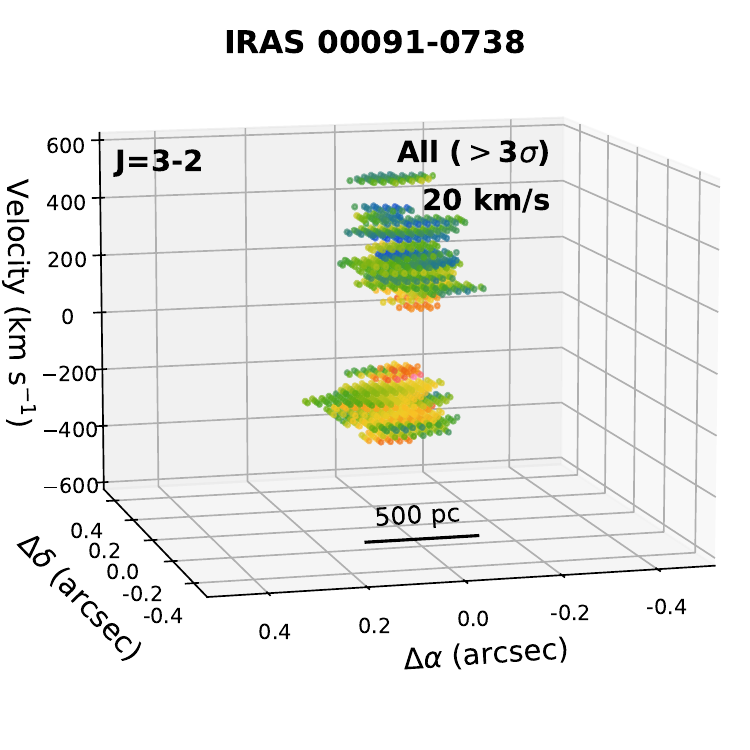} 
\includegraphics[scale=0.45]{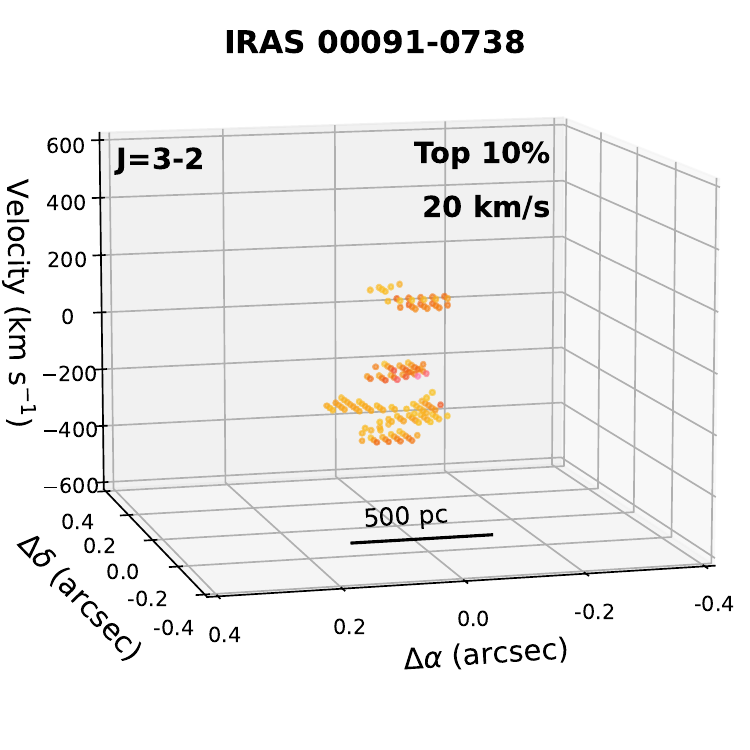} 
\includegraphics[scale=0.45]{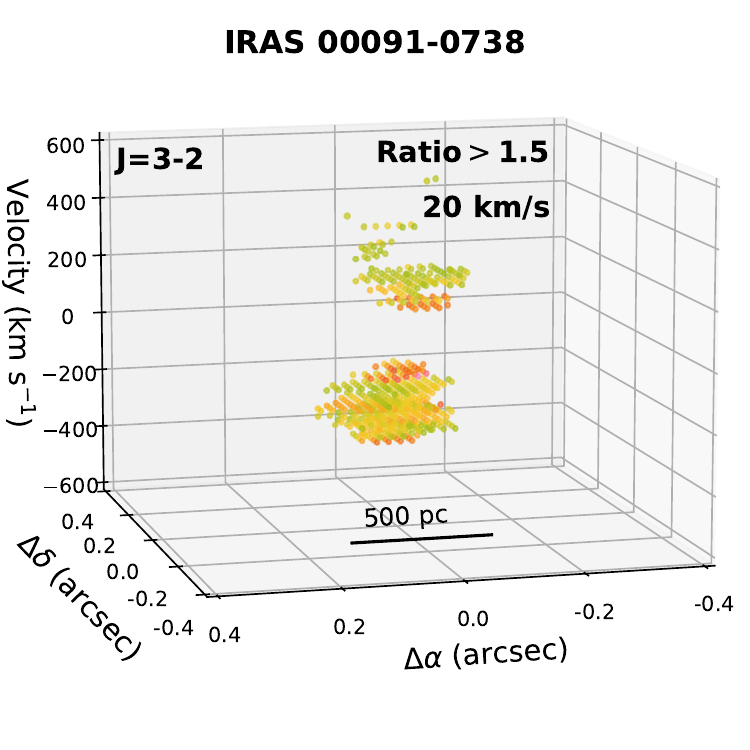} \\
\includegraphics[scale=0.45]{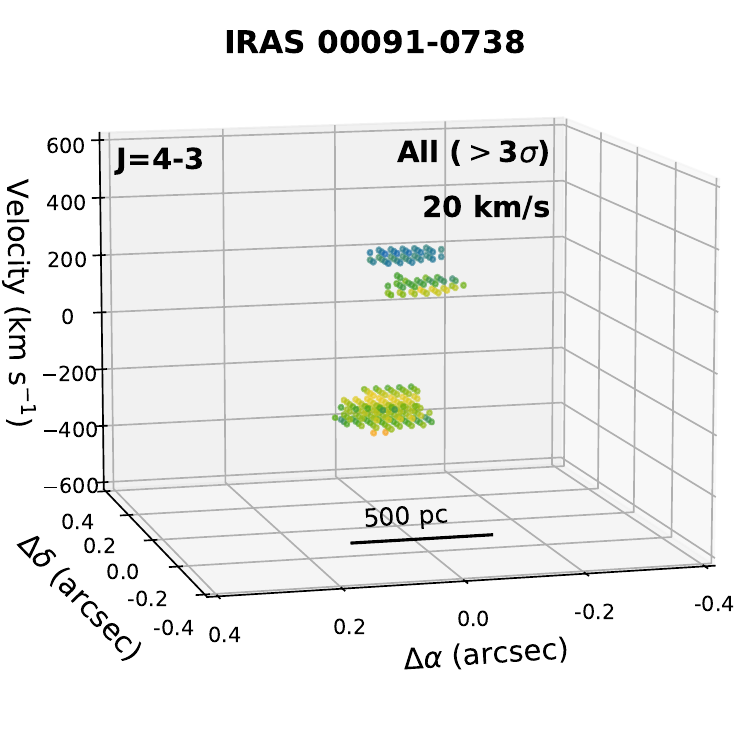} 
\includegraphics[scale=0.45]{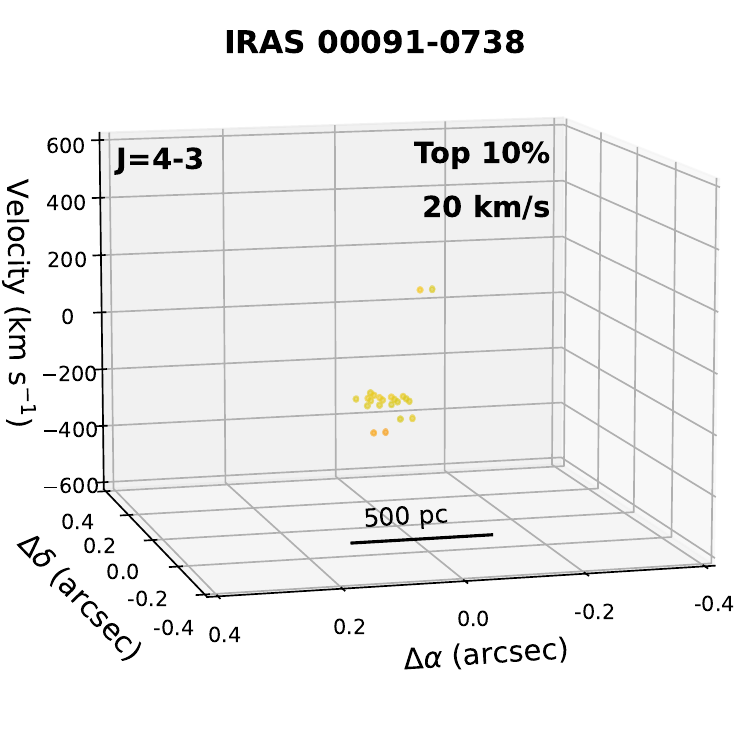} 
\includegraphics[scale=0.45]{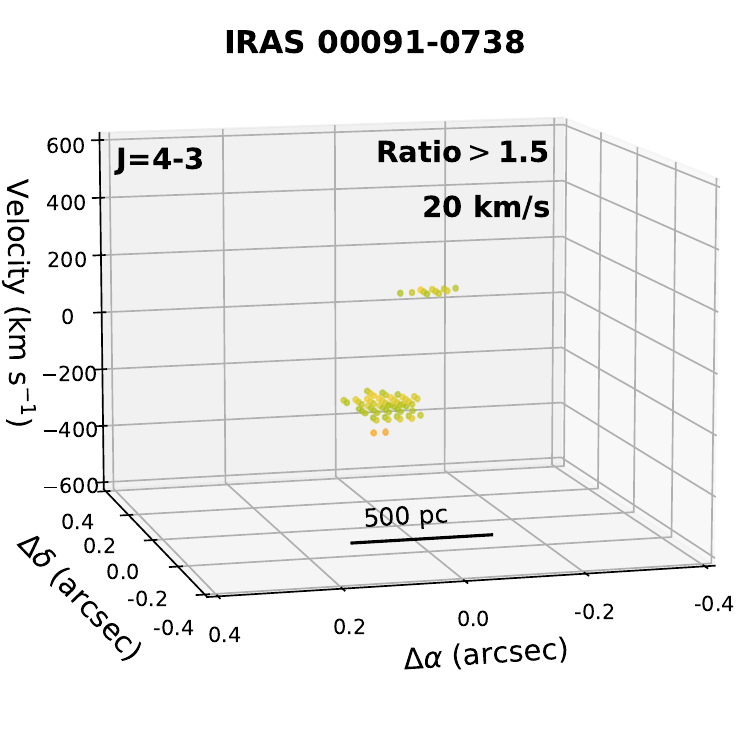} \\ 
\includegraphics[scale=0.45]{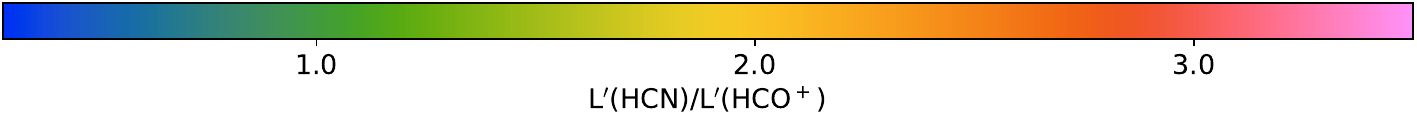} \\
\end{center}
%\vspace{-0.6cm}
\caption{
PPV plot of the (sub)millimeter HCN-to-HCO$^{+}$ flux ratios at
J=2--1, J=3--2, and J=4--3, with a velocity resolution of 20 km
s$^{-1}$ and a pixel scale of 0$\farcs$02 pixel$^{-1}$ for
IRAS~00091$-$0738.   
The HCN-to-HCO$^{+}$ flux ratios are calculated in brightness
temperature (K).  
(This is virtually identical to the ratio in flux density (Jy
$\propto$ K $\times$ $\nu^{2}$), because the HCN and HCO$^{+}$
frequencies differ only by a factor of $\sim$1.006 at the same J
transition.)  
Only data with $>$3$\sigma$ detections in emission for both the HCN and 
HCO$^{+}$ lines are plotted. 
The color bar corresponds to 0.285 (blue) to 3.5 (pink) in the same
way as \citet{nis24}.
\textit{Left}: All $>$3$\sigma$-detected data. 
\textit{Center}: Data with the top 10\% ratios. 
\textit{Right}: Data with ratios $>$1.5. 
East is to the left and west is to the right along the $\Delta\alpha$ axis. 
North is farther and south is nearer along the $\Delta\delta$ axis.
\label{fig:3D1}
}
\end{figure*}

\begin{figure*}[!hbt]
%\vspace*{0.6cm}
\begin{center}
%\hspace{-0.4cm}
\includegraphics[scale=0.45]{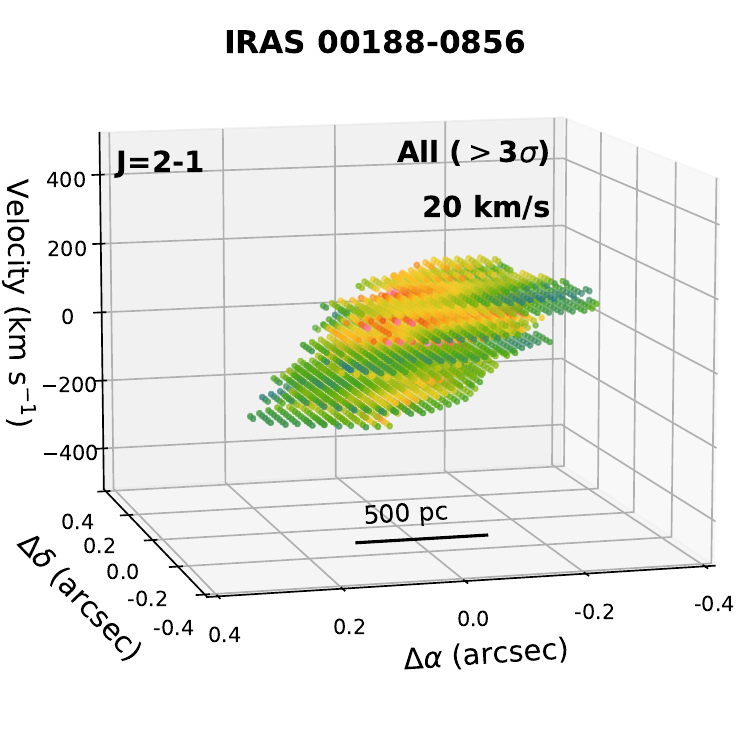} 
\includegraphics[scale=0.45]{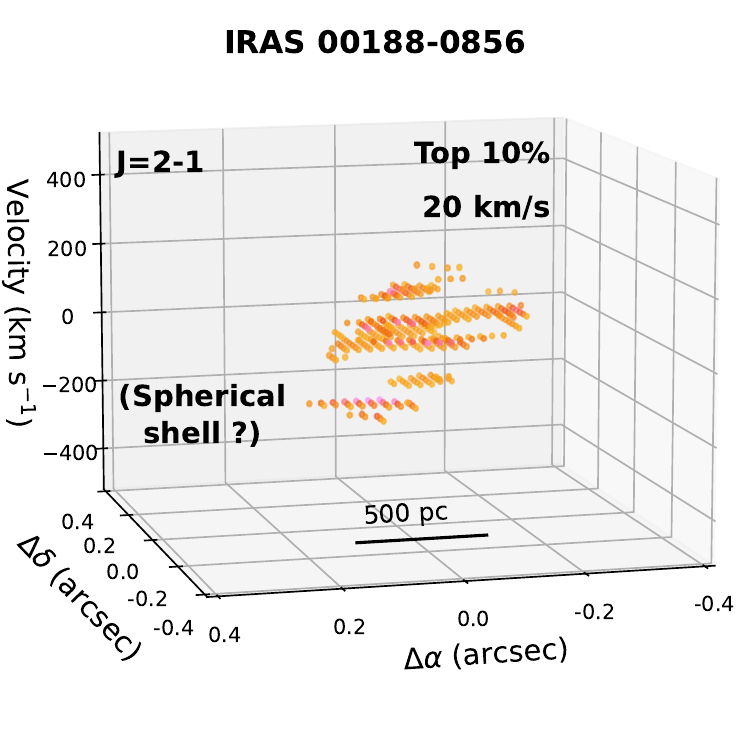} 
\includegraphics[scale=0.45]{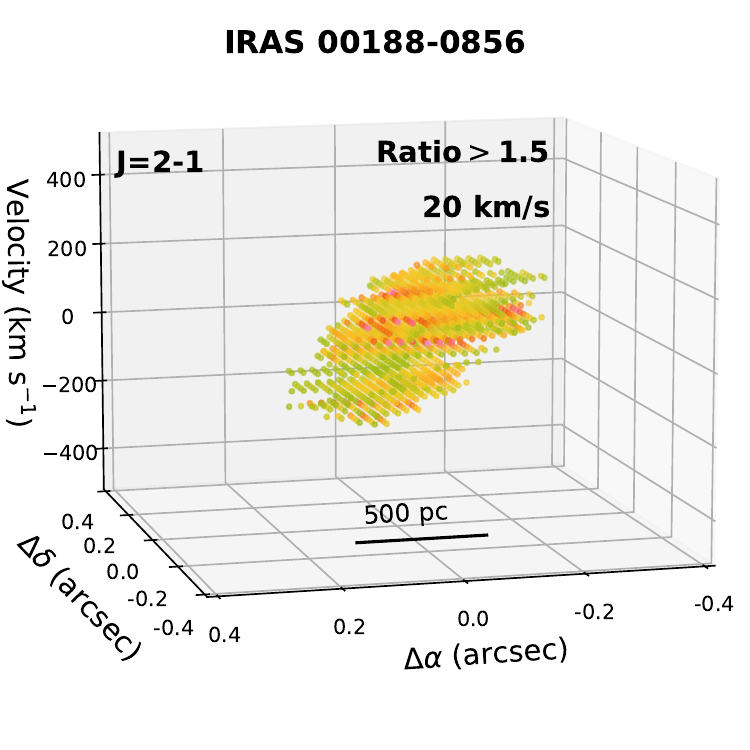} \\
\includegraphics[scale=0.45]{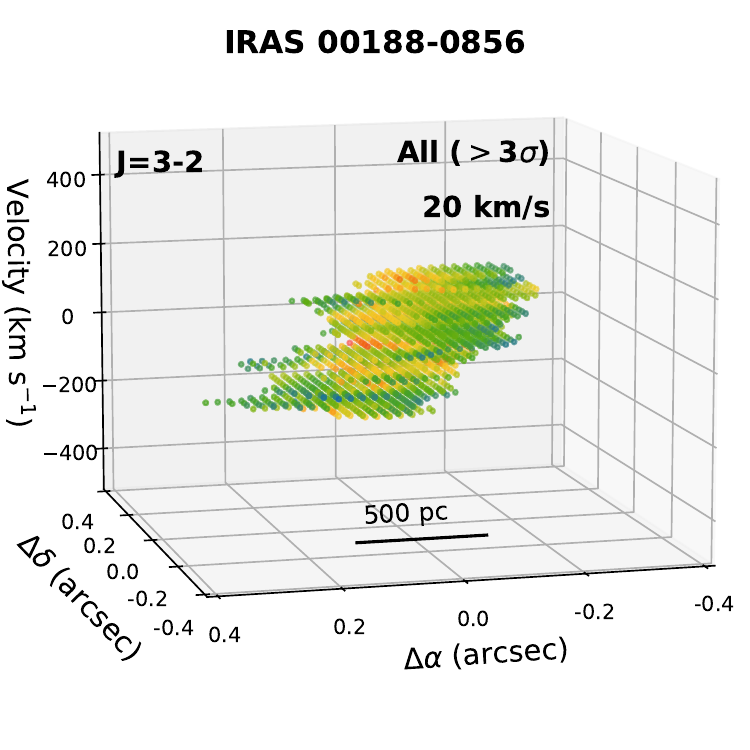} 
\includegraphics[scale=0.45]{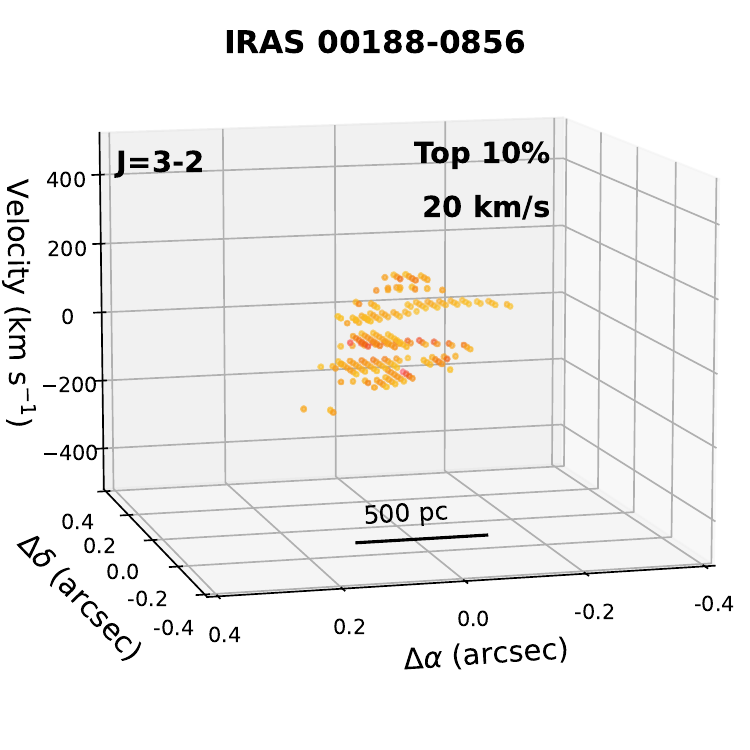} 
\includegraphics[scale=0.45]{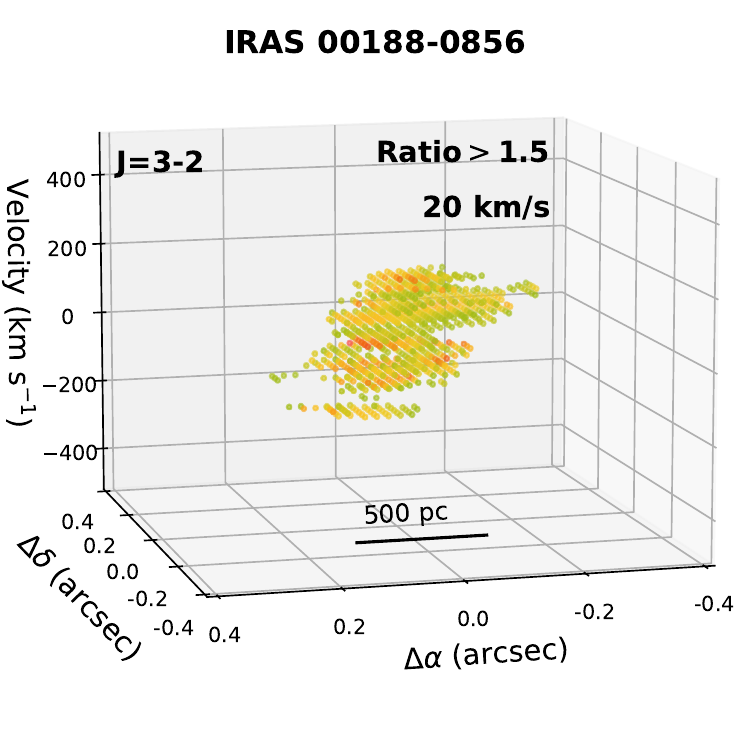} \\
\includegraphics[scale=0.45]{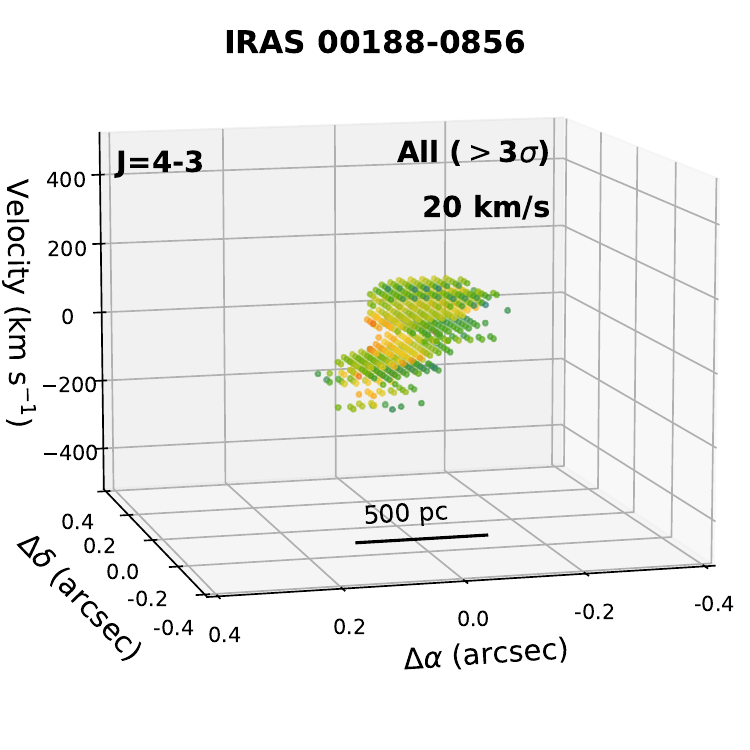} 
\includegraphics[scale=0.45]{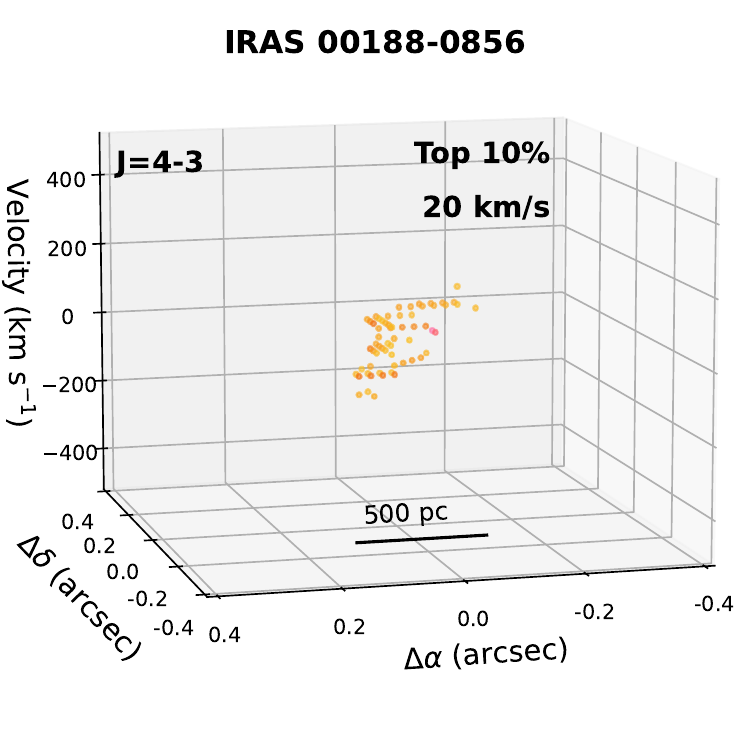} 
\includegraphics[scale=0.45]{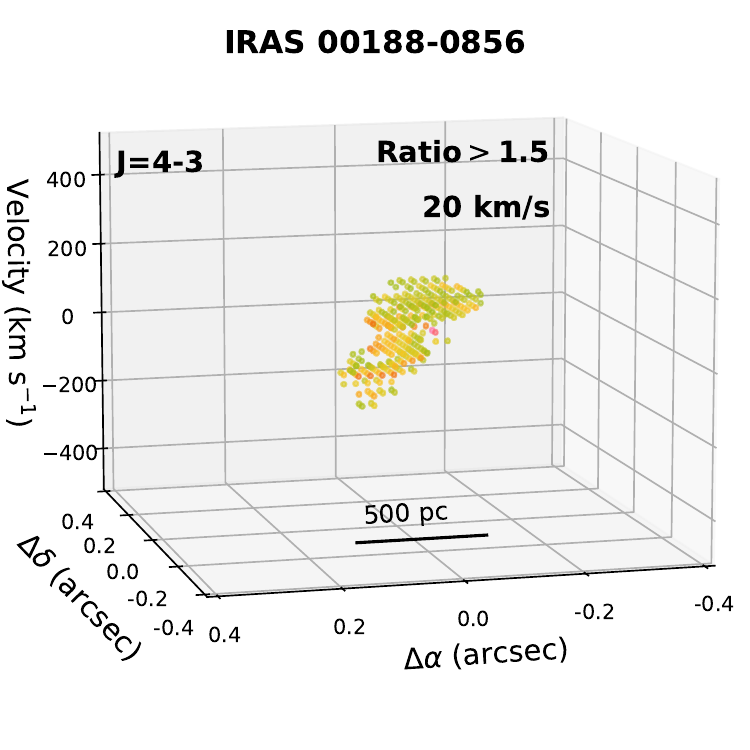} \\
\includegraphics[scale=0.45]{f1-9common.pdf} \\
\end{center}
\vspace{-0.6cm}
\caption{
Same as Figure~\ref{fig:3D1}, but for IRAS~00188$-$0856.
\label{fig:3D2}
}
\end{figure*}

\begin{figure*}[!hbt]
%\vspace*{0.6cm}
\begin{center}
%\hspace{-0.4cm}
\includegraphics[scale=0.45]{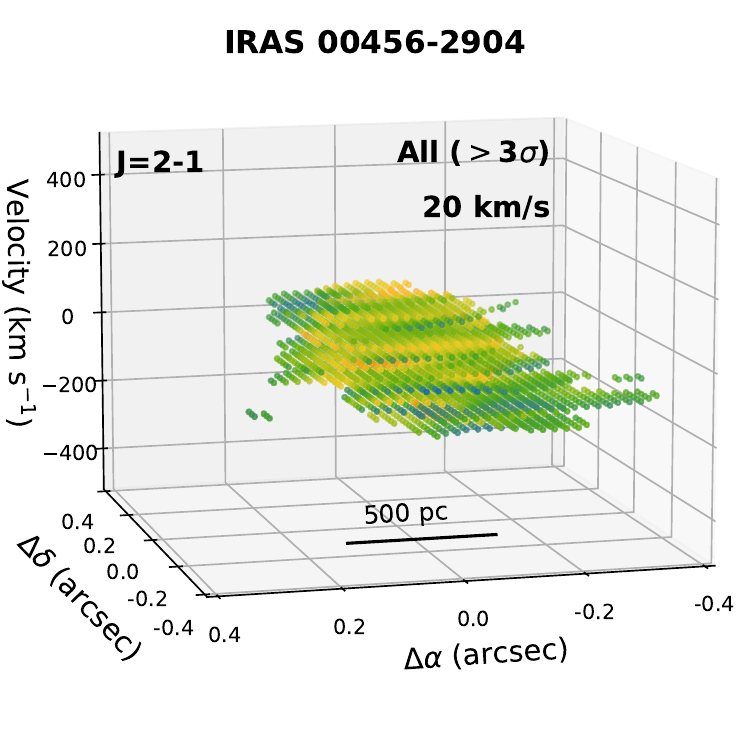} 
\includegraphics[scale=0.45]{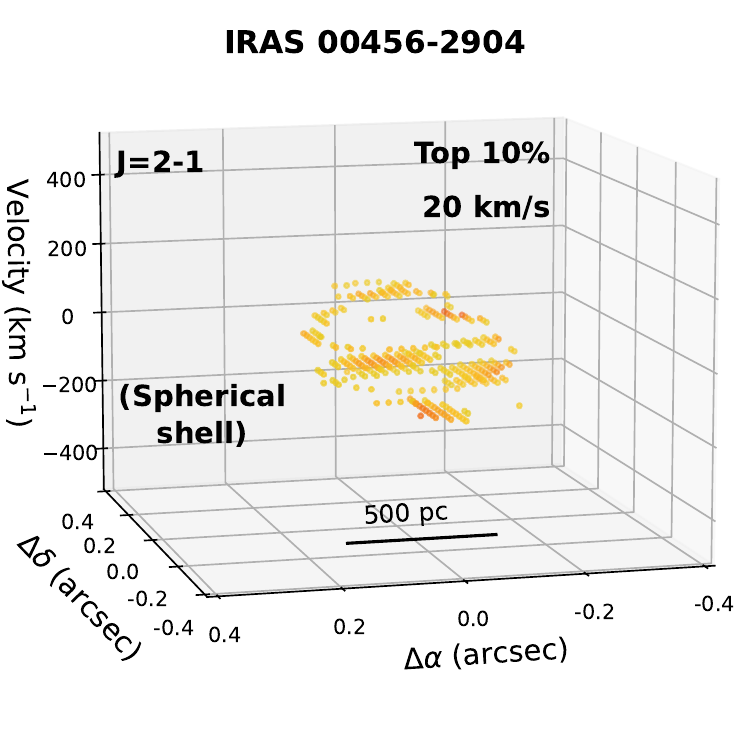} 
\includegraphics[scale=0.45]{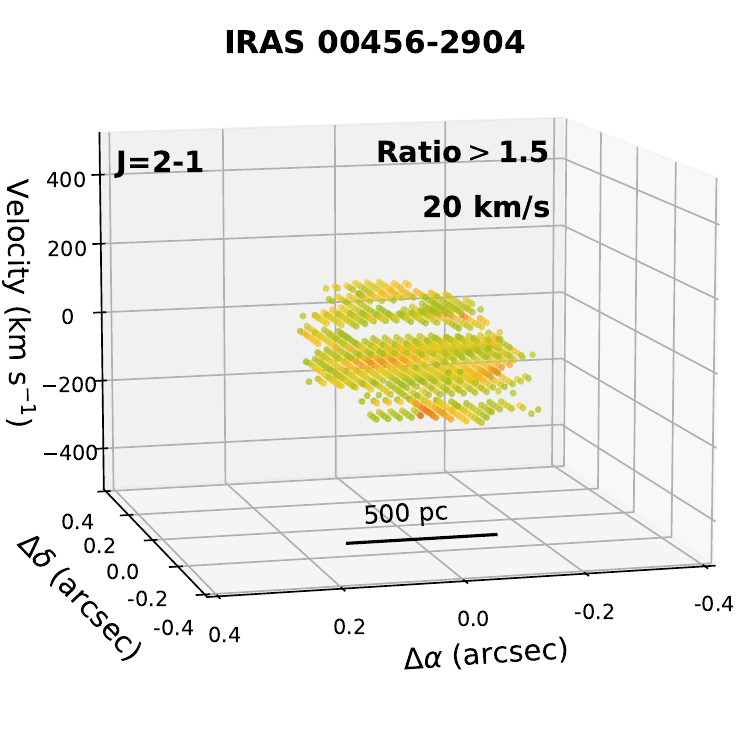} \\
\includegraphics[scale=0.45]{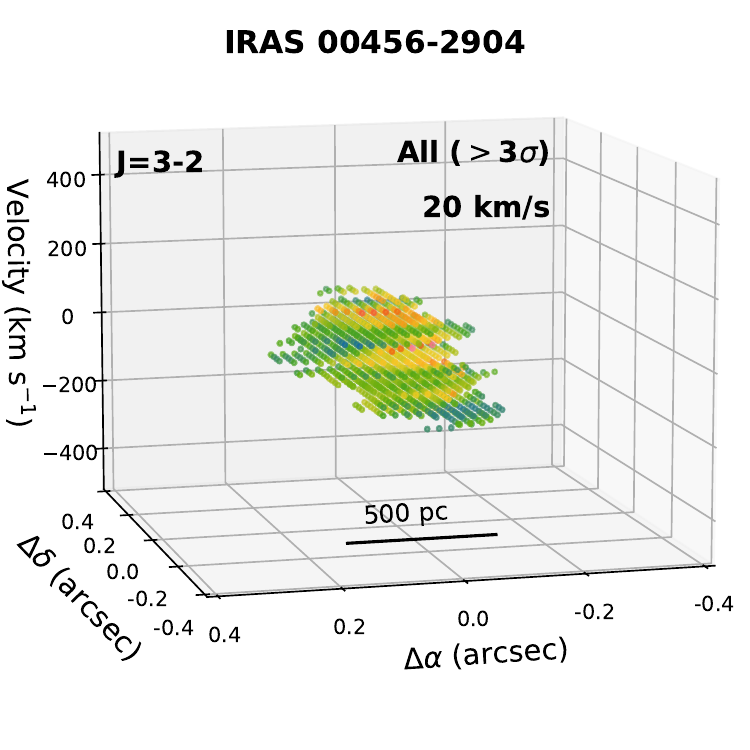} 
\includegraphics[scale=0.45]{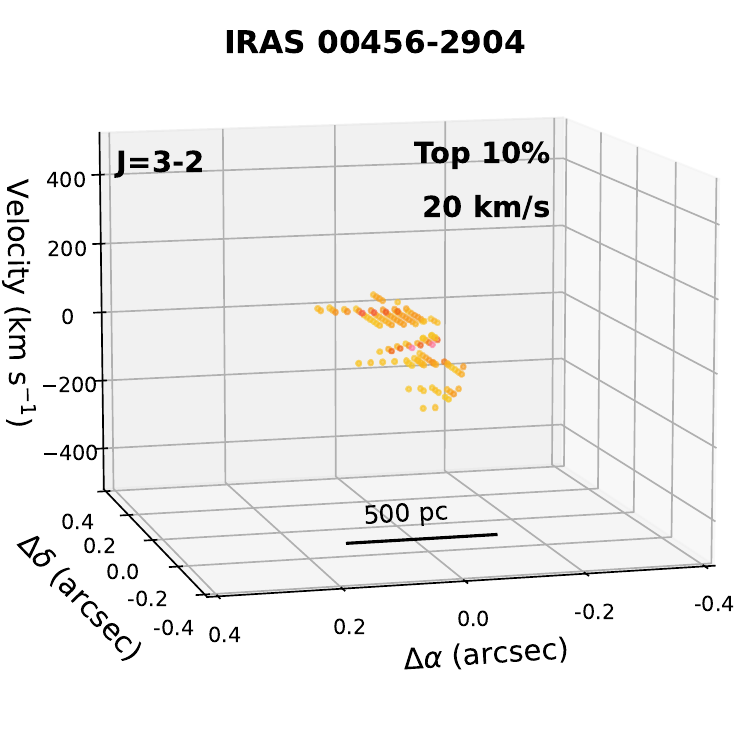} 
\includegraphics[scale=0.45]{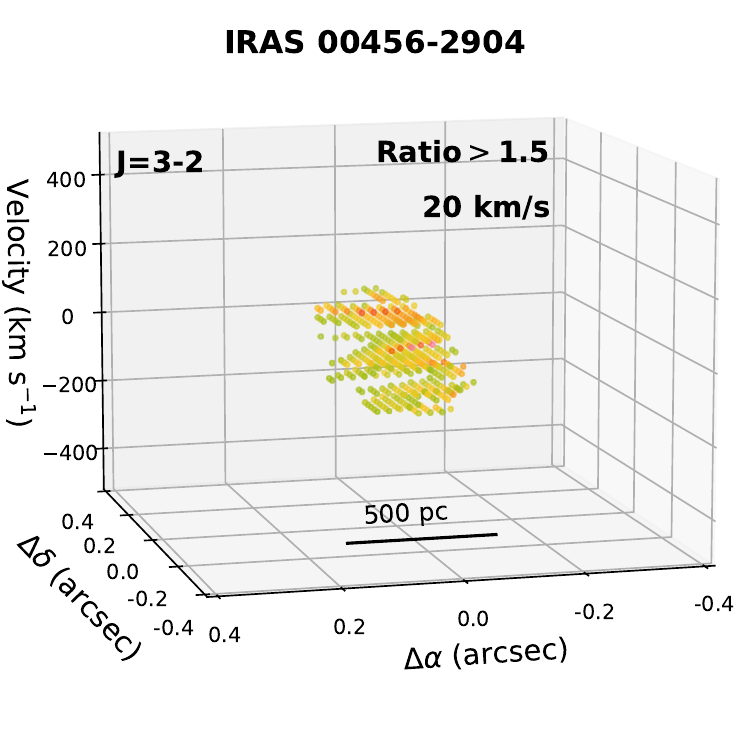} \\
\includegraphics[scale=0.45]{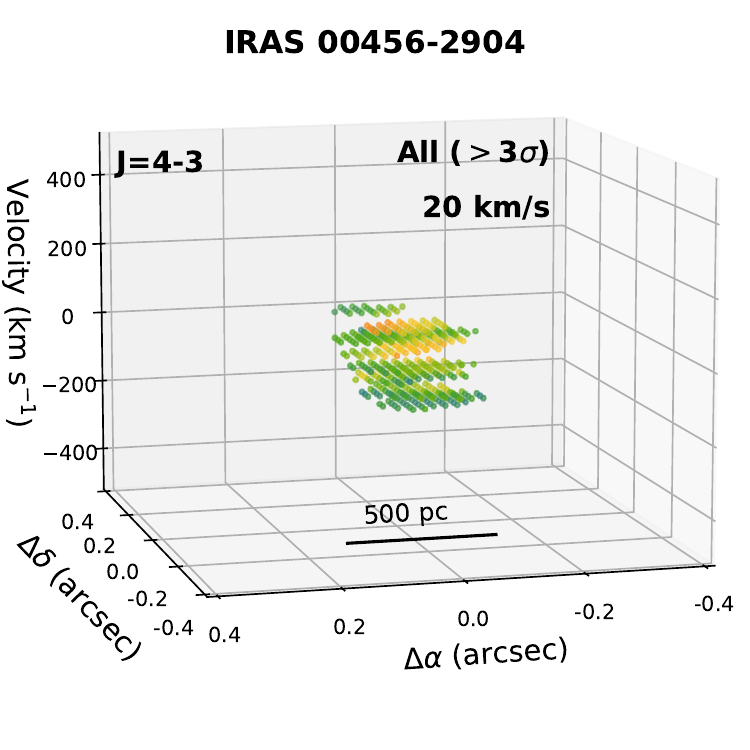} 
\includegraphics[scale=0.45]{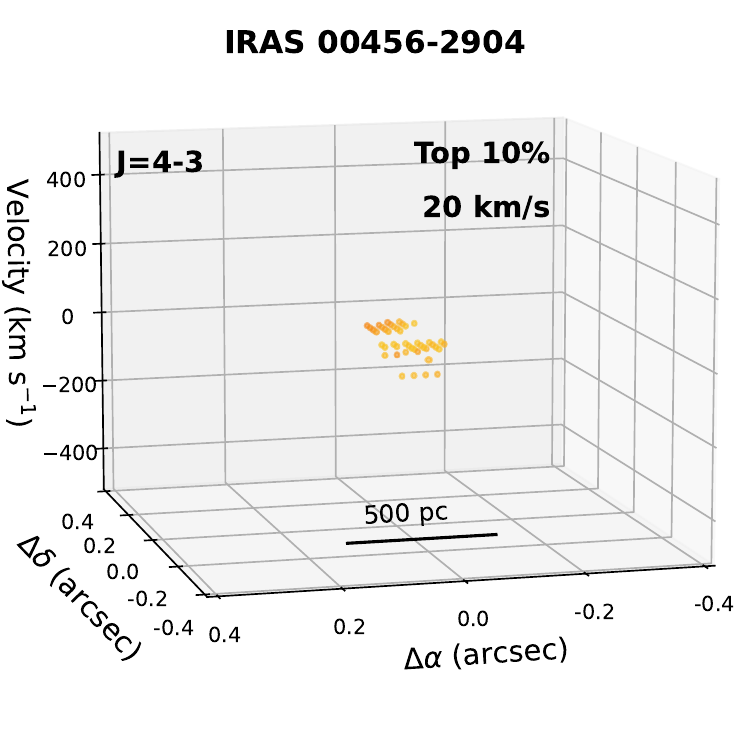} 
\includegraphics[scale=0.45]{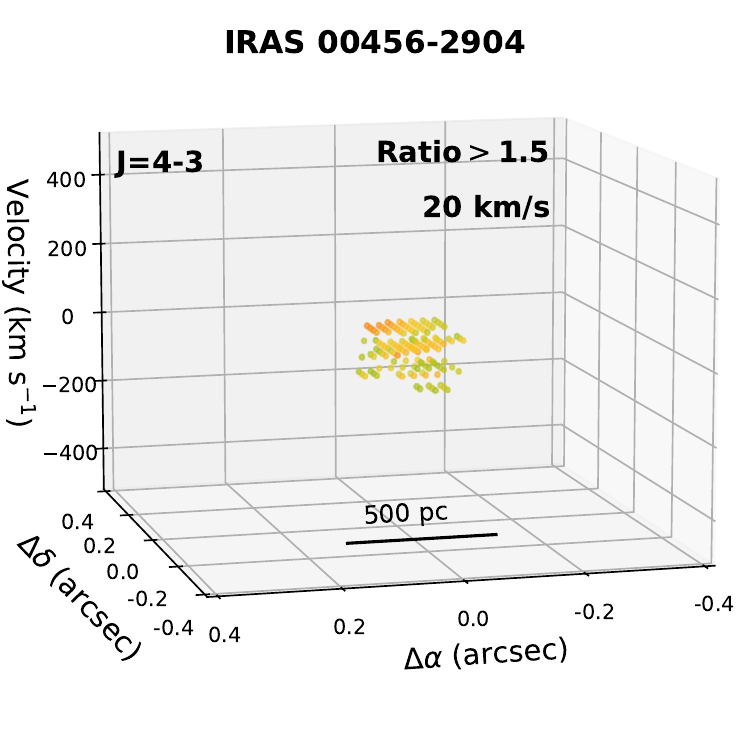} \\
\includegraphics[scale=0.45]{f1-9common.pdf} \\
\end{center}
\vspace{-0.6cm}
\caption{
Same as Figure~\ref{fig:3D1}, but for IRAS~00456$-$2904.
\label{fig:3D3}
}
\end{figure*}

\begin{figure*}[!hbt]
%\vspace*{0.6cm}
\begin{center}
%\hspace{-0.4cm}
\includegraphics[scale=0.45]{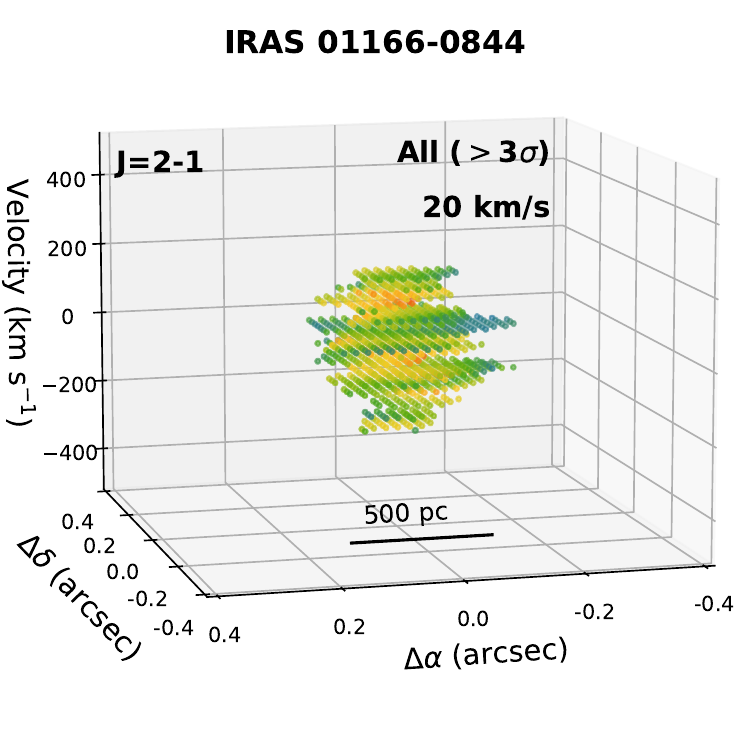} 
\includegraphics[scale=0.45]{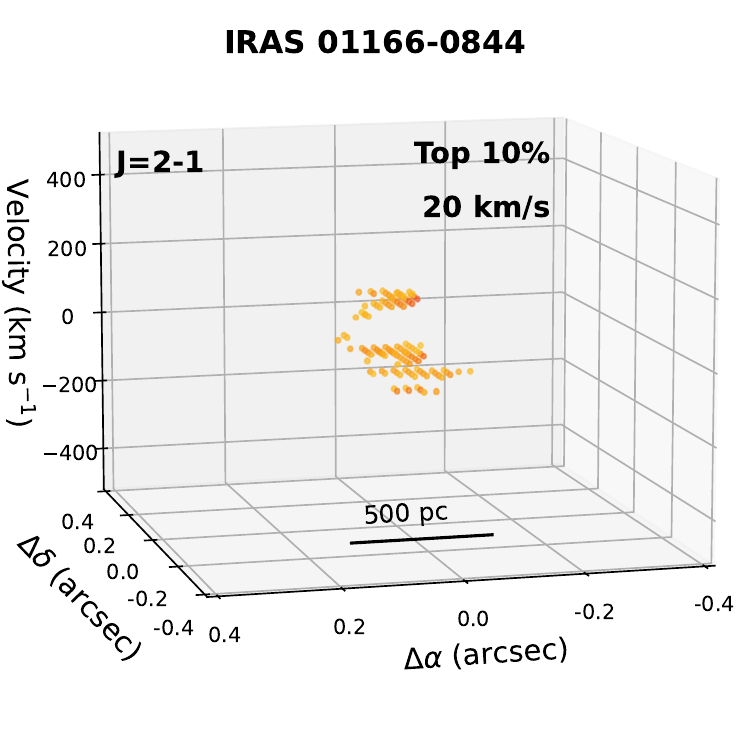} 
\includegraphics[scale=0.45]{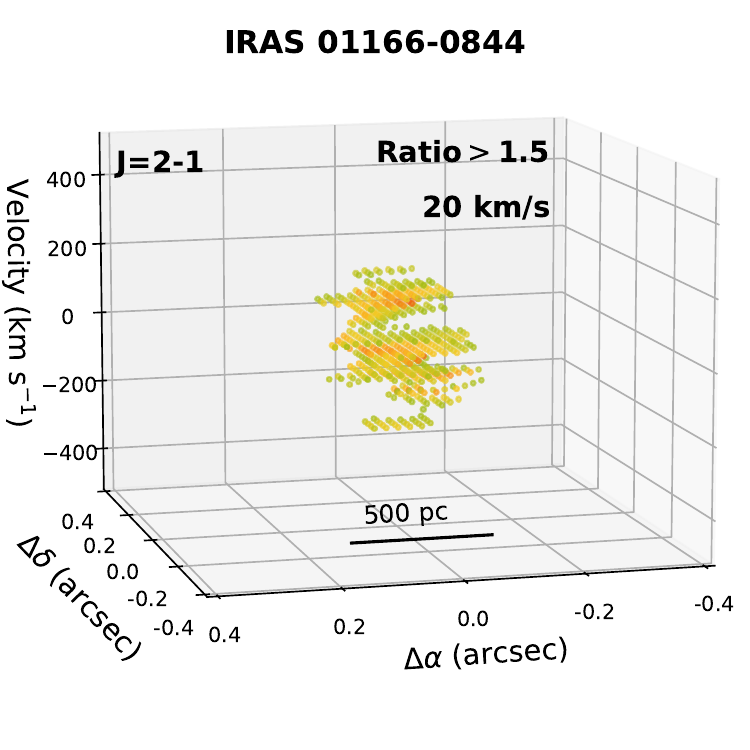} \\
\includegraphics[scale=0.45]{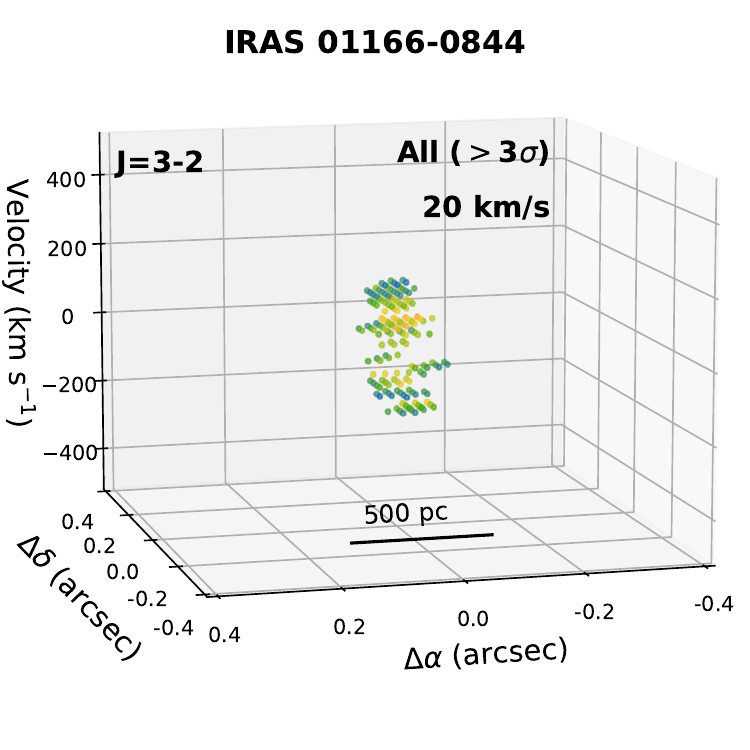} 
\includegraphics[scale=0.45]{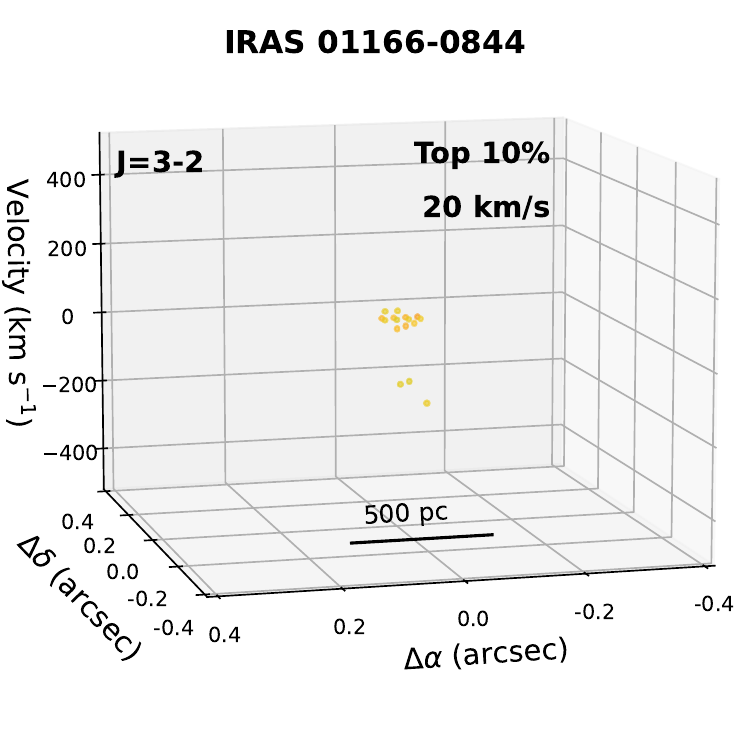} 
\includegraphics[scale=0.45]{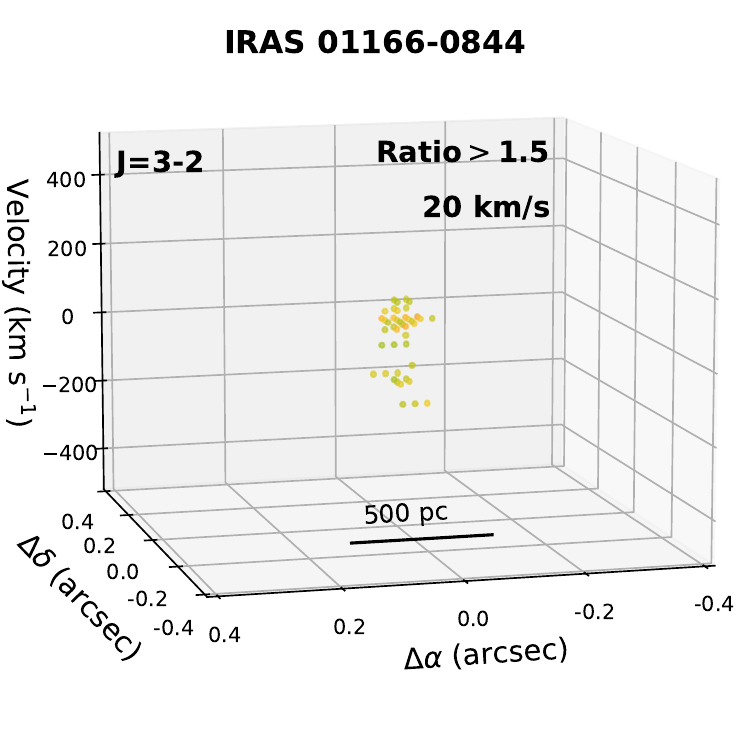} \\
\includegraphics[scale=0.45]{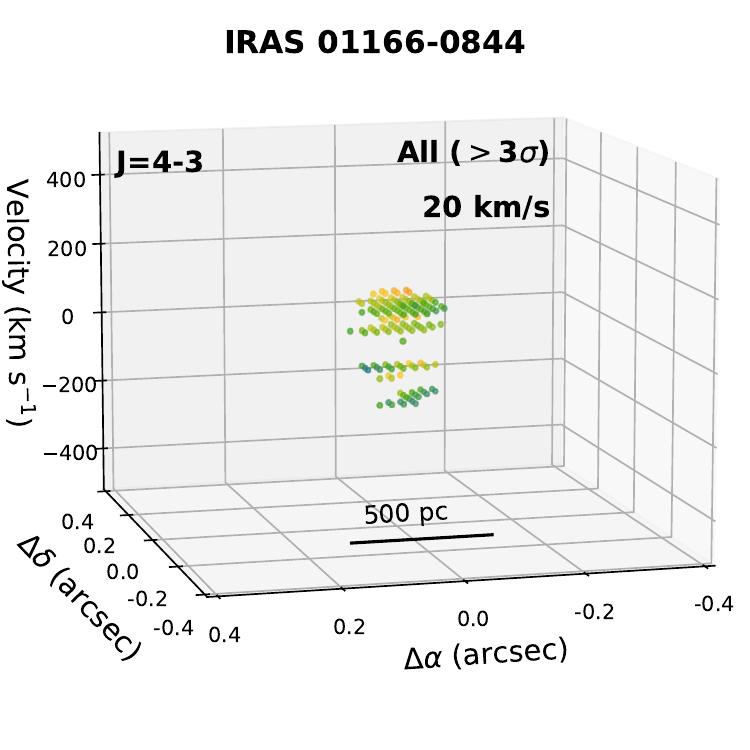} 
\includegraphics[scale=0.45]{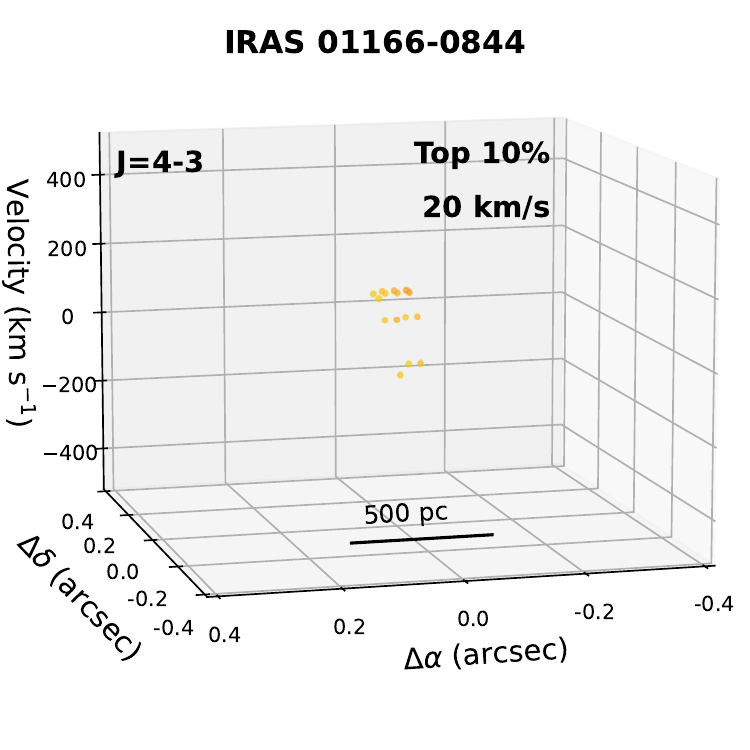} 
\includegraphics[scale=0.45]{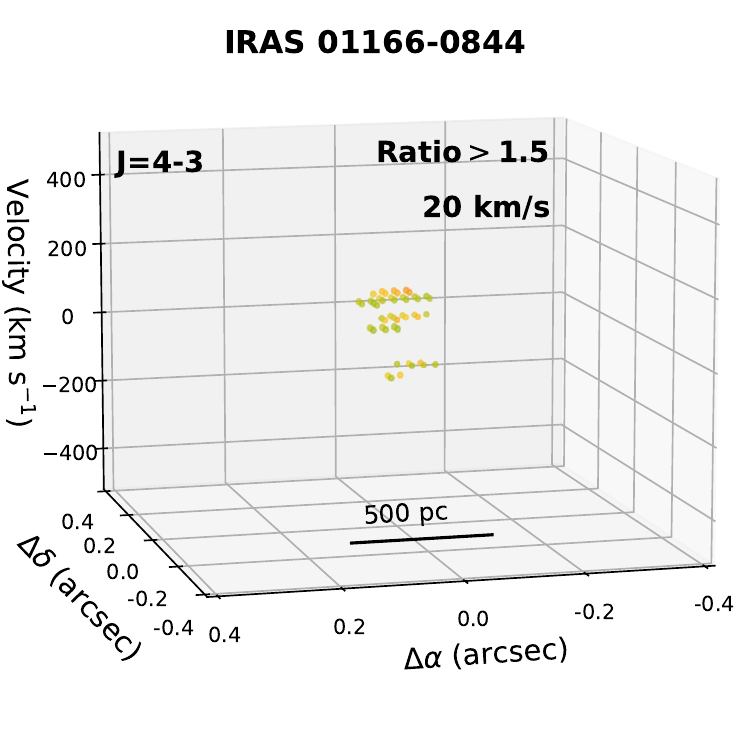} \\
\includegraphics[scale=0.45]{f1-9common.pdf} \\
\end{center}
\vspace{-0.6cm}
\caption{
Same as Figure~\ref{fig:3D1}, but for IRAS~01166$-$0844.
\label{fig:3D4}
}
\end{figure*}

\begin{figure*}[!hbt]
%\vspace*{0.6cm}
\begin{center}
%\hspace{-0.4cm}
\includegraphics[scale=0.45]{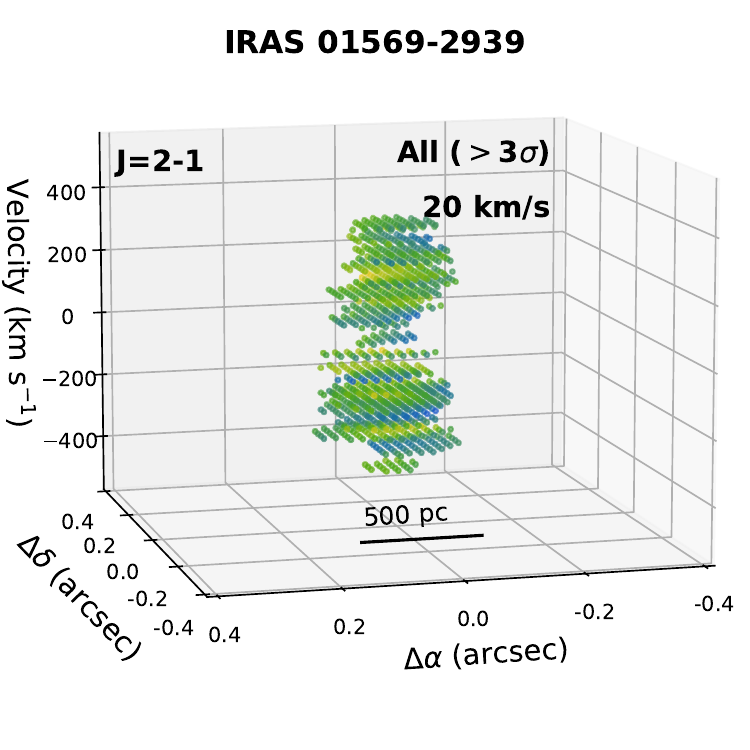} 
\includegraphics[scale=0.45]{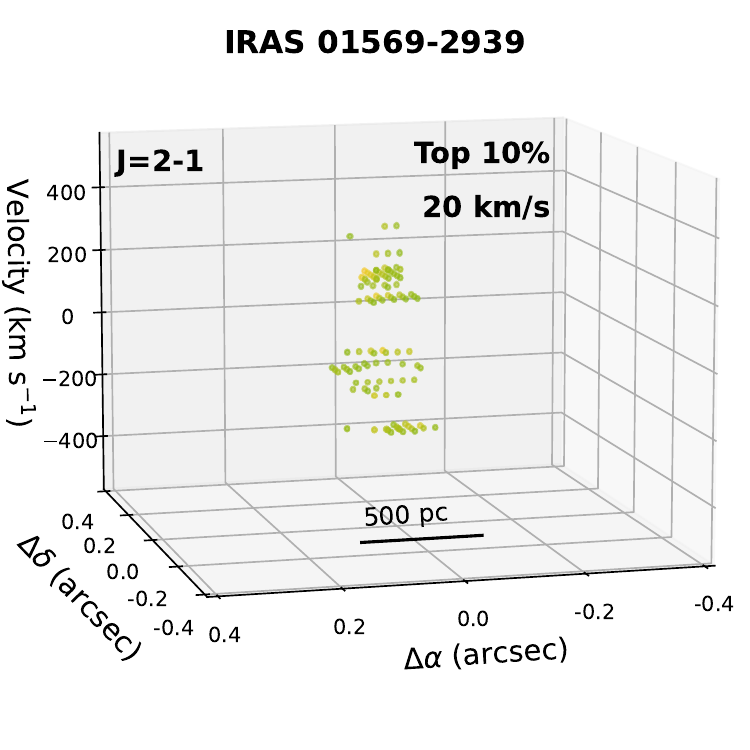} 
\includegraphics[scale=0.45]{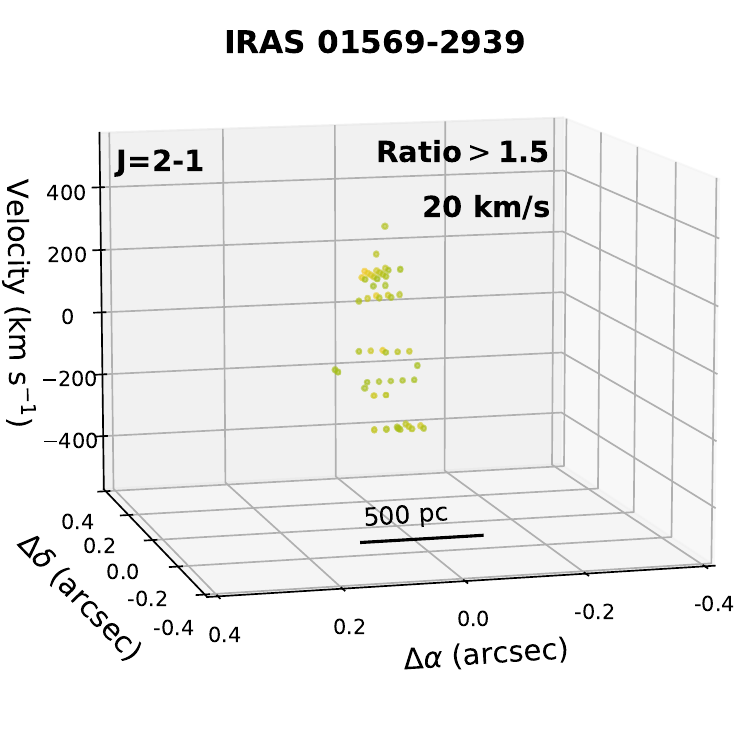} \\
\includegraphics[scale=0.45]{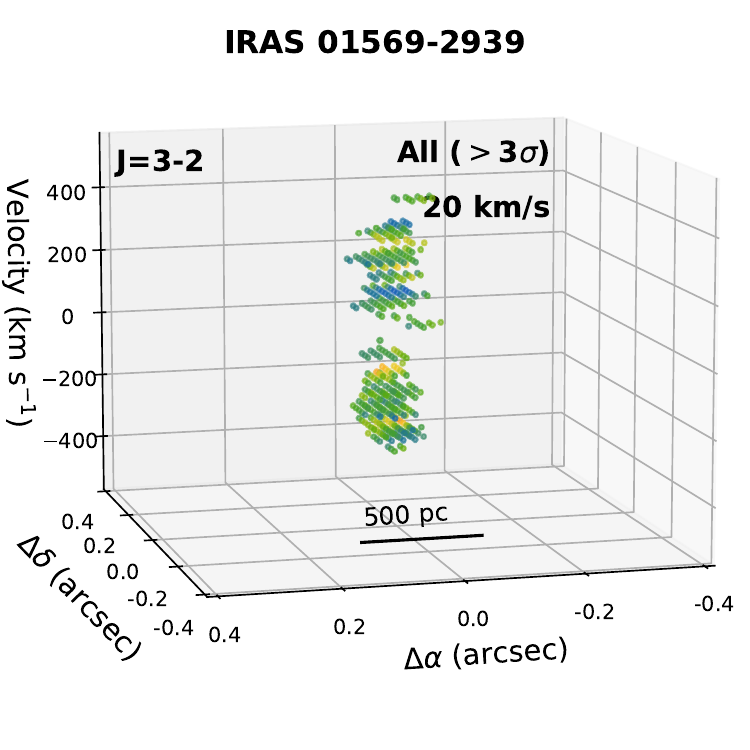} 
\includegraphics[scale=0.45]{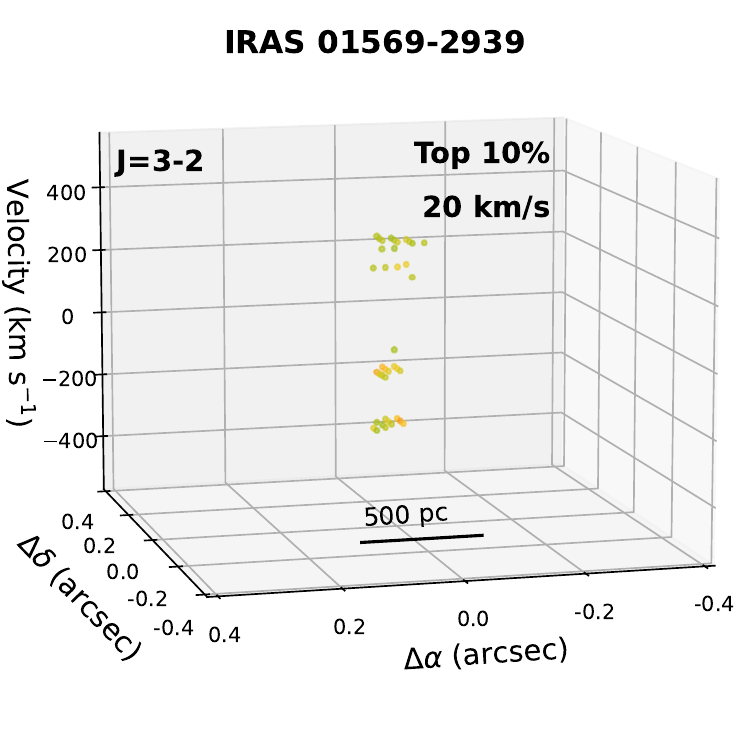} 
\includegraphics[scale=0.45]{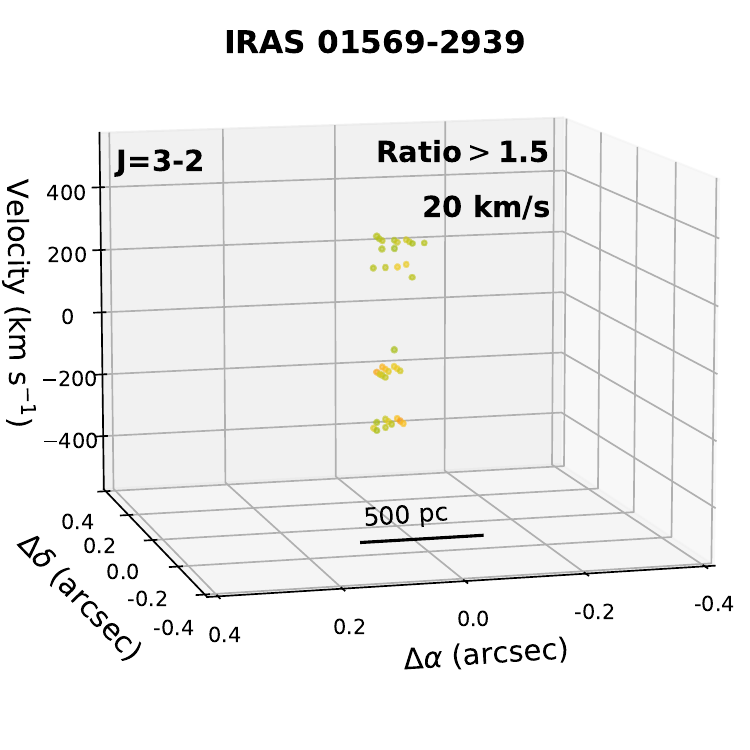} \\
\includegraphics[scale=0.45]{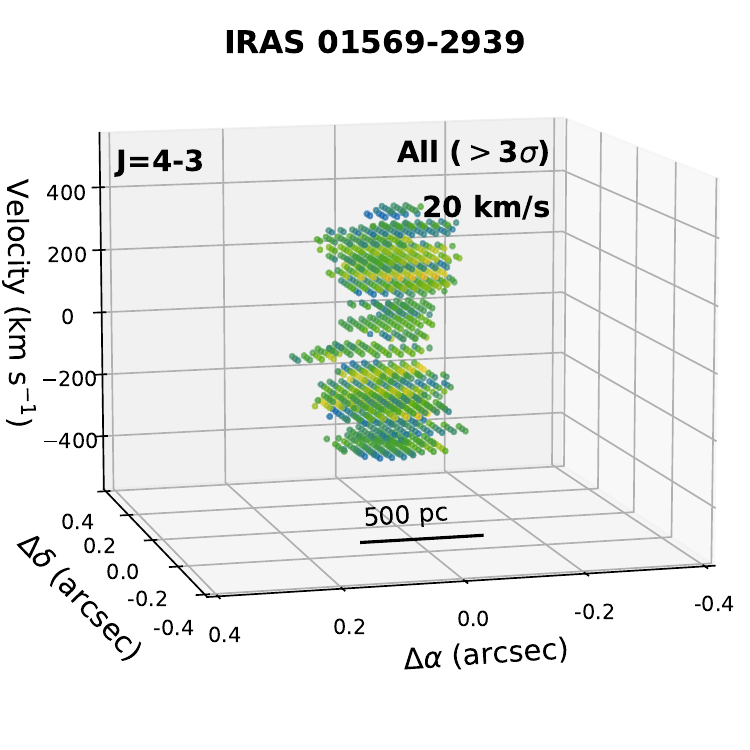} 
\includegraphics[scale=0.45]{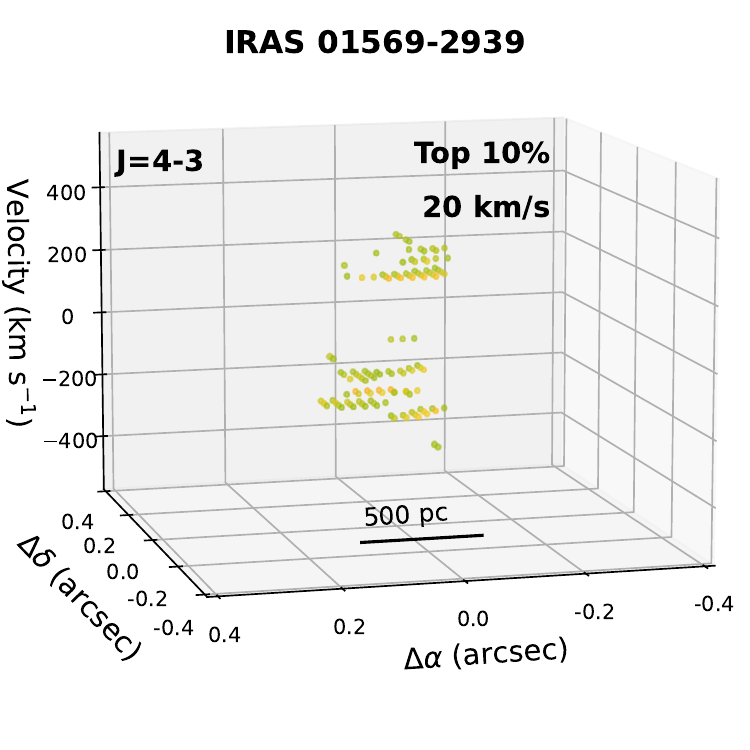} 
\includegraphics[scale=0.45]{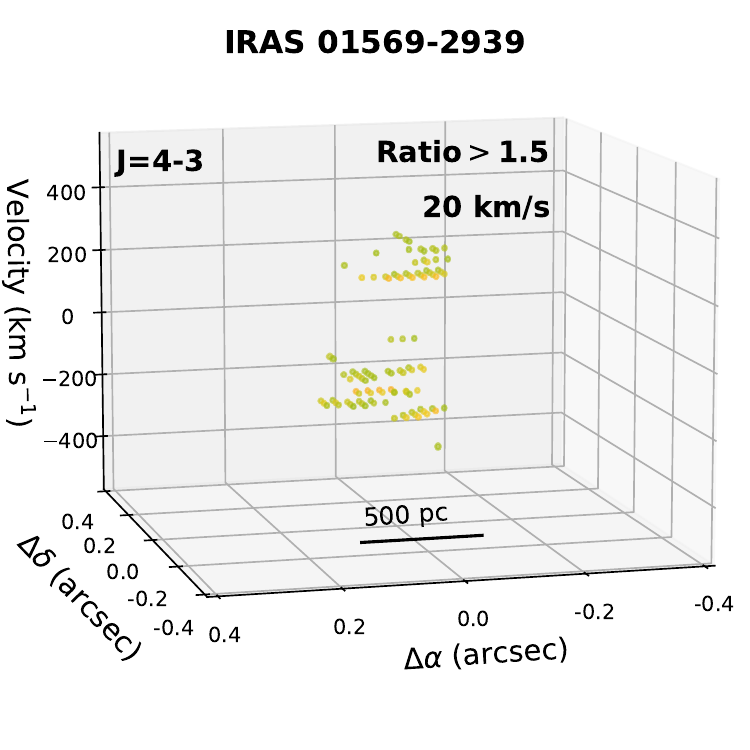} \\
\includegraphics[scale=0.45]{f1-9common.pdf} \\
\end{center}
\vspace{-0.6cm}
\caption{
Same as Figure~\ref{fig:3D1}, but for IRAS~01569$-$2939.
\label{fig:3D5}
}
\end{figure*}

\begin{figure*}[!hbt]
%\vspace*{0.6cm}
\begin{center}
%\hspace{-0.4cm}
\includegraphics[scale=0.45]{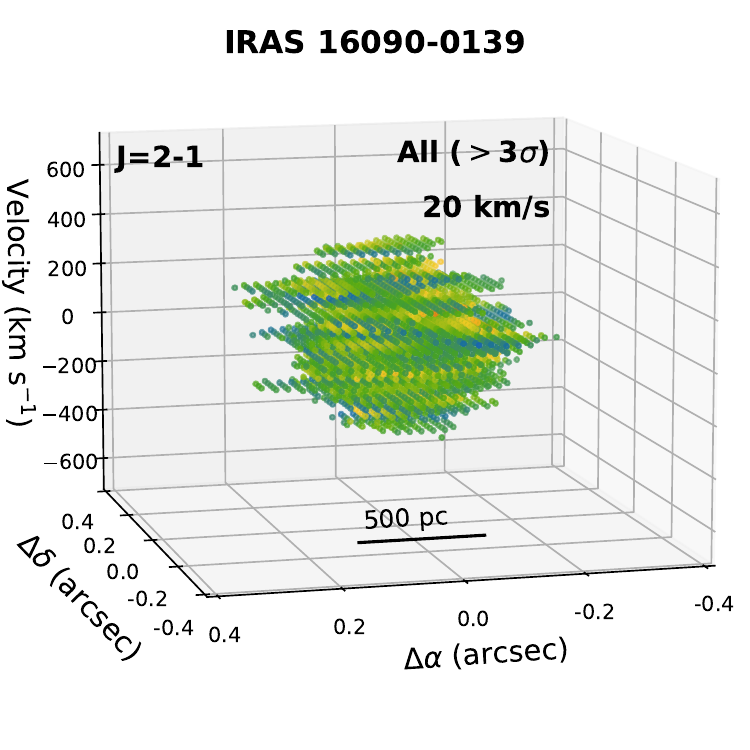} 
\includegraphics[scale=0.45]{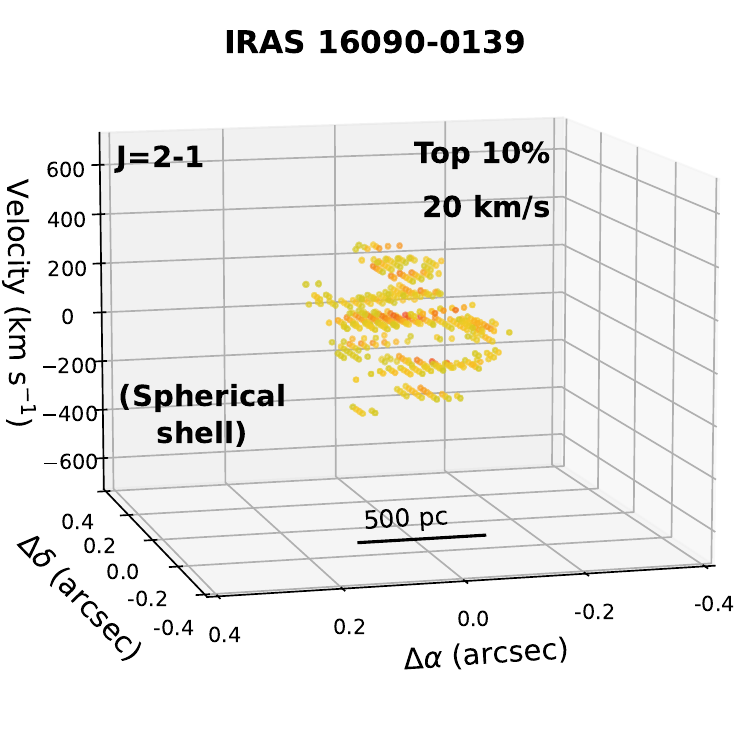} 
\includegraphics[scale=0.45]{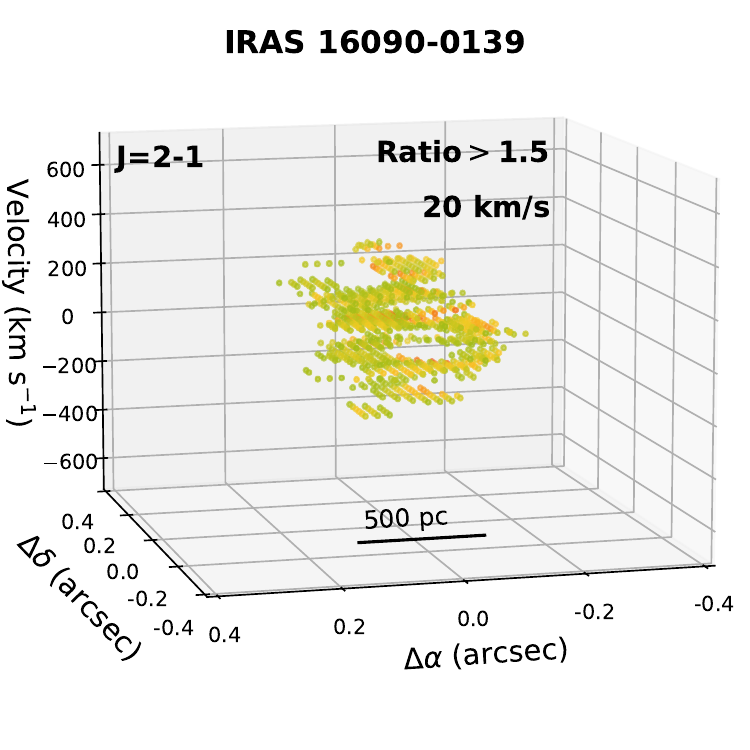} \\
\includegraphics[scale=0.45]{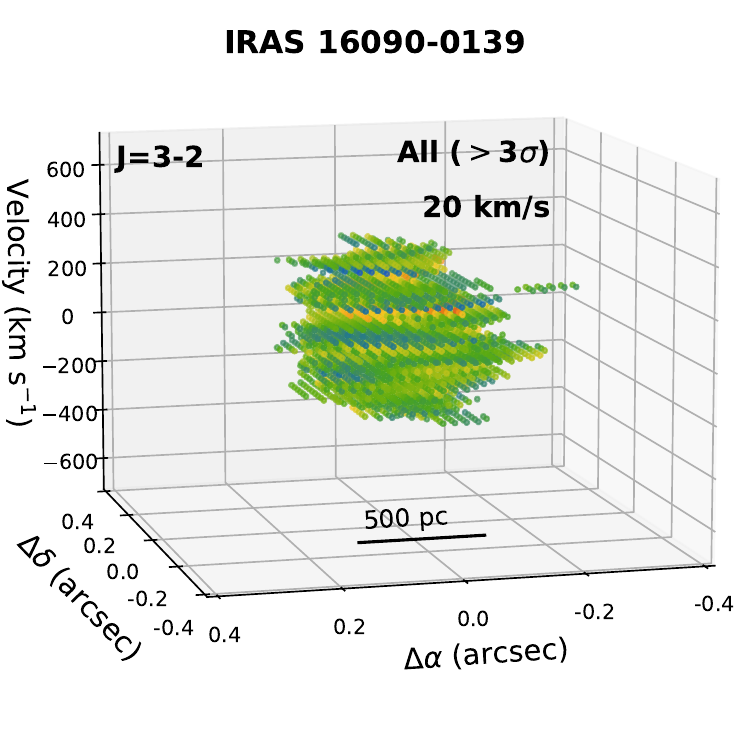} 
\includegraphics[scale=0.45]{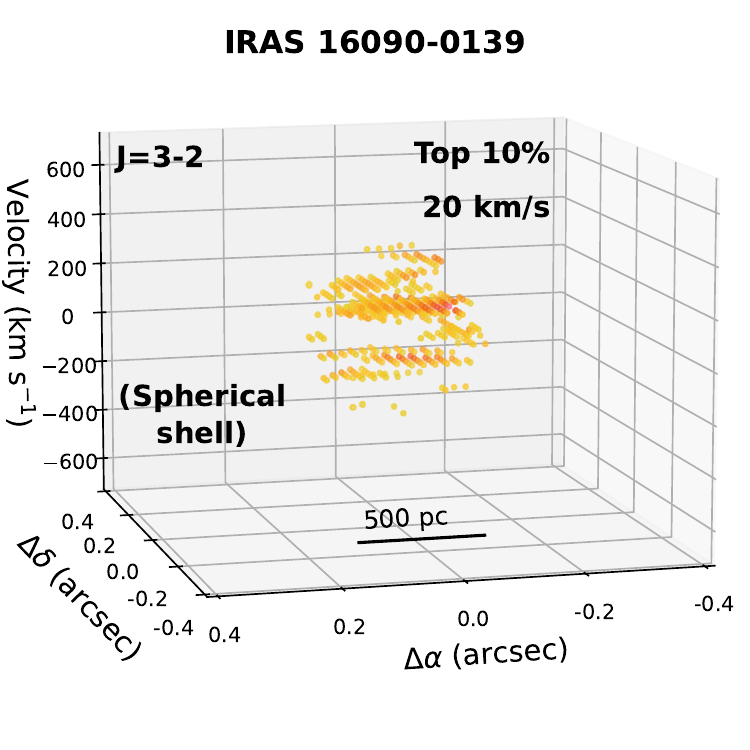} 
\includegraphics[scale=0.45]{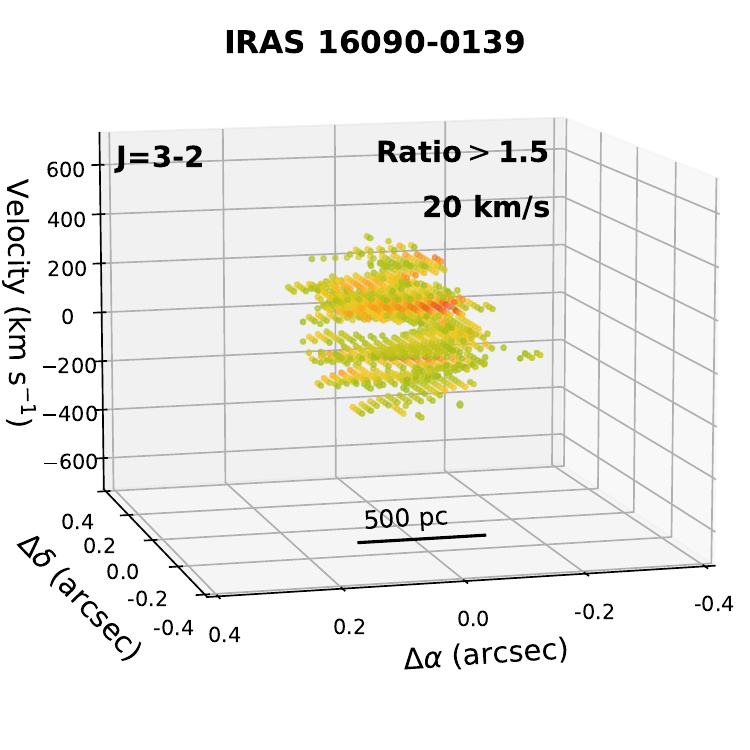} \\
\includegraphics[scale=0.45]{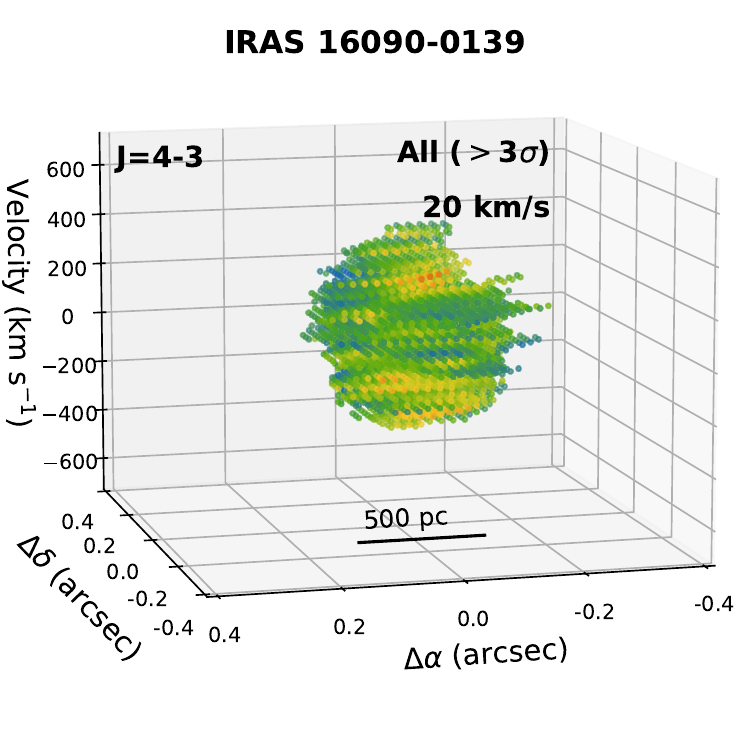} 
\includegraphics[scale=0.45]{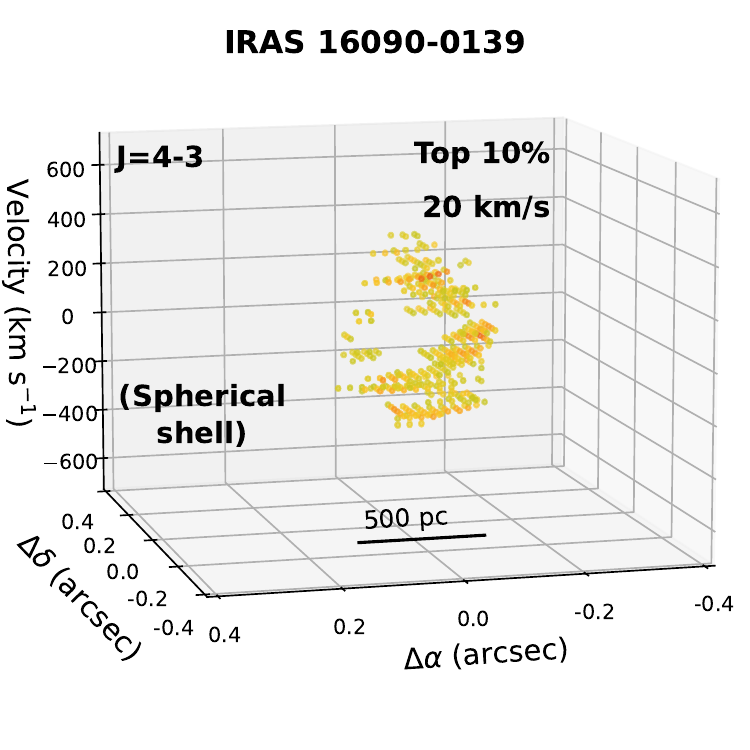} 
\includegraphics[scale=0.45]{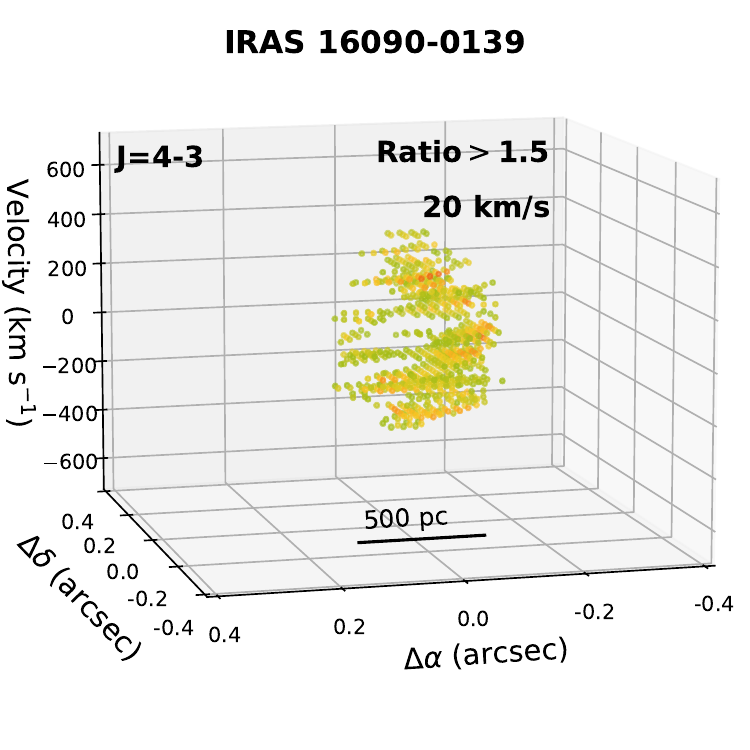} \\
\includegraphics[scale=0.45]{f1-9common.pdf} \\
\end{center}
\vspace{-0.6cm}
\caption{
Same as Figure~\ref{fig:3D1}, but for IRAS~16090$-$0139.
\label{fig:3D6}
}
\end{figure*}

\begin{figure*}[!hbt]
%\vspace*{0.6cm}
\begin{center}
%\hspace{-0.4cm}
\includegraphics[scale=0.45]{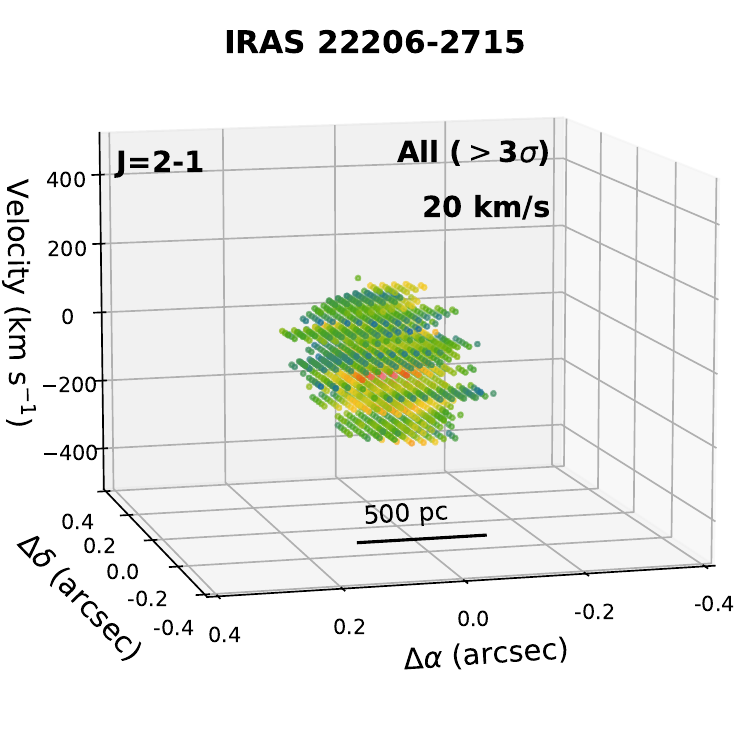} 
\includegraphics[scale=0.45]{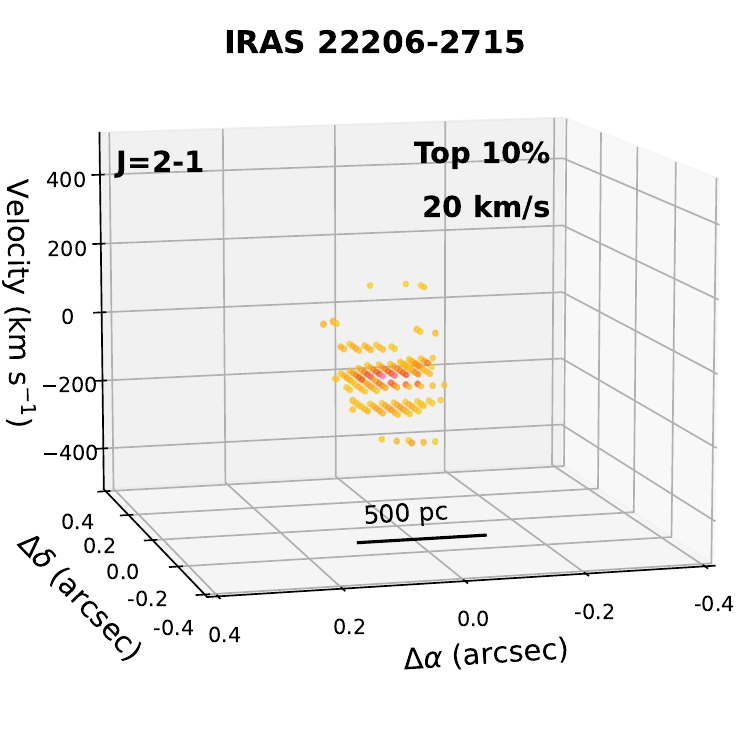} 
\includegraphics[scale=0.45]{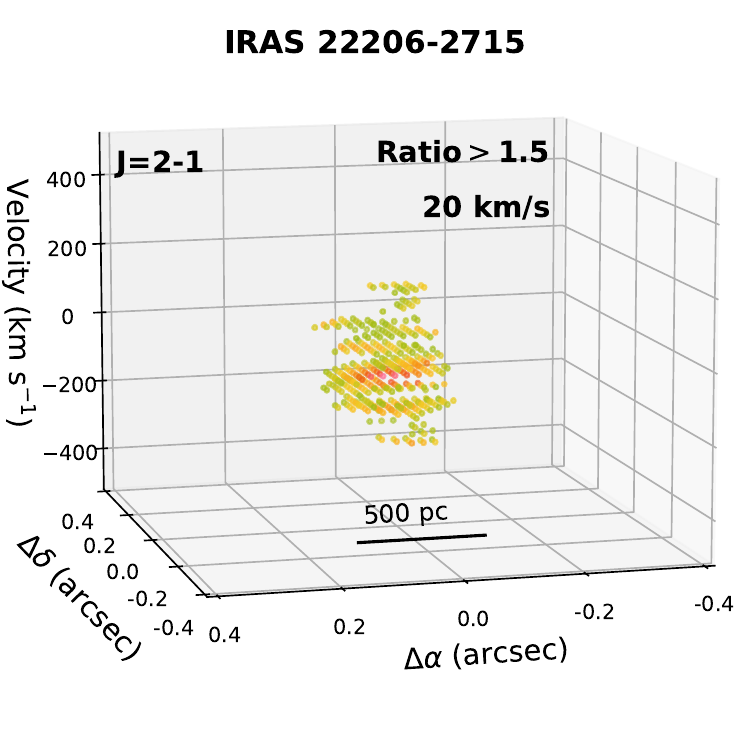} \\
\includegraphics[scale=0.45]{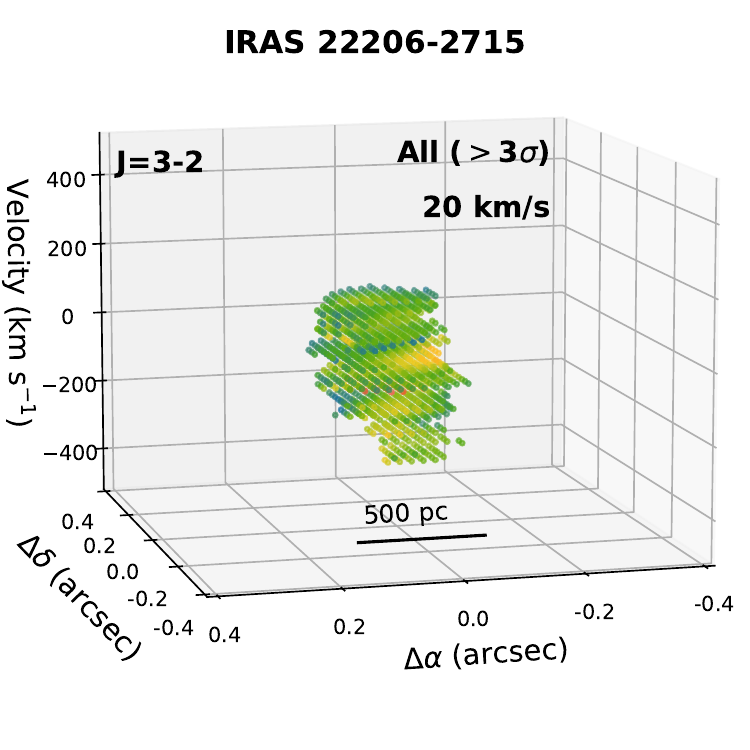} 
\includegraphics[scale=0.45]{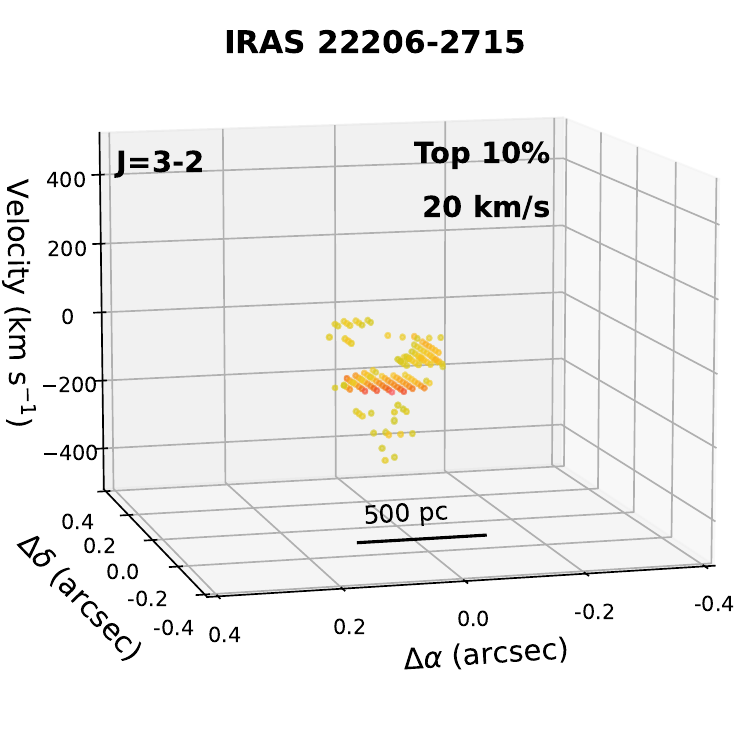} 
\includegraphics[scale=0.45]{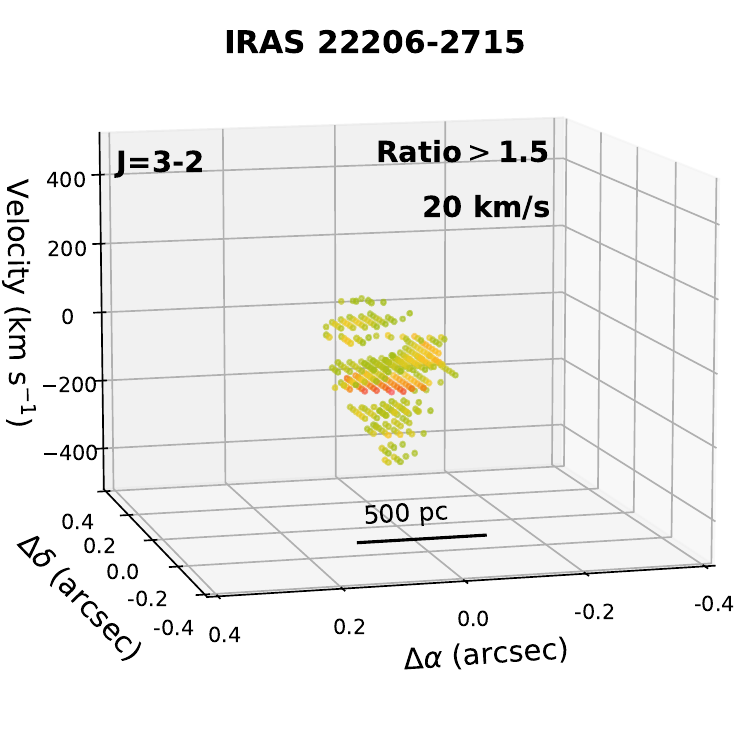} \\
\includegraphics[scale=0.45]{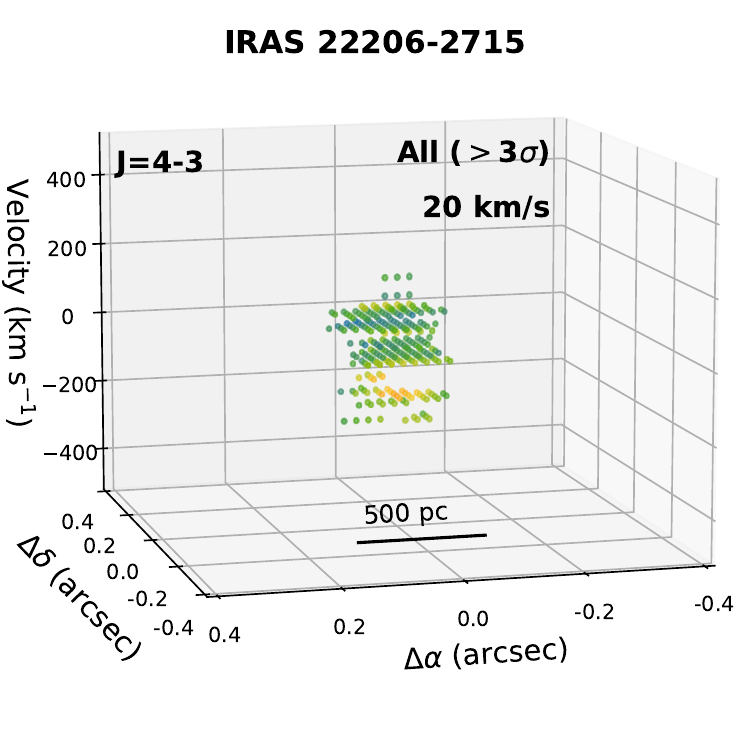} 
\includegraphics[scale=0.45]{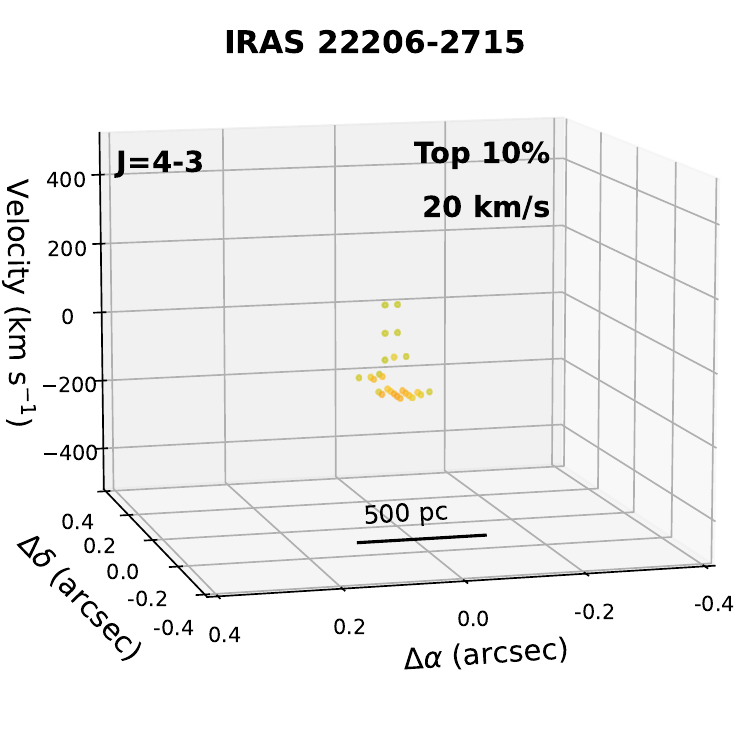} 
\includegraphics[scale=0.45]{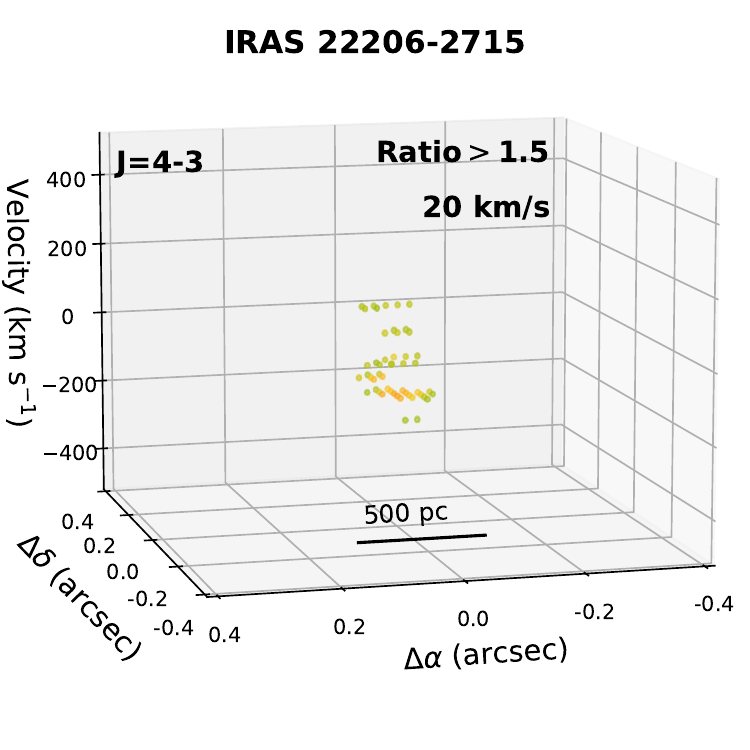} \\
\includegraphics[scale=0.45]{f1-9common.pdf} \\
\end{center}
\vspace{-0.6cm}
\caption{
Same as Figure~\ref{fig:3D1}, but for IRAS~22206$-$2715.
\label{fig:3D7}
}
\end{figure*}

\begin{figure*}[!hbt]
%\vspace*{0.6cm}
\begin{center}
%\hspace{-0.4cm}
\includegraphics[scale=0.45]{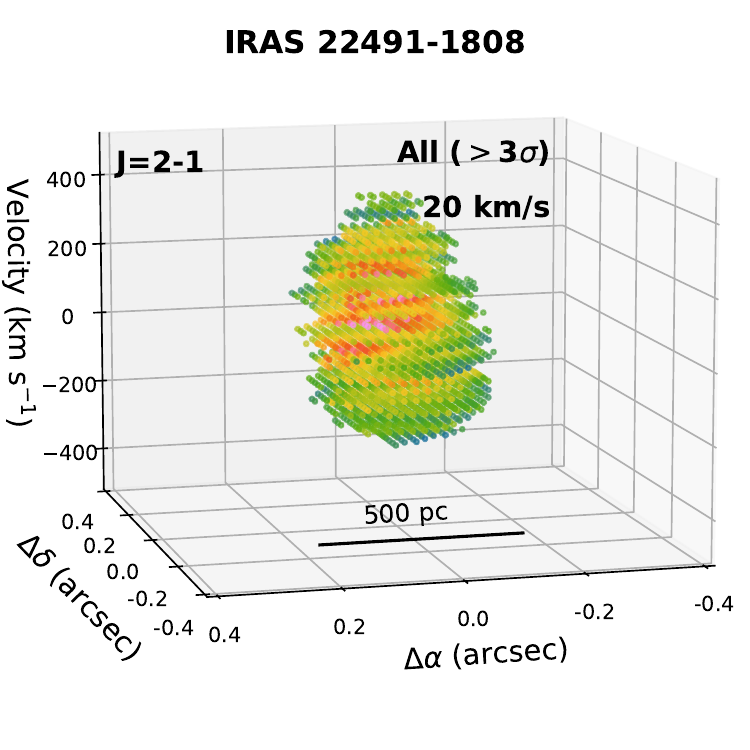} 
\includegraphics[scale=0.45]{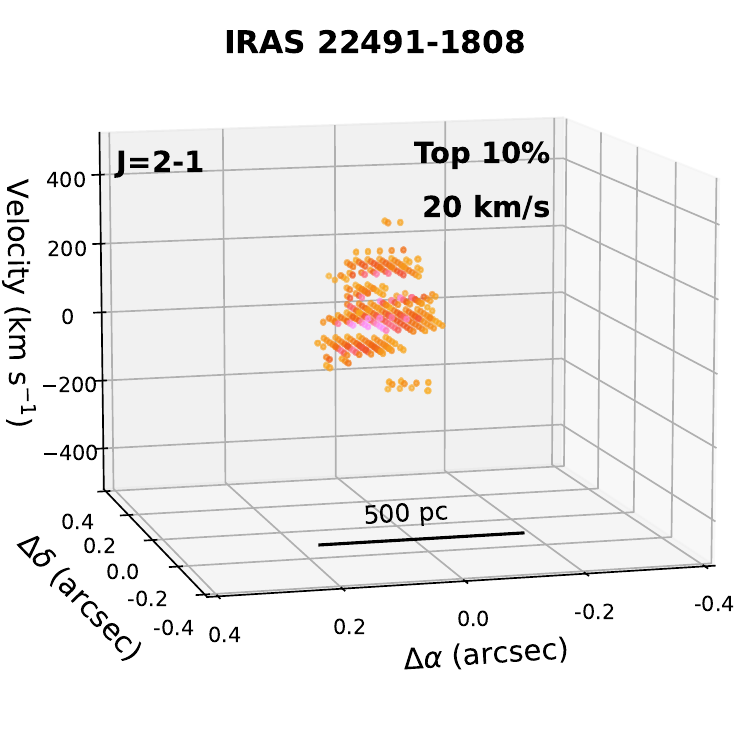} 
\includegraphics[scale=0.45]{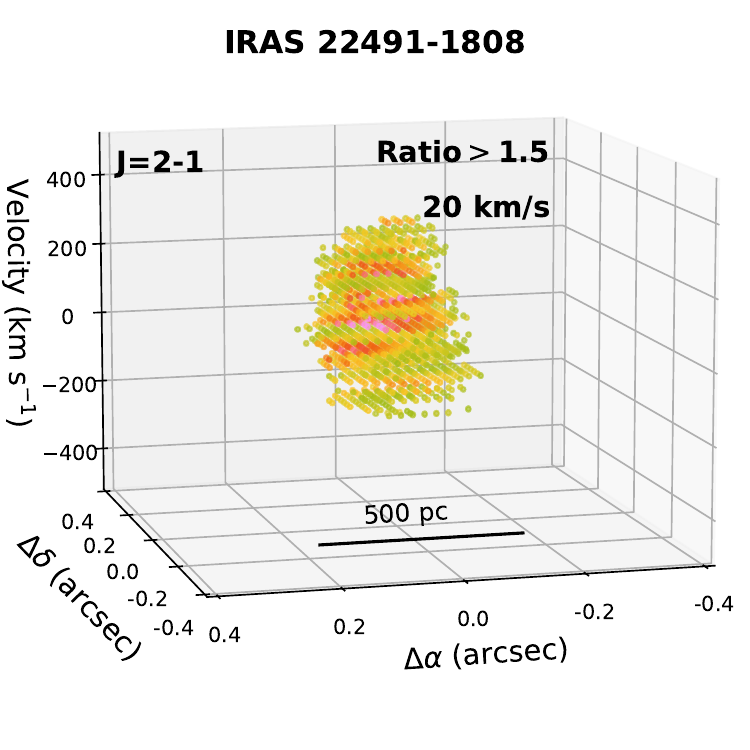} \\
\includegraphics[scale=0.45]{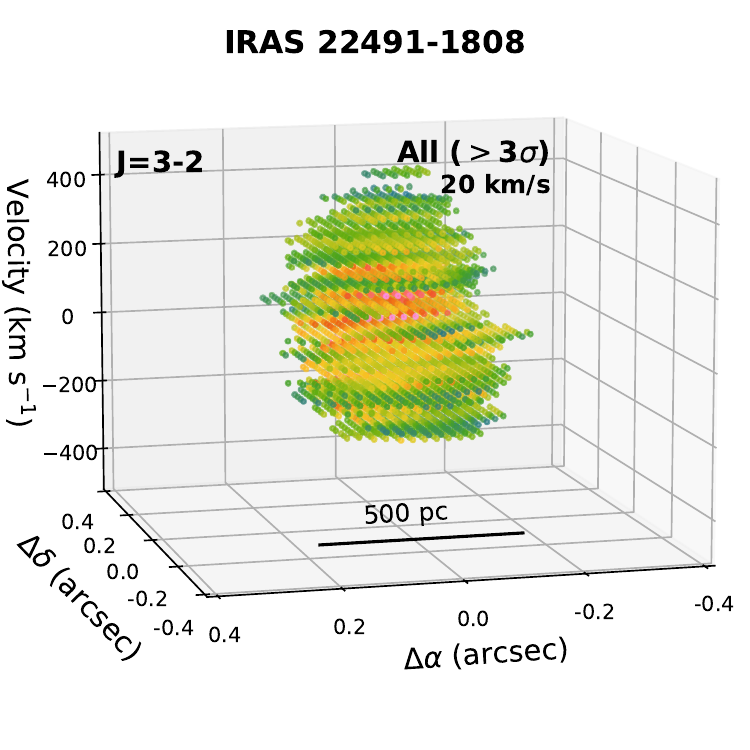} 
\includegraphics[scale=0.45]{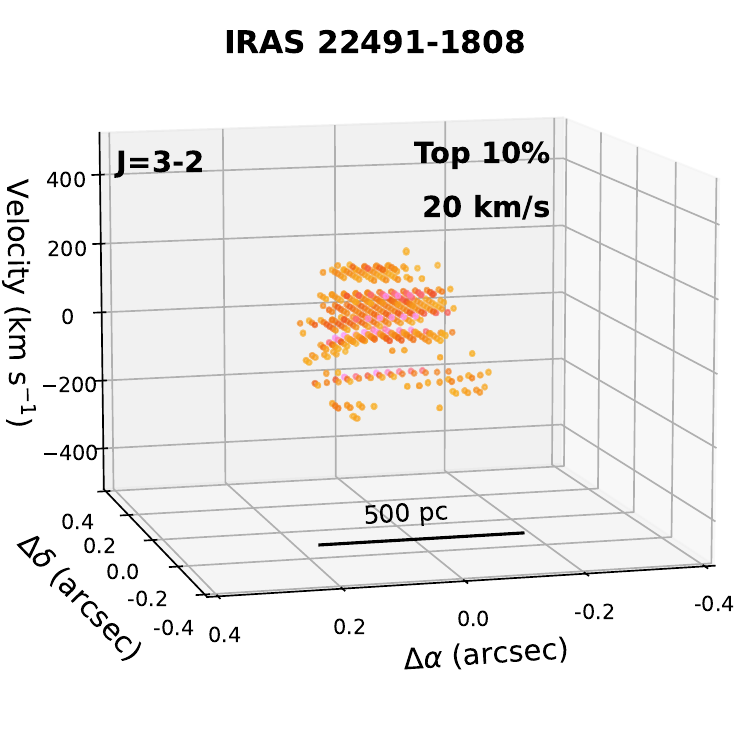} 
\includegraphics[scale=0.45]{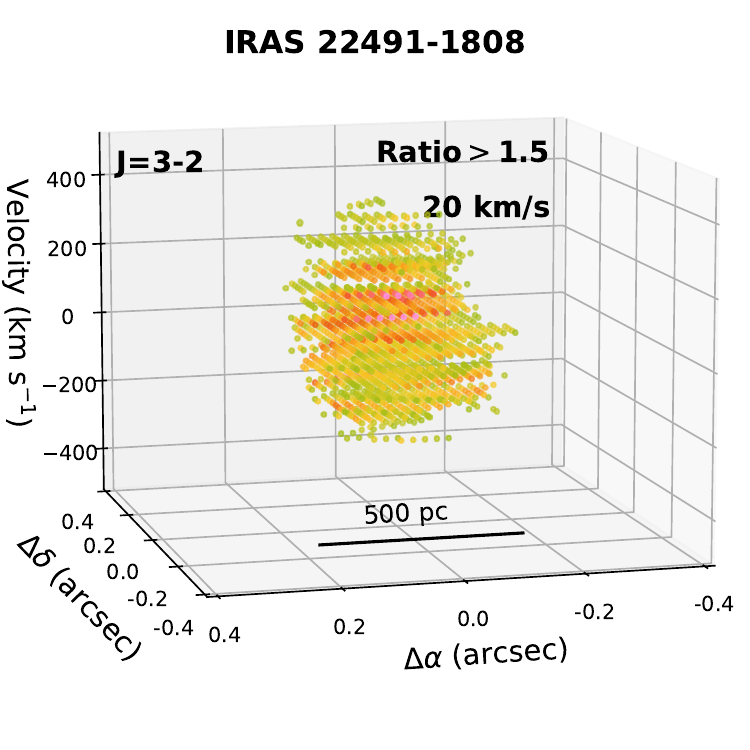} \\
\includegraphics[scale=0.45]{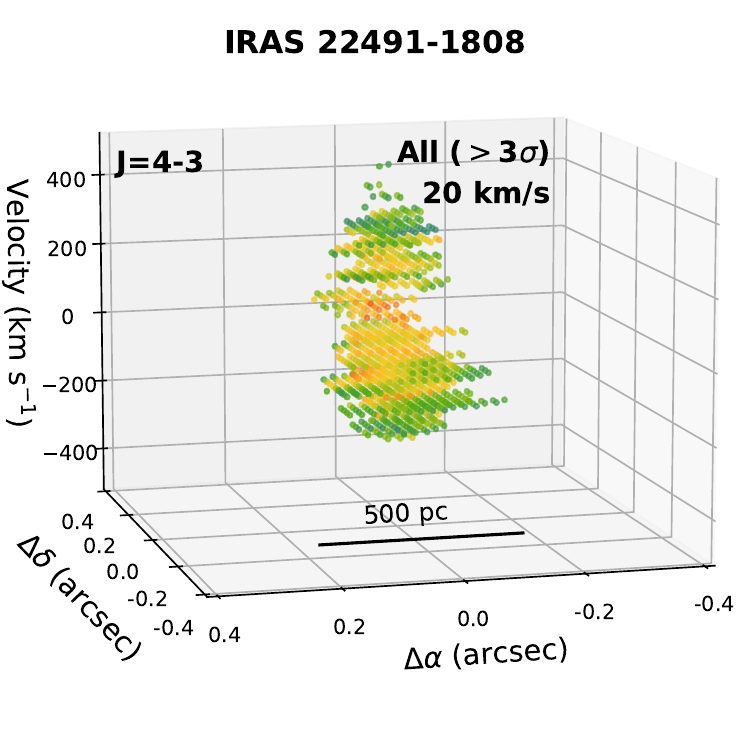} 
\includegraphics[scale=0.45]{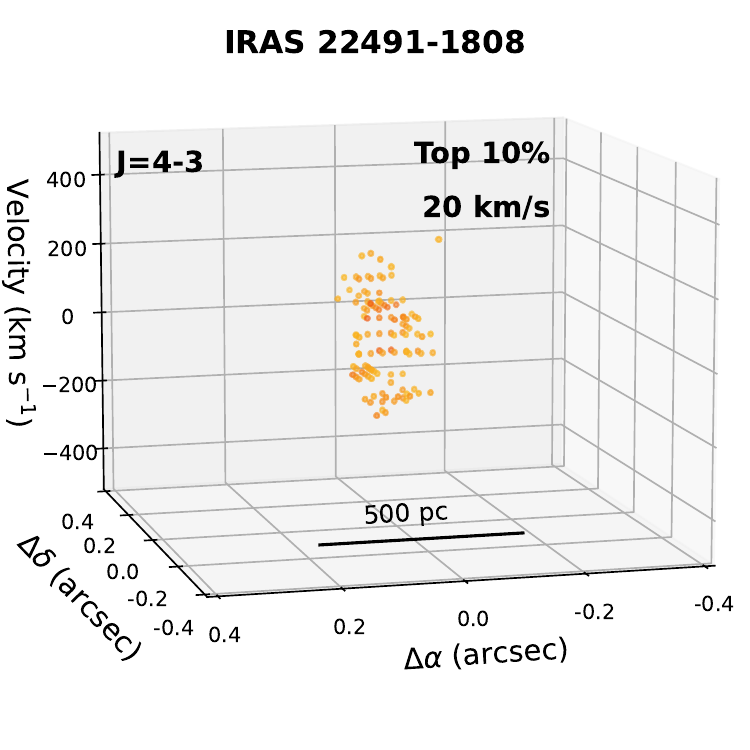} 
\includegraphics[scale=0.45]{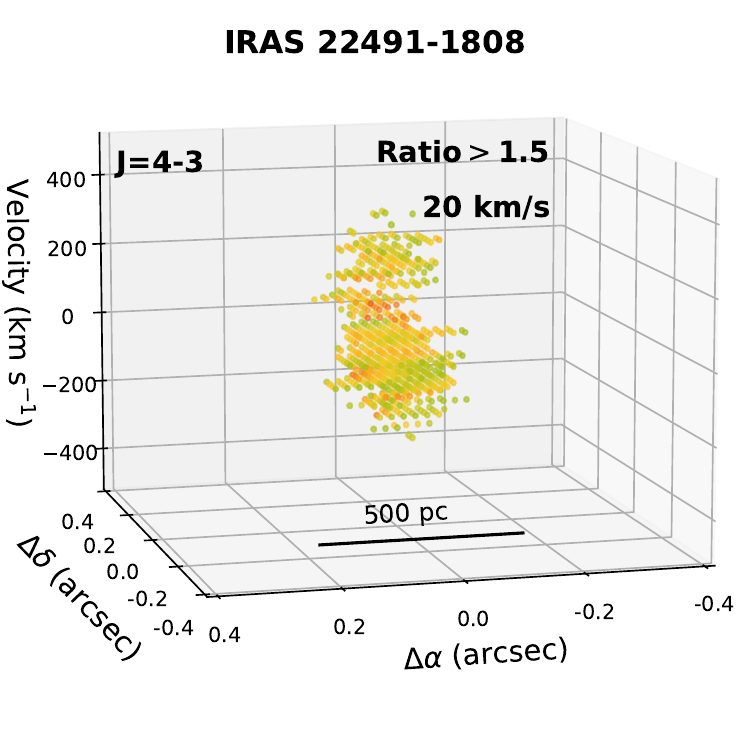} \\
\includegraphics[scale=0.45]{f1-9common.pdf} \\
\end{center}
\vspace{-0.6cm}
\caption{
Same as Figure~\ref{fig:3D1}, but for IRAS~22491$-$1808.
\label{fig:3D8}
}
\end{figure*}

\begin{figure*}
\begin{center}
\includegraphics[scale=0.45]{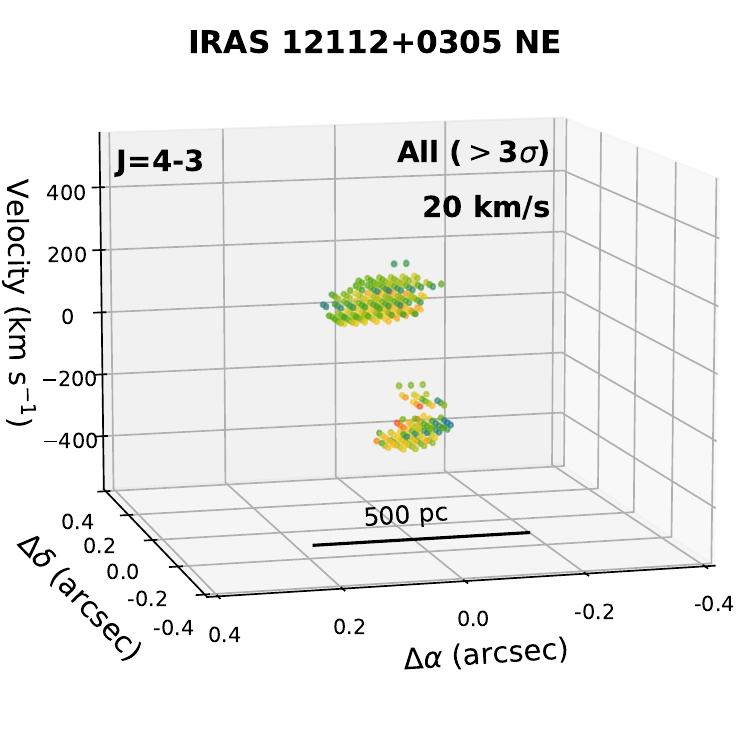} 
\includegraphics[scale=0.45]{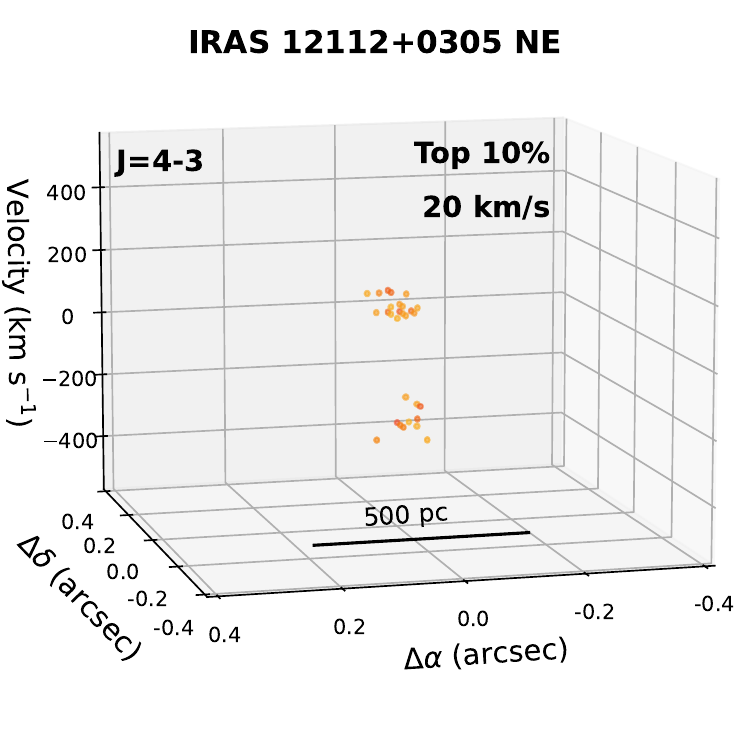} 
\includegraphics[scale=0.45]{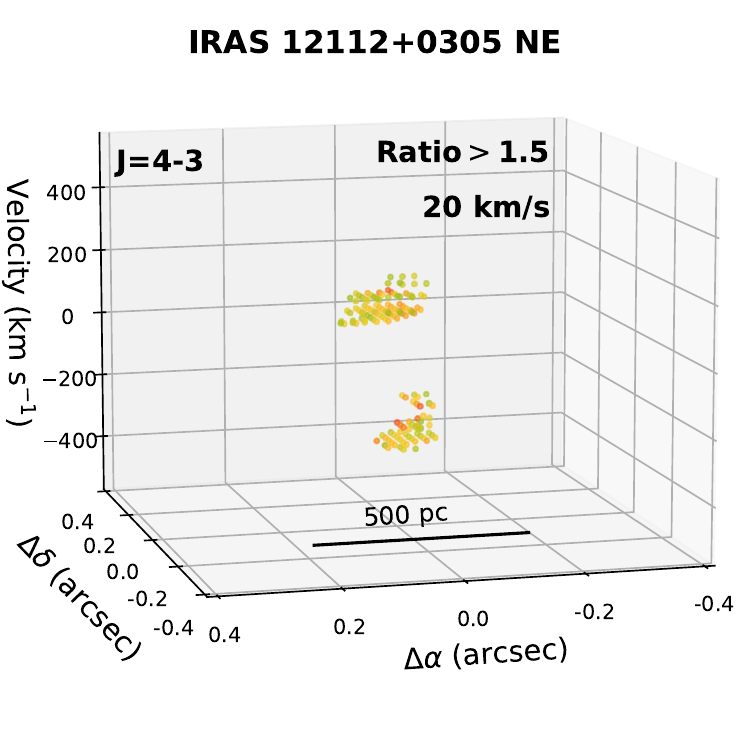} \\
\includegraphics[scale=0.45]{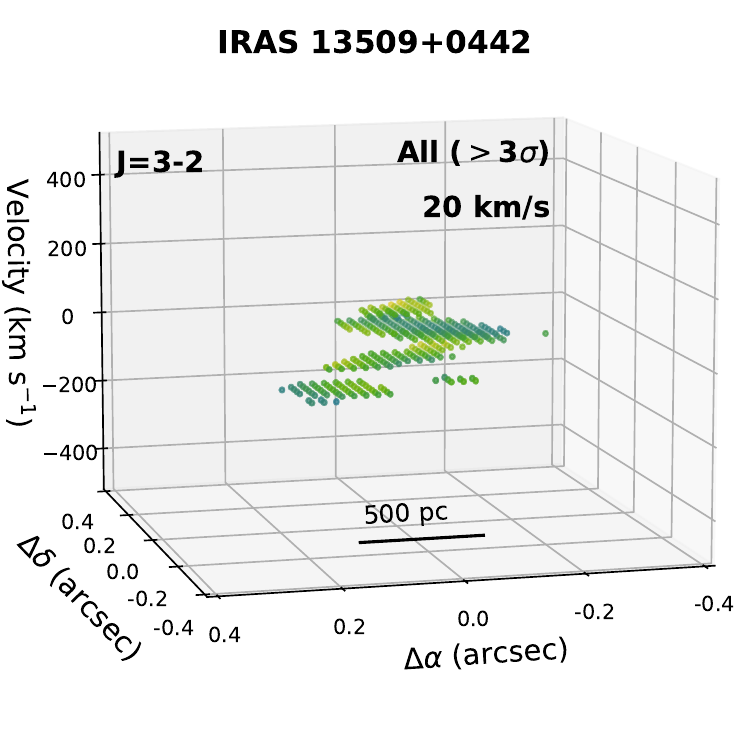} 
\includegraphics[scale=0.45]{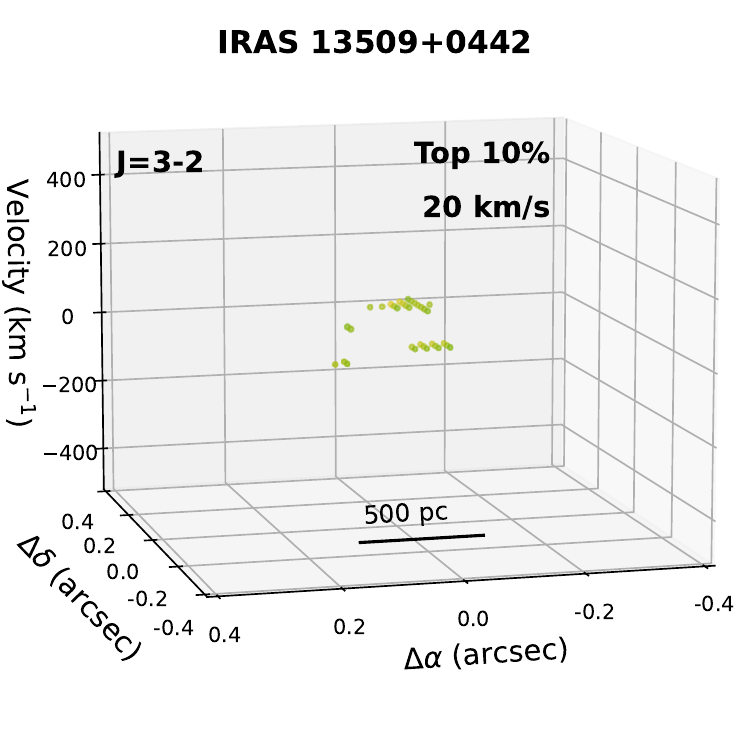} 
\includegraphics[scale=0.45]{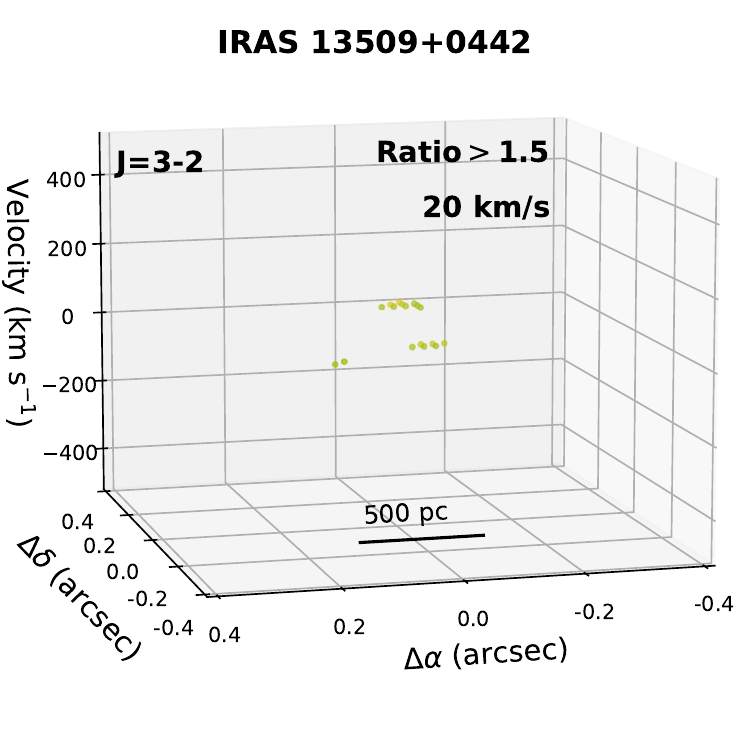} \\
\includegraphics[scale=0.45]{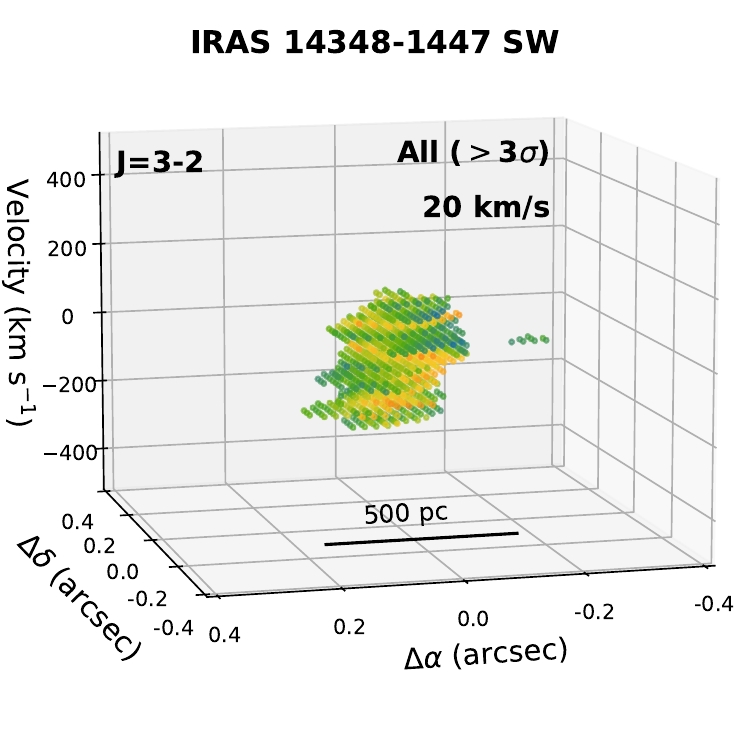} 
\includegraphics[scale=0.45]{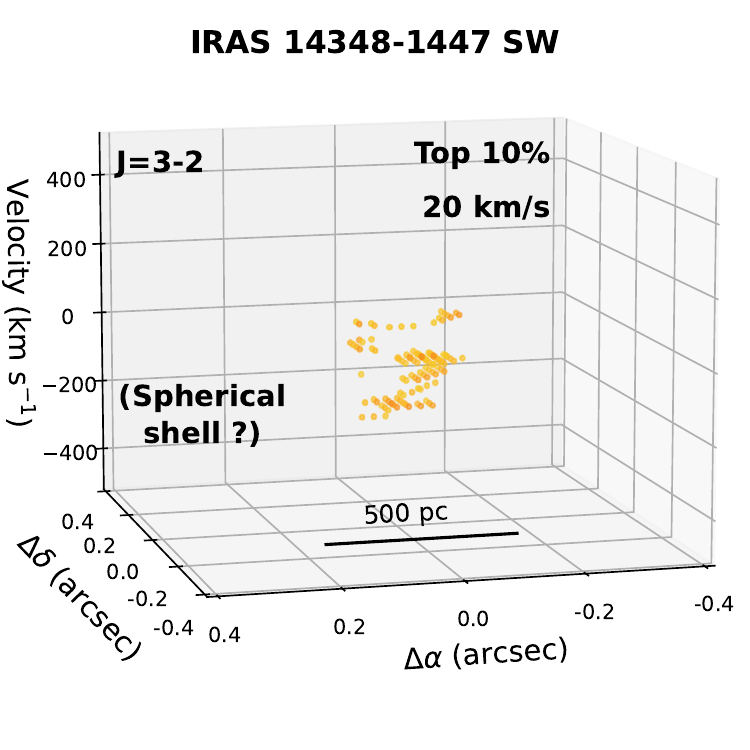} 
\includegraphics[scale=0.45]{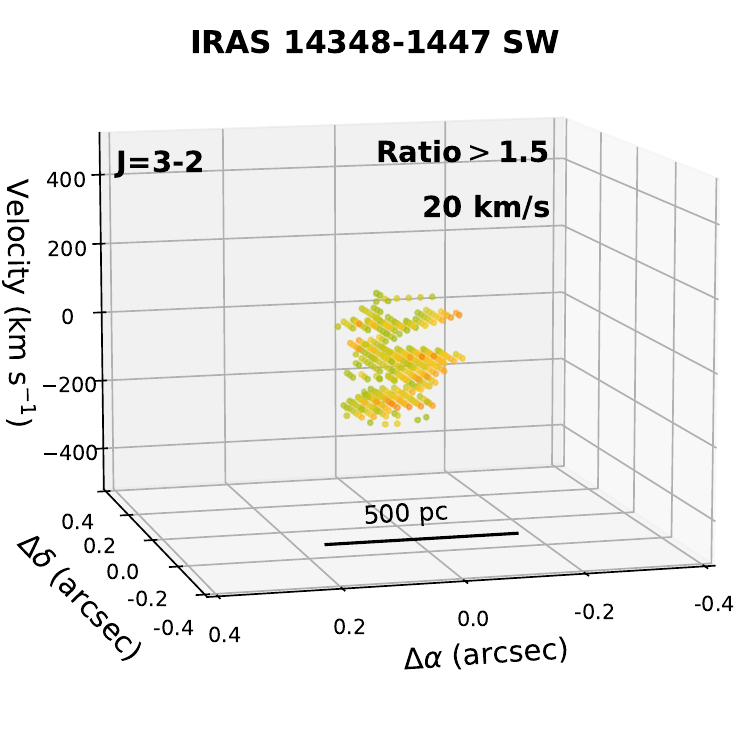} \\
\includegraphics[scale=0.45]{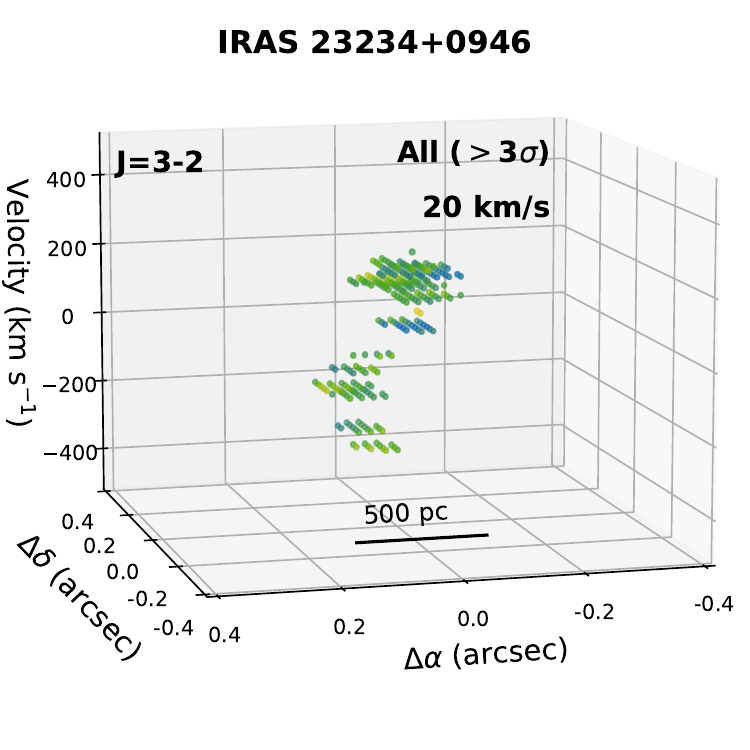} 
\includegraphics[scale=0.45]{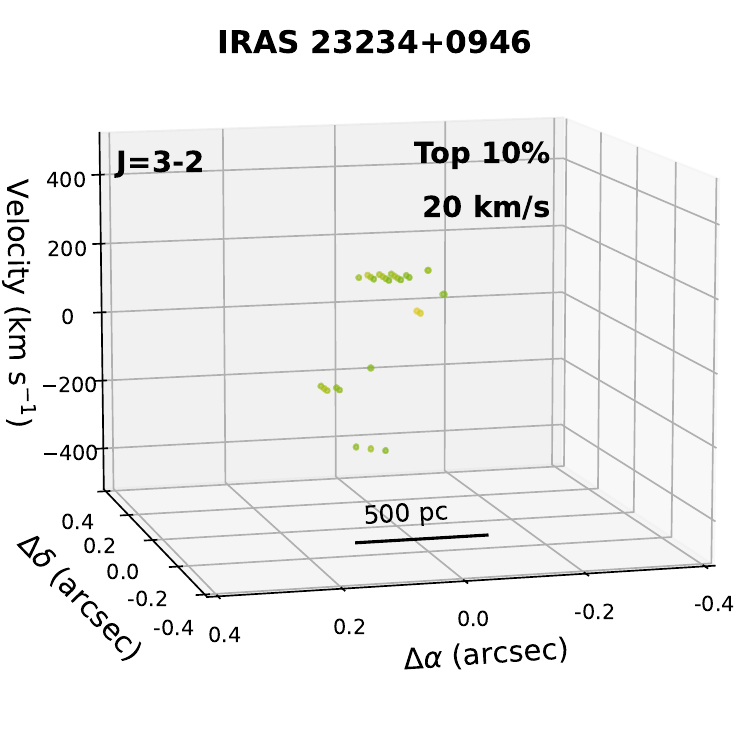} 
\includegraphics[scale=0.45]{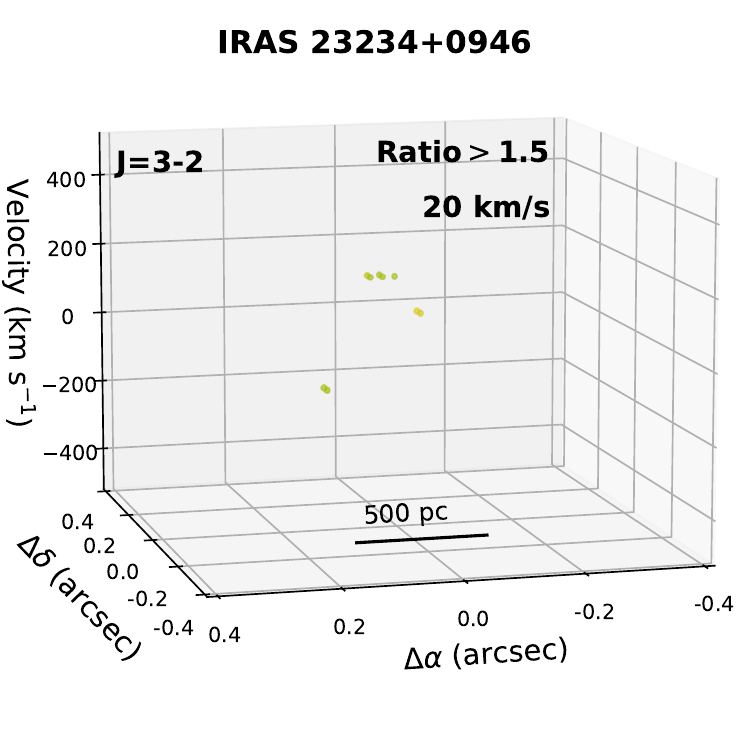}
\\
% Only for Figure 9, last page
\vspace{-0.4cm}
\includegraphics[scale=0.45]{f1-9common.pdf} \\
\end{center}
\vspace{-0.6cm}
\caption{
Same as Figure~\ref{fig:3D1}, but for other ULIRGs for which only 
J=2--1 and/or J=3--2 and/or J=4--3 line data are available 
(IRAS~01004$-$2237, 01298$-$0744, 09039$+$0503, 10190$+$1322,
10378$+$1108, 11095$-$0238, 12112$+$0305, 13509$+$0442, 14348$-$1447, 
and 23234$+$0946). 
The viewing angle is different only for IRAS~10190$+$1322 (see 
Sections~3 and 4).
\label{fig:3D9}
}
\end{figure*}

Figures~\ref{fig:3D1}--\ref{fig:3D9} present the PPV plots for the 18
target ULIRGs.
We adopt a velocity resolution of 20 km s$^{-1}$, following
\citet{nis24}.
Because the (sub)millimeter HCN and HCO$^{+}$ dense molecular emission
lines in the ULIRGs observed in this study are generally fainter and
broader than those in the (U)LIRGs studied by \citet{nis24}, we also
tested coarser velocity resolutions (e.g., 30 km s$^{-1}$ and 40 km s$^{-1}$).
For some ULIRGs with bright HCN and HCO$^{+}$ emission lines
(e.g., IRAS~16090$-$0139 and 22491$-$1808), we also tried finer
velocity resolutions (e.g., 10 km s$^{-1}$ and even 5 km s$^{-1}$).
However, the overall patterns in the PPV plots do not change
significantly.
We therefore display the PPV plots with a velocity resolution of
20 km s$^{-1}$ for all the observed ULIRGs, to enable straightforward
comparison with those in \citet{nis24}.
We also tested several different viewing angles of the PPV plots, but
the overall patterns do not change substantially either.
We thus adopt the same viewing angle for all targets, except for
IRAS~10190$+$1322, for which the angle is set to better visualize the
north-south spaxel distribution (Figure~\ref{fig:3D9}), to investigate
the possible effects of the detected north-south oriented molecular
outflow \citep{lam22}. 

The left panels of Figures~\ref{fig:3D1}--\ref{fig:3D9} show the
HCN-to-HCO$^{+}$ flux ratios for all spaxels with $>$3$\sigma$
detections in both HCN and HCO$^{+}$ emission lines.
Because dense molecular line emission in nearby ULIRGs is usually
concentrated in the compact ($\lesssim$1 kpc) nuclear regions 
\citep[e.g.,][]{ima19,ima23b}, as long as the velocity structure is continuous,
the distribution of all the $>$3$\sigma$-detected spaxels in the PPV plots
is expected to be filled around the nucleus and systemic velocity.
This is the case for the majority of the observed ULIRGs, most clearly
seen in IRAS~16090$-$0139 (Figure~\ref{fig:3D6}).
However, strong HCN and HCO$^{+}$ absorption dips, 
caused by self-absorption, are observed near the systemic
velocity in the beam-sized and/or area-integrated spectra of a few
ULIRG nuclei, such as IRAS~00091$-$0738, 01569$-$2939, and 
12112$+$0305 \citep{ima19,ima23b}.
For these ULIRGs, the distribution of spaxels with $>$3$\sigma$
detections in emission can appear spectrally distinct 
(Figures~\ref{fig:3D1}, \ref{fig:3D5}, and \ref{fig:3D9}).

In the PPV plots, the distribution of spaxels with elevated
HCN-to-HCO$^{+}$ flux ratios (with $>$3$\sigma$ detections for 
both molecular lines) can differ from that of all the
$>$3$\sigma$-detected spaxels, depending on the physical origin of the
elevation.
As explained in Section~1, the geometry of elevated
HCN-to-HCO$^{+}$ flux ratios in the PPV plots can be
(i) ``spherical shell (spectrally and spatially distinct)'' 
(the spaxels are separated in velocity and position space) if a
spatially resolved outflow is responsible for the elevation \citep{nis24}.
The geometry can be (ii) ``spectrally distinct (and spatially compact)''
(the spaxels are separated in velocity space, but are
clustered to the nuclear region in position space) if a
spatially unresolved outflow is important, or  
(iii) ``filled'' (the spaxels are clustered close to the
systemic velocity and the nuclear region)
in the case of a spatially confined molecular outflow
in which the blueshifted and redshifted velocity components are not
clearly separated. 
On the other hand, the geometry of the elevated HCN-to-HCO$^{+}$ flux
ratios caused by AGN effects can be 
(ii) ``spectrally distinct (and spatially compact)'' 
if dense molecular gas in the vicinity of, and largely affected by, a
luminous AGN is rotation-dominated, or 
(iii) ``filled (spectrally filled and spatially compact)'' if such gas
is random-motion-dominated (i.e., has a continuous velocity
distribution).
The distribution of spaxels with the top 10\% of the HCN-to-HCO$^{+}$
flux ratios (i.e., ``relatively'' high ratios) and that with
HCN-to-HCO$^{+}$ flux ratios $>$1.5 (i.e., ``absolutely'' high ratios)
are displayed in the middle and right panels of
Figures~\ref{fig:3D1}--\ref{fig:3D9}, respectively. 
Because the synthesized beam sizes presented in this paper
($\lesssim$0$\farcs$2) are generally smaller than those of most
ULIRGs with reported molecular outflow detections through the
millimeter CO broad emission-line wings ($\gtrsim$0$\farcs$2) 
\citep[e.g.,][]{per18,flu19,lam22}, our PPV plots may provide new
morphological information on the detected molecular outflows.

We note that AGN-origin elevation of the HCN-to-HCO$^{+}$ flux ratios is
expected to be found in the nuclear region where dense molecular line
emission is usually bright. However, spatially extended outflows with
faint dense molecular line emission may be missed by our criterion of
$>$3$\sigma$ detections for both HCN and HCO$^{+}$ emission lines.
Furthermore, very compact outflows ($\lesssim$100 pc) may not
be spatially resolved properly with our data at physical resolutions of
$\sim$100--500 pc (Table~\ref{tab:object}).
Thus, the detected fraction of spatially resolved outflows as the origin 
of elevated HCN-to-HCO$^{+}$ flux ratios in our PPV analysis can be a
lower limit.  

%%%%%%%%%% Figure 1 (3D1) %%%%%%%%%
%%%%%%%%%%%%%%%%%%%%%%%%%%%%%%%%%%

%%%%%%%%%% Figure 2 (3D2) %%%%%%%%%
%%%%%%%%%%%%%%%%%%%%%%%%%%%%%%%%%%

%%%%%%%%%% Figure 3 (3D3) %%%%%%%%%
%%%%%%%%%%%%%%%%%%%%%%%%%%%%%%%%%%

%%%%%%%%%% Figure 4 (3D4) %%%%%%%%%
%%%%%%%%%%%%%%%%%%%%%%%%%%%%%%%%%%

%%%%%%%%%% Figure 5 (3D5) %%%%%%%%%
%%%%%%%%%%%%%%%%%%%%%%%%%%%%%%%%%%

%%%%%%%%%% Figure 6 (3D6) %%%%%%%%%
%%%%%%%%%%%%%%%%%%%%%%%%%%%%%%%%%%

%%%%%%%%%% Figure 7 (3D7) %%%%%%%%%
%%%%%%%%%%%%%%%%%%%%%%%%%%%%%%%%%%

%%%%%%%%%% Figure 8 (3D8) %%%%%%%%%
%%%%%%%%%%%%%%%%%%%%%%%%%%%%%%%%%%

%%%%%%%%%% Figure 9 (3D9) %%%%%%%%%
\begin{figure*}[!hbt]
%\vspace*{0.6cm}
\begin{center}
%\hspace{-0.4cm}
\includegraphics[scale=0.45]{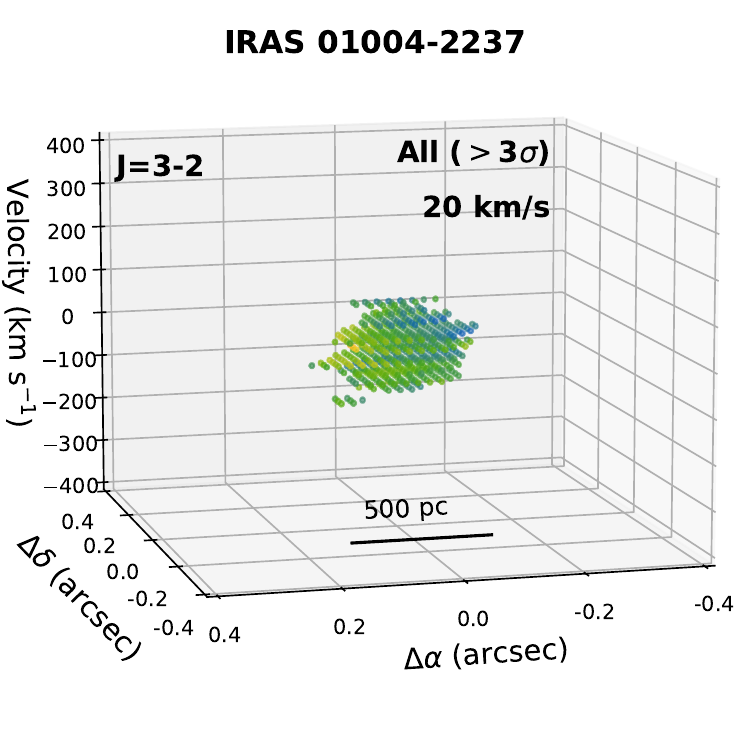} 
\includegraphics[scale=0.45]{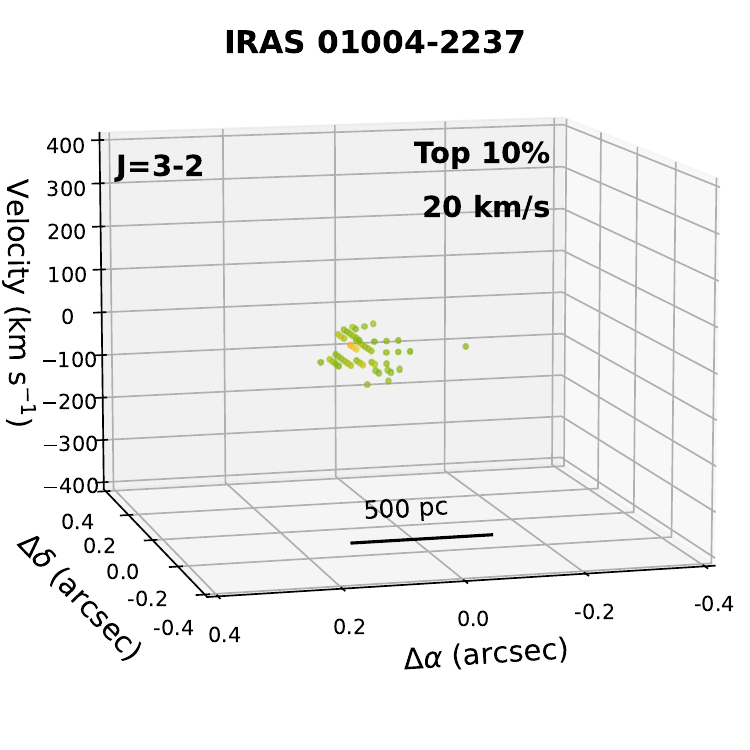} 
\includegraphics[scale=0.45]{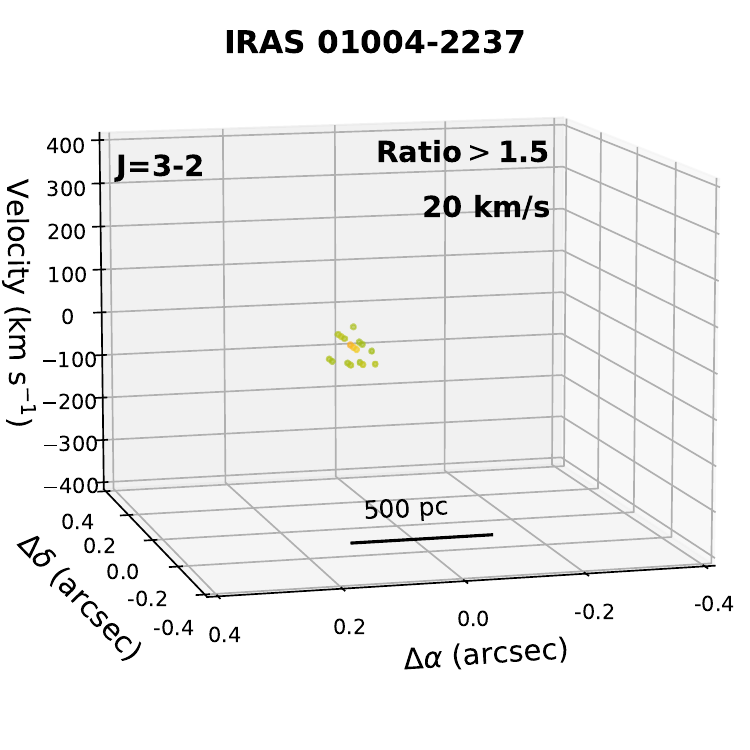} \\
\includegraphics[scale=0.45]{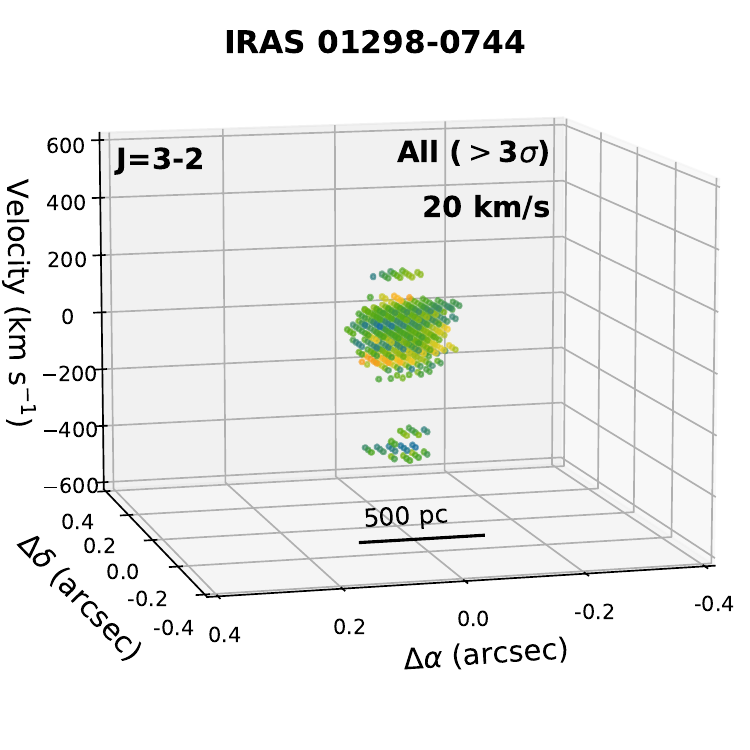} 
\includegraphics[scale=0.45]{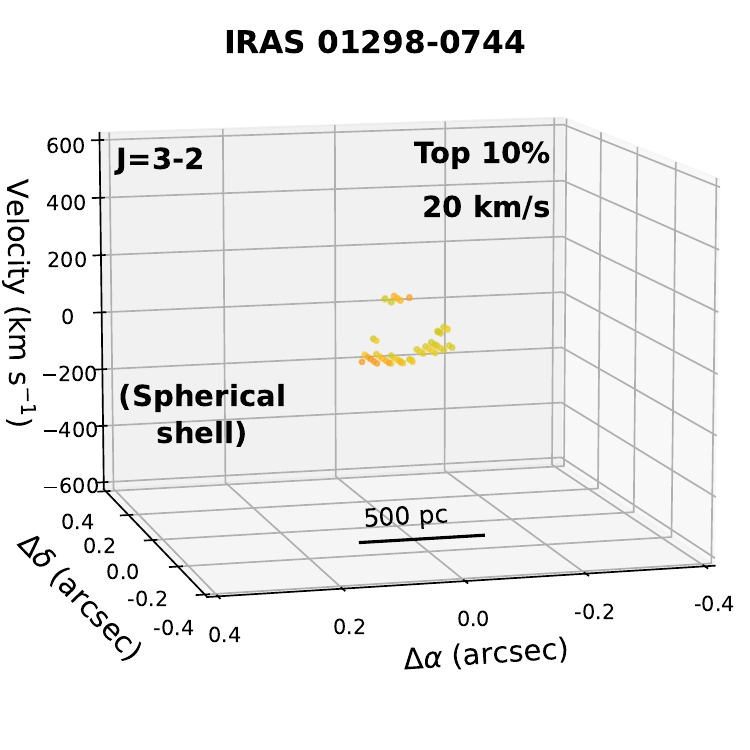} 
\includegraphics[scale=0.45]{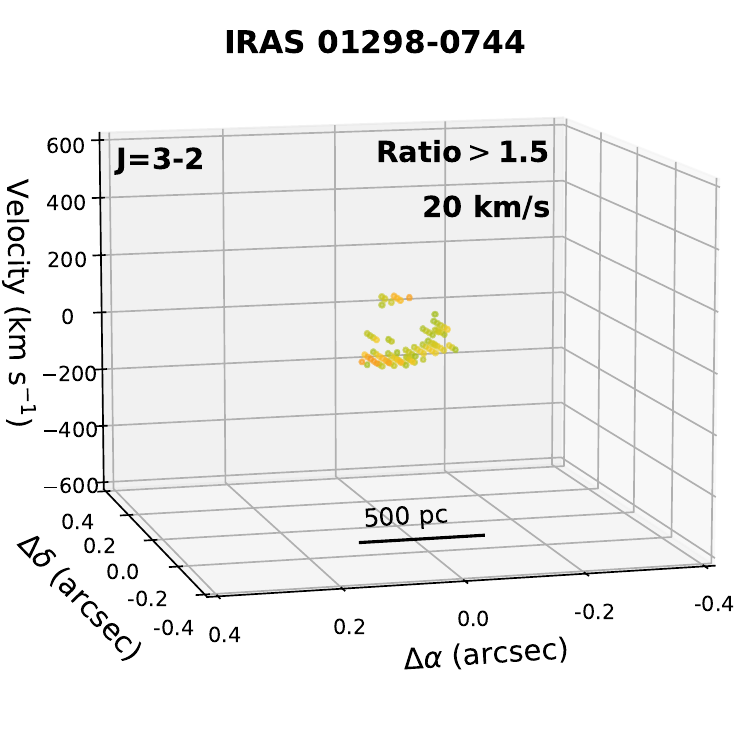} \\
\includegraphics[scale=0.45]{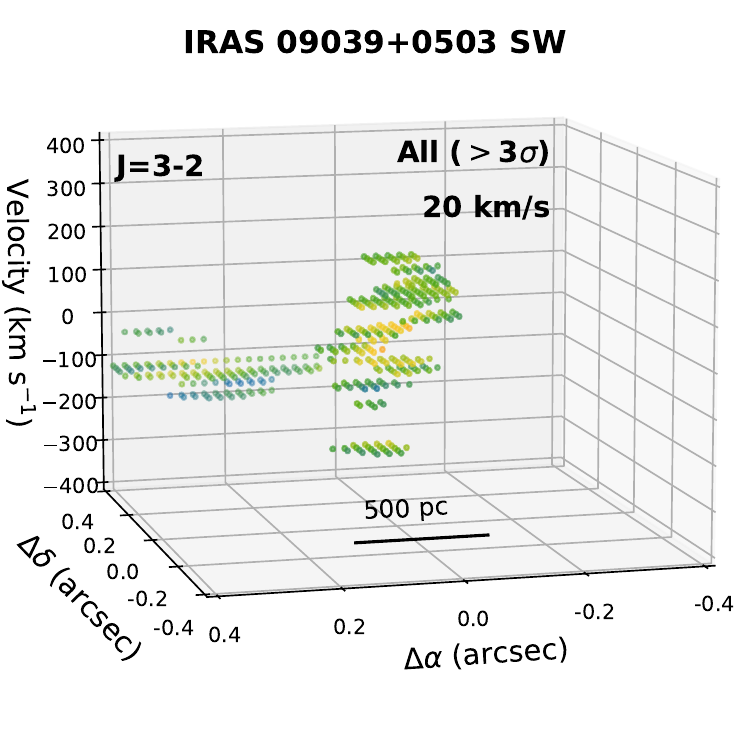} 
\includegraphics[scale=0.45]{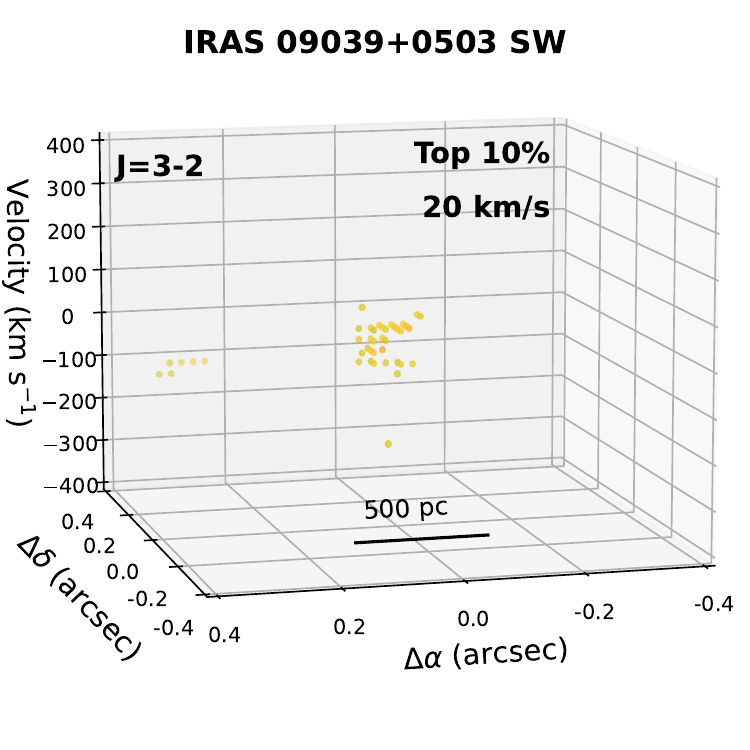} 
\includegraphics[scale=0.45]{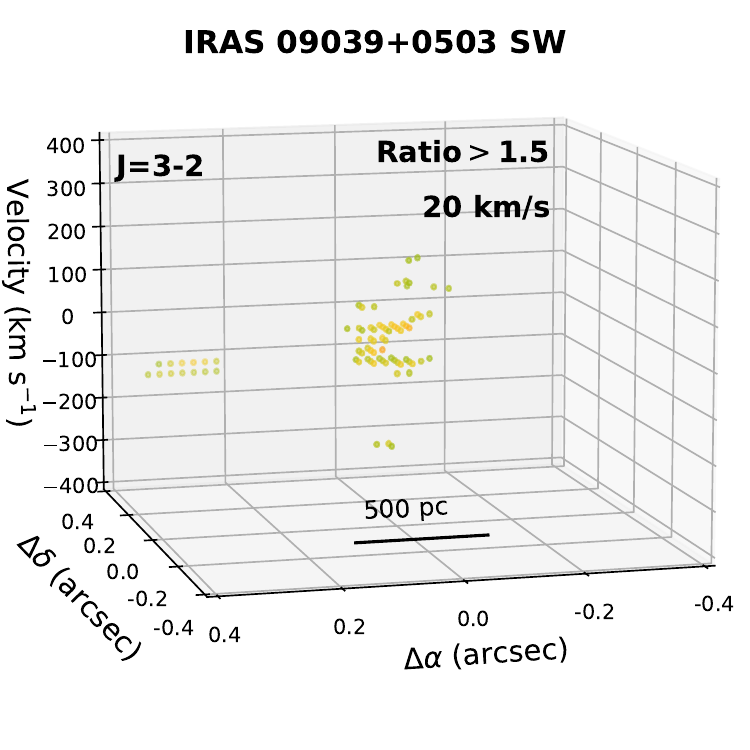} \\
\includegraphics[scale=0.45]{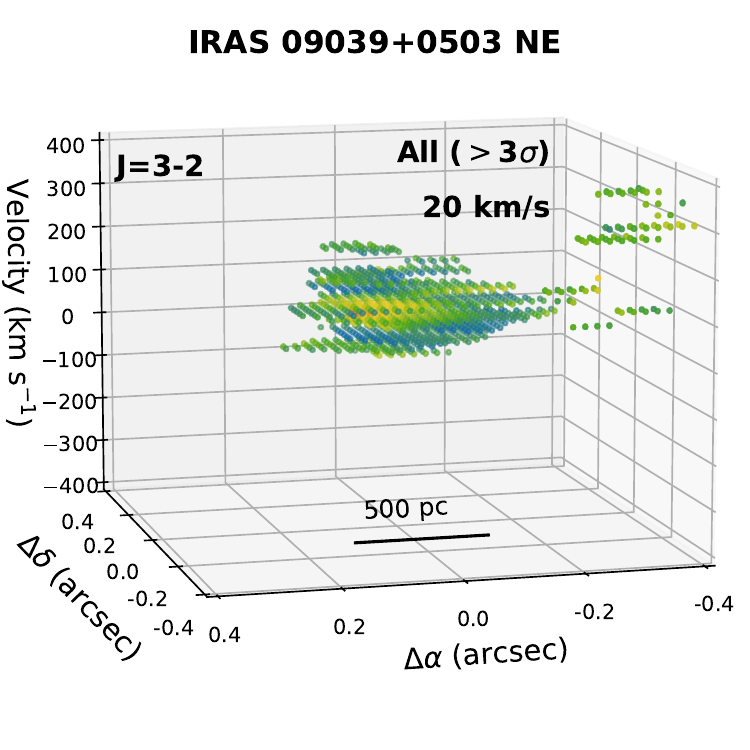} 
\includegraphics[scale=0.45]{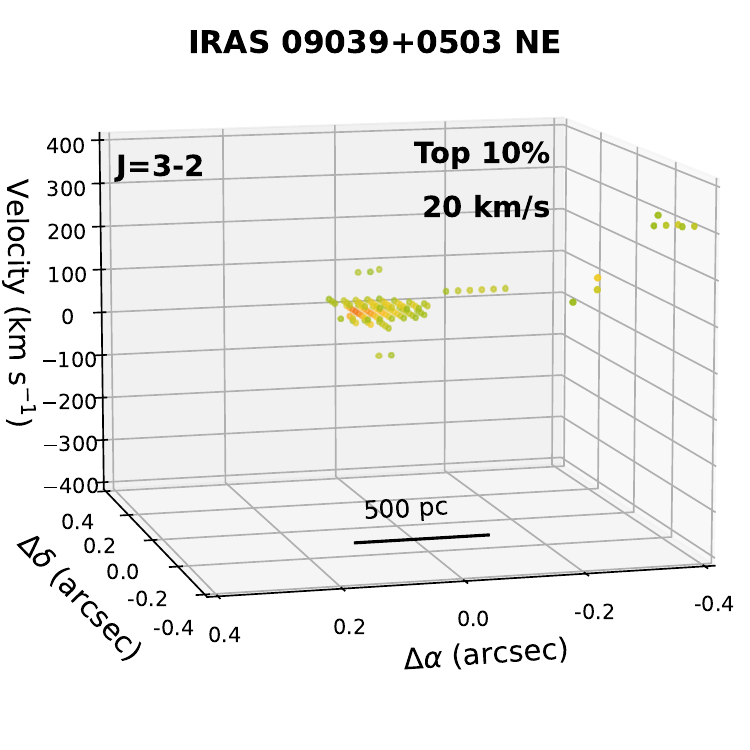} 
\includegraphics[scale=0.45]{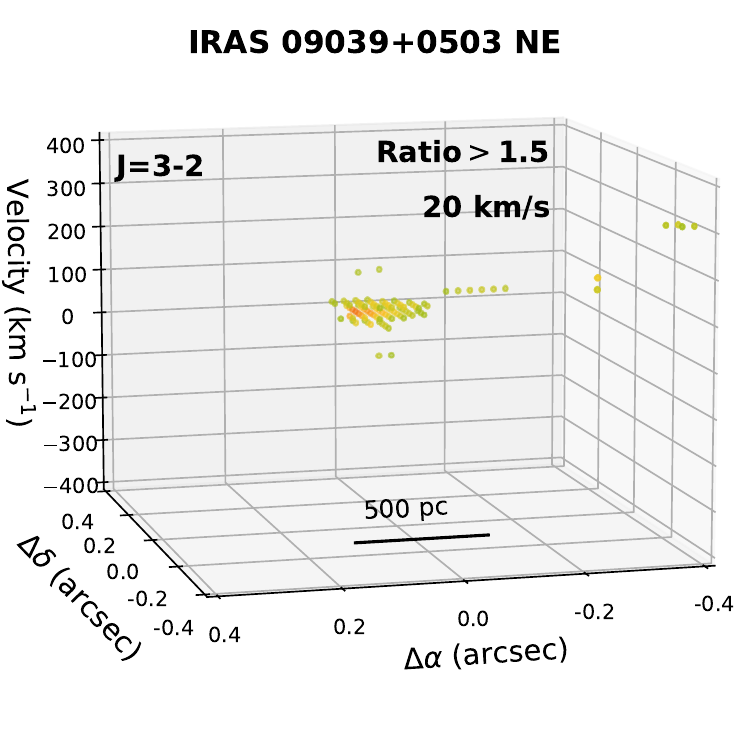} \\
\includegraphics[scale=0.45]{f1-9common.pdf} \\
\end{center}
\end{figure*}

%\clearpage

\begin{figure*}
\begin{center}
\includegraphics[scale=0.446]{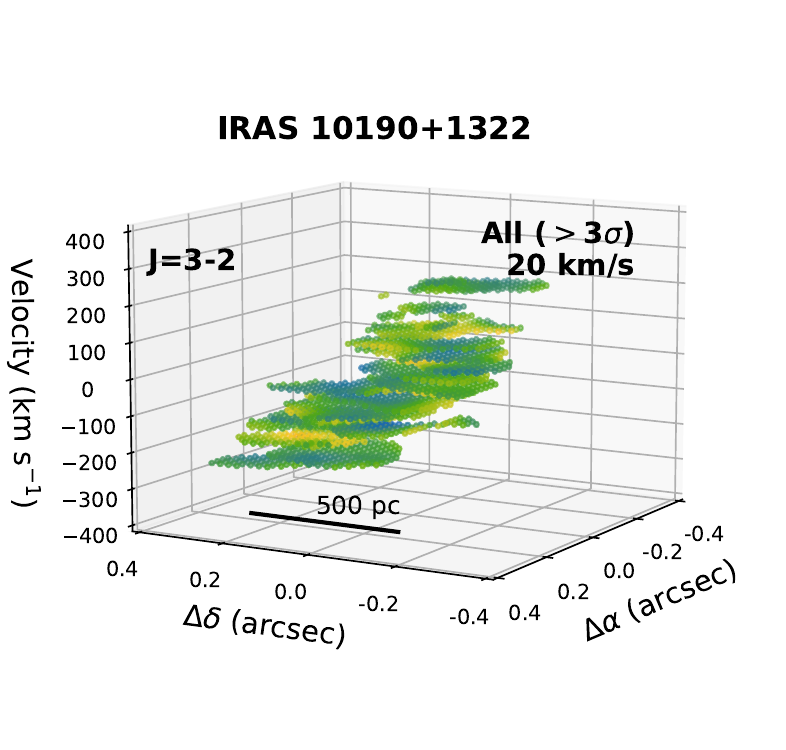}
\includegraphics[scale=0.446]{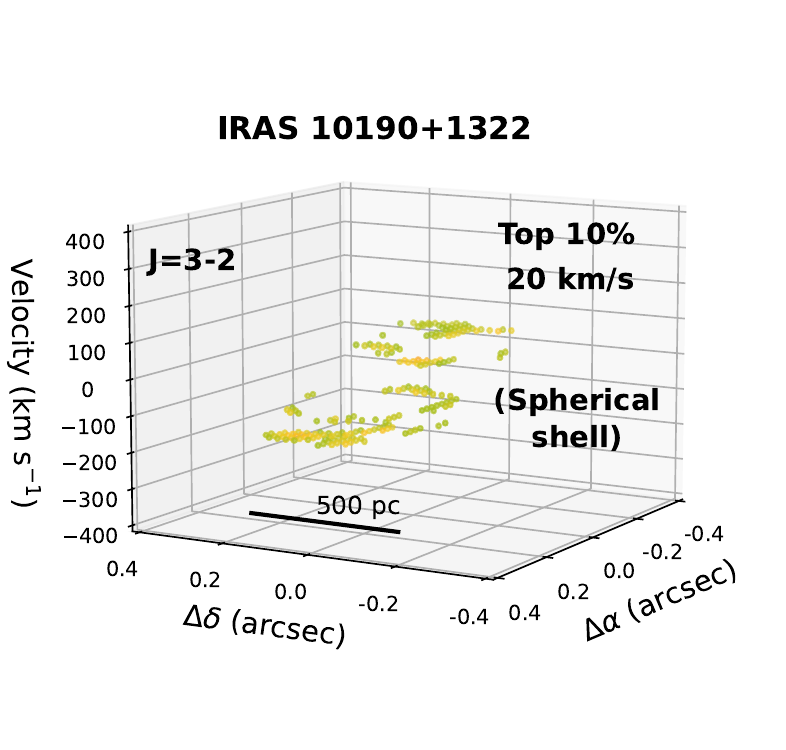} 
\includegraphics[scale=0.446]{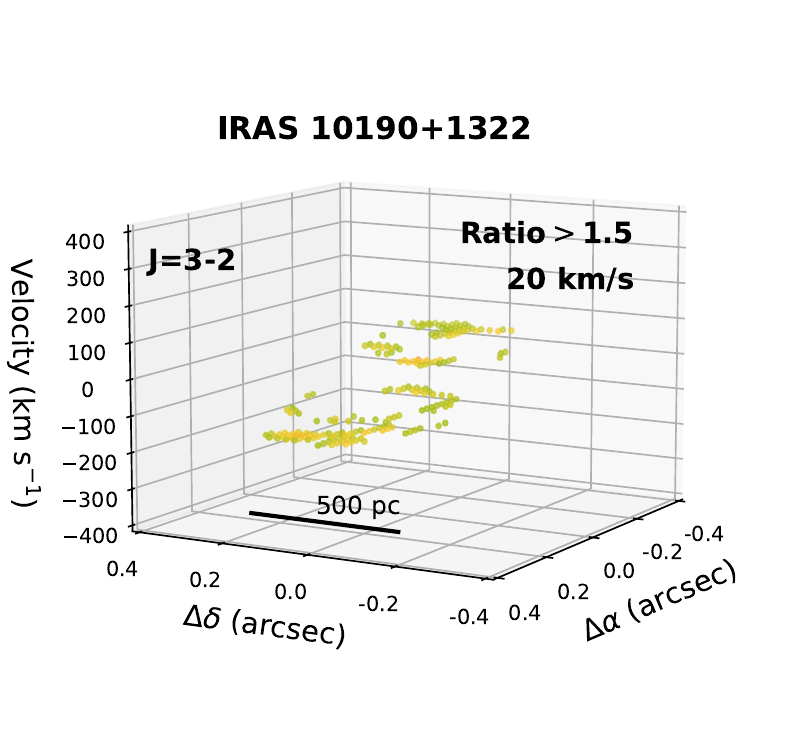} \\
\includegraphics[scale=0.45]{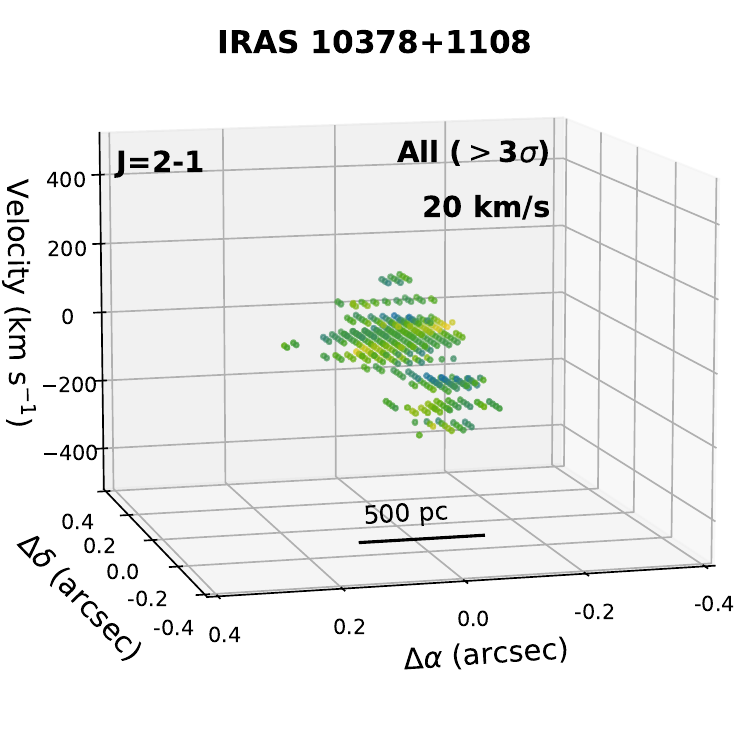} 
\includegraphics[scale=0.45]{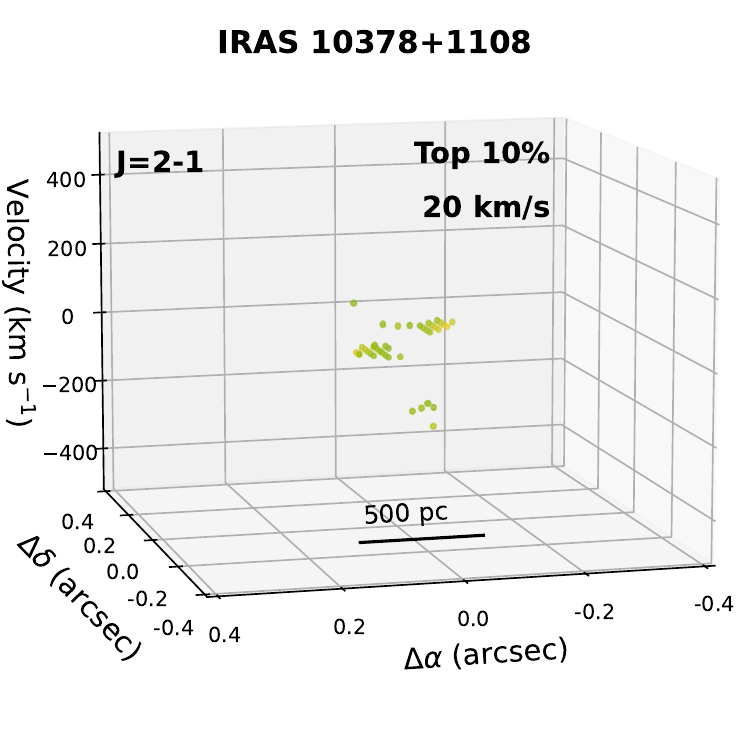} 
\includegraphics[scale=0.45]{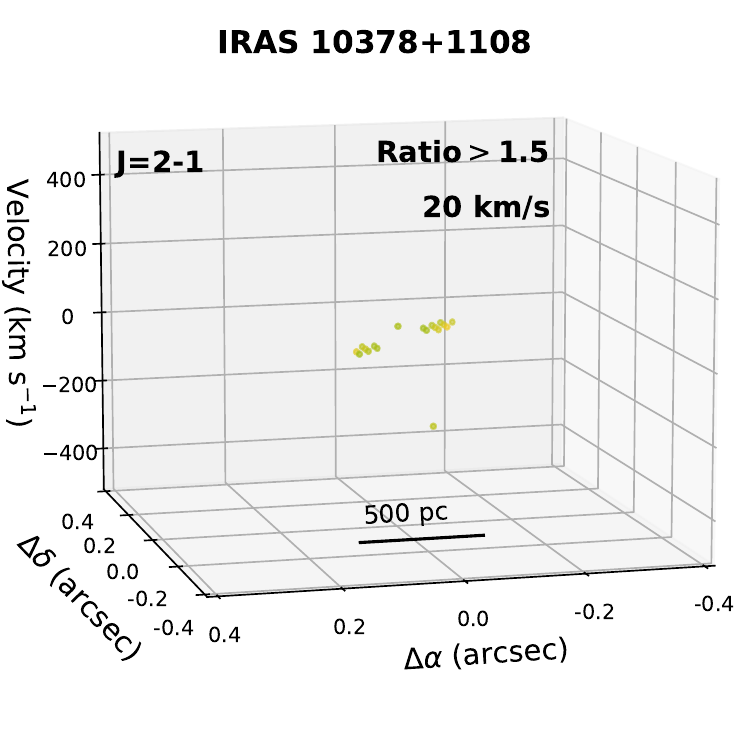} \\
\includegraphics[scale=0.45]{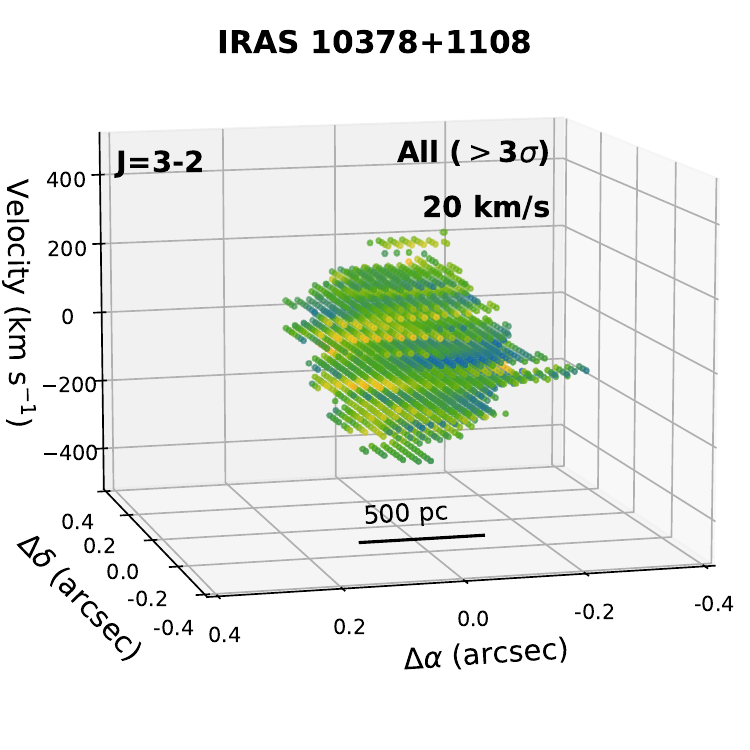} 
\includegraphics[scale=0.45]{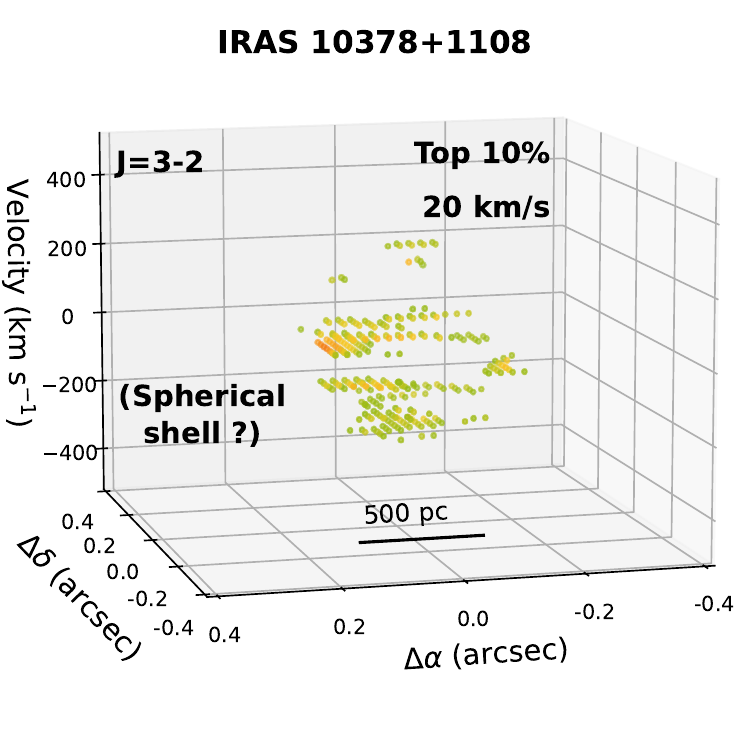} 
\includegraphics[scale=0.45]{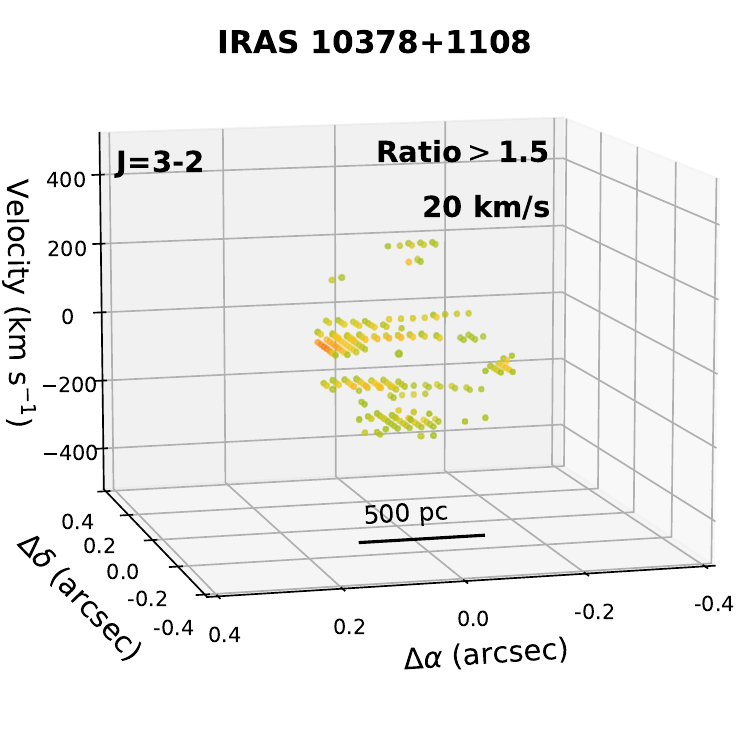} \\
\includegraphics[scale=0.45]{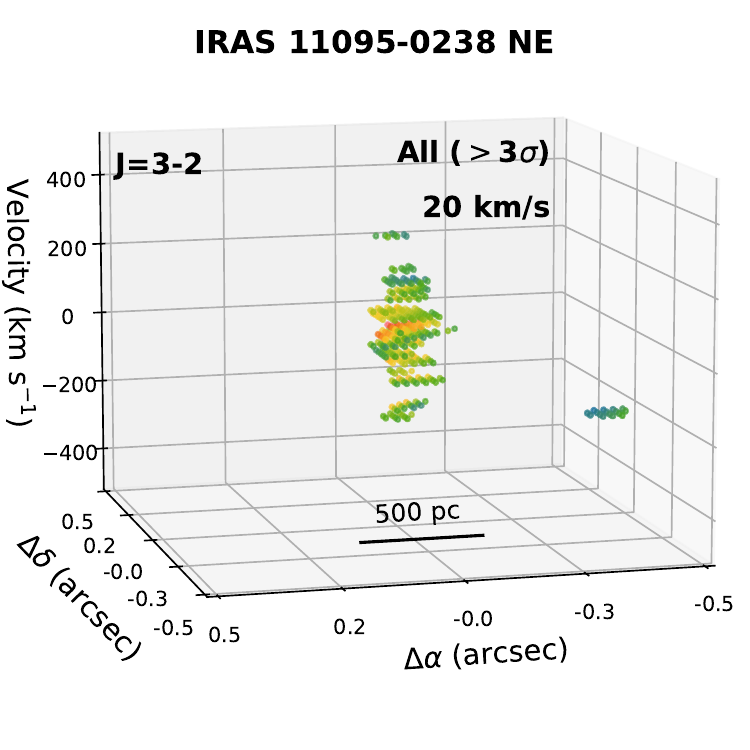} 
\includegraphics[scale=0.45]{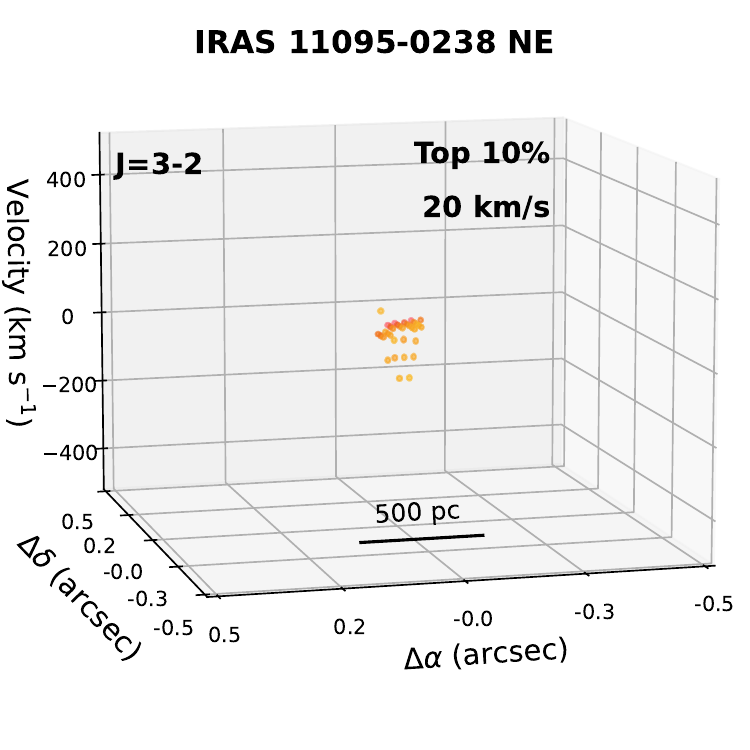} 
\includegraphics[scale=0.45]{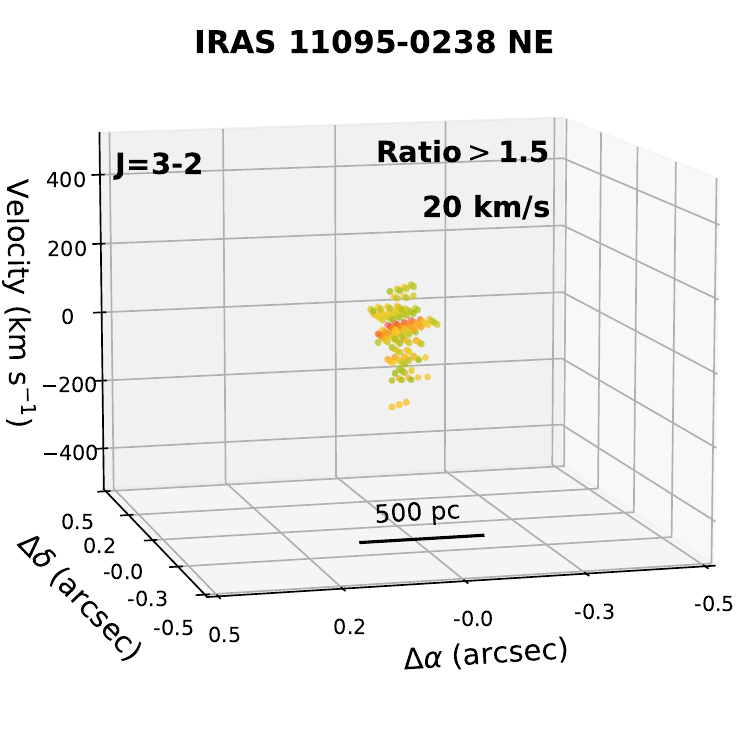} \\
\includegraphics[scale=0.45]{f1-9common.pdf} \\
\end{center}
\end{figure*}

%\clearpage

%%%%%%%%%%%%%%%%%%%%%%%%%%%%%%%%%%

\section{Discussion}

\subsection{ULIRGs with elevated velocity-integrated HCN-to-HCO$^{+}$ 
flux ratios}

Velocity-integrated (sub)millimeter HCN-to-HCO$^{+}$ flux ratios
within the nuclear $\lesssim$500 pc regions \citep{ima19,ima23b} are
summarized in Table~\ref{tab:result} (column 3).   
Among the 18 ULIRGs investigated in this paper, 14 ULIRGs, 
except IRAS 01569$-$2939, 01004$-$2237, 13509$+$0442, and
23234$+$0946, show velocity-integrated HCN-to-HCO$^{+}$ flux ratios $>$1.0.  
We visually classify these 14 ULIRGs mainly into 
(i) ``spherical shell'',
(ii) ``spectrally distinct (and spatially compact)'', and 
(iii) ``filled (spectrally filled and spatially compact)'', 
from the PPV plots in Figures~\ref{fig:3D1}--\ref{fig:3D9}.
Because our ULIRGs are generally farther away than the (U)LIRGs
studied by \citet{nis24}, spaxels with elevated HCN-to-HCO$^{+}$ flux
ratios (with $>$3$\sigma$ detections for both HCN and HCO$^{+}$) 
are more compact in angular scale and fewer in number. 
Nevertheless, signatures of the above three different kinds of
geometries can be seen in the observed ULIRGs.

\subsubsection{Spherical shell geometry}

First, signatures of the (i) ``spherical shell'' geometry are
seen for spaxels with elevated HCN-to-HCO$^{+}$ flux ratios 
(top 10\% and $>$1.5) in several ULIRGs, particularly clearly in the
J=2--1 line of IRAS~00456$-$2904 (Figure~\ref{fig:3D3}), 
the J=4--3 line of IRAS~16090$-$0139 (Figure~\ref{fig:3D6}), and 
the J=3--2 line of IRAS~01298$-$0744 (Figure~\ref{fig:3D9}).  
For IRAS~16090$-$0139, this ``spherical shell'' geometry is seen  
in the J=2--1 and J=3--2 lines as well (Figure~\ref{fig:3D6}).
This geometry for these three ULIRGs (IRAS~00456$-$2904, 16090$-$0139,
and 01298$-$0744) can naturally be explained by the elevation of the 
HCN-to-HCO$^{+}$ flux ratios due to spatially resolved outflow
activity (Section~3).  
The label ``(spherical shell)'' is added for these corresponding 
PPV plots in Figures~\ref{fig:3D3}, \ref{fig:3D6}, and \ref{fig:3D9}.

For IRAS~16090$-$0139, the presence of a molecular outflow has been
argued through the detection of (1) a blueshifted far-infrared 119
$\mu$m OH absorption line with a maximum velocity of $\sim$1400 km
s$^{-1}$ \citep{spo13}, and (2) a 2.6 mm CO J=2--1 broad emission-line
wing on both the blueshifted and redshifted sides at $\pm$350--600
km s$^{-1}$ from the systemic velocity \citep{lam22}.   
The spatial separation of the blueshifted and redshifted outflow 
components in the CO J=2--1 integrated-intensity map is 
$\sim$200--250 pc ($\sim$0$\farcs$1) along the southeast-to-northwest
direction in $\sim$0$\farcs$2-resolution data \citep{lam22}.
The spatial separation of the observed spherical shell in the PPV plots
is $\sim$250--500 pc ($\sim$0$\farcs$1--0$\farcs$2; Figure~\ref{fig:3D6}),
slightly larger than the above separation of $\sim$200--250 pc.
The velocity separation of the observed spherical shell in the PPV
plot is $\sim$200--500 km s$^{-1}$ (Figure~\ref{fig:3D6}), suggesting
that the spaxels with elevated HCN-to-HCO$^{+}$ flux ratios trace 
modest-velocity components of the molecular outflow compared to the
maximum outflow velocity ($\sim$1400 km s$^{-1}$).  
In Figure~\ref{fig:3D6}, the spaxels with elevated
HCN-to-HCO$^{+}$ flux ratios are extended to higher velocity in the 
J=4--3 line than in the J=2--1 and J=3--2 lines, suggesting that
higher J-transitions lines can better probe higher velocity outflow
components than lower J-transition lines.  
The geometry in the right panels (HCN-to-HCO$^{+}$ flux ratios $>$1.5) is
more like ``filled,'' and the ``spherical shell'' geometry is less clear
than in the middle panels (top 10\% ratios). 
This suggests that, in addition to spatially resolved molecular
outflow activity, AGN effects and/or spatially unresolved outflow 
activity also play a role to elevate the
velocity-integrated HCN-to-HCO$^{+}$ flux ratios ($\sim$1.2--1.4) in
the nuclear $\lesssim$500 pc region of IRAS~16090$-$0139.
 
For IRAS~01298$-$0744, \citet{wu24} found molecular-outflow signatures
with a $\sim$300 km s$^{-1}$ blueshifted velocity, with respect to the
1.1 mm HCN J=3--2 emission line, and a $\sim$0$\farcs$05 ($\sim$120 pc)
spatial extent, through VLBI 0$\farcs$012 $\times$ 0$\farcs$005 (30 pc
$\times$ 12 pc) resolution observations of the 18 cm OH emission line.
These authors related this OH-detected molecular outflow to the
elevation of the millimeter HCN-to-HCO$^{+}$ J=3--2 flux ratios.    
The spatial separation of the spherical shell in the PPV plots is
$\sim$300--400 pc (Figure~\ref{fig:3D9}), larger than the
$\sim$120 pc indicated by the VLBI 18 cm OH line observations.
The spherical-shell geometry of the spaxels with elevated
HCN-to-HCO$^{+}$ flux ratios for IRAS~01298$-$0744 (Figure
\ref{fig:3D9}) may originate in a more spatially extended portion of 
the OH-detected molecular outflow.

For IRAS~00456$-$2904, the geometry of the elevated HCN-to-HCO$^{+}$
flux ratios for the J=3--2 and J=4--3 lines is more like ``filled,''
and the ``spherical shell'' geometry is less clear than for the
J=2--1 line (Figure~\ref{fig:3D3}).
This suggests that in addition to spatially resolved outflow activity,
AGN effects and/or spatially unresolved outflows can also contribute to 
elevating the velocity-integrated HCN-to-HCO$^{+}$ flux ratios ($\sim$1.5--1.8) 
in the nuclear $\lesssim$500 pc region of IRAS~00456$-$2904.
IRAS~00456$-$2904 is not included in representative papers that
systematically investigate molecular outflows through a millimeter CO
broad emission-line wing \citep[e.g.,][]{cic14,per18,flu19,lut20,lam22}
and/or far-infrared 60--120 $\mu$m OH absorption lines
\citep[e.g.,][]{spo13,vei13,gon17}.
Thus, no reference reporting a clear molecular-outflow detection is
found.  
In the ALMA archive, no molecular-line data other than our HCN and
HCO$^{+}$ observations \citep{ima19,ima23b} are found. 
Future ALMA millimeter CO line observations of IRAS~00456$-$2904 
will be interesting to search for the presence of a CO broad
emission-line wing caused by a molecular outflow.

Although less clear, possible signatures of the (i) ``spherical
shell'' geometry may also be seen for the spaxels in the top 10\% of
HCN-to-HCO$^{+}$ flux ratios in the J=2--1 line of IRAS~00188$-$0856
(Figure~\ref{fig:3D2}), and the J=3--2 line of IRAS~10378$+$1108 and
14348$-$1447 (Figure~\ref{fig:3D9}). 
The label ``(spherical shell ?)'' is added for these corresponding 
PPV plots in Figures~\ref{fig:3D2} and \ref{fig:3D9}.

For IRAS~14348$-$1447, detection of a molecular outflow has been
argued based on both (1) blueshifted OH absorption lines at
60--120 $\mu$m \citep{vei13,gon17} and (2) a CO J=2--1 broad emission-line
wing at 2.6 mm in $\sim$0$\farcs$3-resolution data \citep{per18,lam22}. 
The maximum outflow velocity is estimated to be $\sim$900 km s$^{-1}$
from the OH blueshifted absorption study \citep{vei13}.  
The CO J=2--1 observations also detect an emission-line wing on both
the blueshifted and redshifted sides at 250--800 km s$^{-1}$ from the
systemic velocity \citep{per18,lam22}. 
The velocity difference of the observed spherical shell in the PPV
plot is $\sim$200 km s$^{-1}$ (Figure~\ref{fig:3D9}), suggesting that
spaxels with elevated HCN-to-HCO$^{+}$ flux ratios probe modest-velocity
outflow components compared to the maximum outflow velocity 
($\sim$900 km s$^{-1}$), as in the case of IRAS~16090$-$0139.
%In fact, several chemical models predict abundance enhancement of HCN,
%relative to HCO$^{+}$, in modest-velocity ($\gtrsim$30--40 km
%s$^{-1}$) shocks \citep[e.g.,][]{bac97,vit14,nis24}, which can
%naturally explain the elevated HCN-to-HCO$^{+}$ flux ratios. 
The spatial separation of the blueshifted and redshifted outflow
components in the CO J=2--1 integrated-intensity map is 200--500 pc
(0$\farcs$13--0$\farcs$3) along an almost east--west direction
\citep{per18,lam22}. 
The spatial separation of the observed spherical shell in the PPV plot
is $\sim$200--300 pc (Figure~\ref{fig:3D9}), roughly comparable to the
above 200--500 pc separation suggested by the CO broad emission-line
wing study. 
For IRAS~14348$-$1447, the geometry in the right panel 
(HCN-to-HCO$^{+}$ flux ratios $>$1.5) is more like ``filled,''
and the ``spherical shell'' geometry is less clear than in the
middle panel (top 10\% ratios).
It is suggested that in addition to spatially resolved outflows, 
AGN effects and/or spatially unresolved outflows can also
be responsible for the elevation of the velocity-integrated 
HCN-to-HCO$^{+}$ flux ratios ($\sim$1.5) in the nuclear $\lesssim$500 pc
region of IRAS~14348$-$1447.  

For IRAS~00188$-$0856, a molecular outflow is argued to be present based
on the detection of a blueshifted 119 $\mu$m OH absorption line with a
maximum velocity of $\gtrsim$1700 km s$^{-1}$ \citep{spo13}. 
The velocity separation of the observed spherical shell in the PPV
plot is $\lesssim$200--300 km s$^{-1}$ (Figure~\ref{fig:3D2}).
As in the cases of IRAS~16090$-$0139 and 14348$-$1447, this
suggests that spaxels with elevated HCN-to-HCO$^{+}$ flux ratios in
IRAS~00188$-$0856 trace modest-velocity molecular outflow
components with respect to the maximum outflow velocity 
($\gtrsim$1700 km s$^{-1}$).
For IRAS~00188$-$0856, the geometry of the J=3--2 and J=4--3 lines 
(top 10\% ratios) and that of the spaxels with HCN-to-HCO$^{+}$ flux
ratios $>$1.5 are more like ``filled'' rather 
than ``spherical shell'' (Figure~\ref{fig:3D2}), suggesting
contributions not only from spatially resolved outflows, but also from 
AGN and/or spatially unresolved outflows, 
to the elevated velocity-integrated HCN-to-HCO$^{+}$ flux ratios
($\sim$1.8--1.9) in the nuclear $\lesssim$500 pc region.   
 
For IRAS~10378$+$1108, a molecular outflow is argued to be present
based on the detection of a blueshifted 119 $\mu$m OH absorption line
with a maximum velocity of $\sim$1300 km s$^{-1}$ \citep{spo13}. 
The velocity separation of the observed spherical shell in the PPV
plot is $\lesssim$200--300 km s$^{-1}$ (Figure~\ref{fig:3D9}), again
suggesting that spaxels with elevated HCN-to-HCO$^{+}$ flux ratios
trace modest-velocity molecular outflow components.

For IRAS~10190$+$1322, the geometry of all the $>$3$\sigma$-detected
spaxels in the PPV plot is filled in velocity but spatially extended
beyond $\sim$500 pc (Figure~\ref{fig:3D9}, left).  
That of the elevated HCN-to-HCO$^{+}$ flux ratios (top 10\% and
$>$1.5) is spectrally distinct and spatially {\it extended}
(Figure~\ref{fig:3D9}, middle and right), namely, can be classified into  
(i) ``spherical shell'' geometry, caused by 
a spatially resolved molecular outflow.
In fact, \citet{lam22} detected a broad emission-line wing for the 2.6
mm CO J=2--1 line on both the blueshifted and redshifted sides at
$\pm$300--400 km s$^{-1}$ from the systemic velocity.
These blueshifted and redshifted outflow components are separated 
by $\sim$500 pc along the north-south direction \citep{lam22}.
We therefore display the PPV plots of IRAS~10190$+$1322 in such a way that
the north-south extent of spaxels is more easily visible 
(Figure~\ref{fig:3D9}).
The spatial and velocity separations of the ``spherical shell'' in 
Figure~\ref{fig:3D9} are $\sim$500 pc and $\sim$250--300 km s$^{-1}$,
respectively.
The former spatial separation of $\sim$500 pc roughly agrees with
the above estimate from the CO J=2--1 emission-line wing ($\sim$500 pc). 
However, the latter small velocity separation of $\sim$250--300 km
s$^{-1}$, relative to $\sim$600--800 km s$^{-1}$ between the
blueshifted and redshifted outflow-origin CO J=2--1 emission-wing
components, again suggests that modest-velocity outflow components are
probed by the spaxels with elevated HCN-to-HCO$^{+}$ flux ratios in 
IRAS~10190$+$1322.  
  
In summary, signatures of elevated HCN-to-HCO$^{+}$ flux ratios are 
found in modest-velocity outflows, compared to the maximum outflow
velocity, in five ULIRGs (IRAS~16090$-$0139, 14348$-$1447,
00188$-$0856, 10378$+$1108, and 10190$+$1322). 
Several chemical models predict abundance enhancement of HCN, relative
to HCO$^{+}$, in modest-velocity ($\gtrsim$30--40 km s$^{-1}$) shocks, 
with a time scale of 10$^{3-5}$ years
\citep[e.g.,][]{bac97,vit14,nis24}.
The dynamical time scale of the detected outflow in our PPV analysis 
(r $\sim$ 200--500 pc and v $\sim$ 200--500 km s$^{-1}$) is r/v $\sim$
10$^{6}$ years, longer than the above chemical reaction time scale. 
Thus, our results can naturally be explained.
However, we note that our results do not necessarily mean that the 
HCN-to-HCO$^{+}$ flux ratios are not elevated in high-velocity outflow 
components.
The elevation is investigated only for spaxels with $>$3$\sigma$
detections for both HCN and HCO$^{+}$ emission lines. 
High-velocity outflow components generally show smaller emission line
flux than modest-velocity outflow components
\citep[e.g.,][]{cic14,ima17,per18,flu19,lut20,lam22} and thus may not
be sufficiently probed with our PPV analysis.

\subsubsection{Spectrally distinct (and spatially compact) geometry}

IRAS~01166$-$0844 shows a ``filled'' geometry for all the
$>$3$\sigma$-detected spaxels, but a ``spectrally distinct (and
spatially compact)'' geometry for the spaxels in the top 10\% of
HCN-to-HCO$^{+}$ flux ratios in the J=2--1 line (Figure~\ref{fig:3D4}). 
Similar geometry is seen also for the spaxels with top 10\% flux
ratios in the J=3--2 and J=4--3 lines, although less clear than in the
J=2--1 line.  
The geometry of the spaxels with HCN-to-HCO$^{+}$ flux ratios $>$1.5
is more like ``filled'' particularly in the J=2--1 line.

IRAS~22206$-$2715 also shows signatures of 
``spectrally distinct (and spatially compact)'' geometry for the
spaxels in the top 10\% of HCN-to-HCO$^{+}$ flux ratios in the J=3--2
line (Figure~\ref{fig:3D7}).
However, the geometry for the spaxels with HCN-to-HCO$^{+}$ flux
ratios $>$1.5 is more like ``filled'' in the J=2--1 and J=3--2 lines. 

For these two ULIRGs (IRAS~01166$-$0844 and 22206$-$2715), 
the elevated velocity-integrated HCN-to-HCO$^{+}$ flux
ratios ($\sim$1.3--1.8) in the nuclear $\lesssim$500 pc regions
(Table~\ref{tab:result}) can be caused by a spatially unresolved
outflow and/or AGN effects.
For both ULIRGs, we have found no reference reporting a
molecular-outflow detection, nor observational data in the ALMA
archive to investigate the presence of an outflow-origin millimeter CO 
broad emission-line wing.   

\subsubsection{Filled (spectrally filled and spatially compact) geometry}

IRAS~22491$-$1808, 09039$+$0503, and 11095$-$0238 display 
(iii) ``filled (spectrally filled and spatially compact)'' geometry
for all the $>$3$\sigma$-detected spaxels and for spaxels with elevated
HCN-to-HCO$^{+}$ flux ratios (top 10\% and $>$1.5) 
(Figures~\ref{fig:3D8} and \ref{fig:3D9}).    
AGN effects can explain the elevated ($\sim$1.1--1.9)
velocity-integrated HCN-to-HCO$^{+}$ flux ratios in the nuclear
$\lesssim$500 pc regions of these three ULIRGs (Table~\ref{tab:result}).  
This is also the case for the six ULIRGs that display the 
(iii) ``filled'' geometry in some J-transitions lines 
(IRAS~16090$-$0139, 00456$-$2904, 14348$-$1447, 00188$-$0856,
01166$-$0844, and 22206$-$2715), discussed in Sections~4.1.1 and 4.1.2. 

For IRAS~22491$-$1808 and 11095$-$0238, detection of a CO J=2--1
broad emission-line wing is argued on both the blueshifted and redshifted
sides at $\pm$200--600 km s$^{-1}$ from the systemic velocity, which
can be of molecular-outflow origin, but the blueshifted and redshifted
outflow components are not clearly separated spatially in
$\sim$0$\farcs$3--0$\farcs$4-resolution data \citep{per18,lam22}.
It is possible that the HCN-to-HCO$^{+}$ flux ratios for 
IRAS~22491$-$1808 and 11095$-$0238 are elevated by a molecular outflow
that is still spatially confined to a very compact region, with
outflow-origin blueshifted and redshifted components not yet clearly
separated.
For IRAS~09039$+$0503, there are no millimeter CO line observational
data in the ALMA archive, nor any reference reporting the detection of 
an outflow-origin CO broad emission-line wing.   
The same possibility applies to IRAS~09039$+$0503 (and some of the 
six ULIRGs) if such an outflow exists.

\subsubsection{Other geometry}

IRAS~00091$-$0738 and 12112$+$0305 display spectrally distinct (and
spatially compact) geometry for all the $>$3$\sigma$-detected spaxels 
(Figures~\ref{fig:3D1} and \ref{fig:3D9}), 
because the HCN and HCO$^{+}$ emission lines display double-peaked
profiles with strong central dips, due to self-absorption, 
in the beam-sized and/or nuclear area-integrated spectra
\citep{ima19,ima23b}.   
For these two ULIRGs, it is difficult to determine whether 
the spaxels with elevated HCN-to-HCO$^{+}$ flux ratios
are spectrally distinct or filled, because only the $>$3$\sigma$-detected
spaxels in emission are used in this analysis. 
The physical origin of the elevated ($\sim$1.3--1.9)
velocity-integrated HCN-to-HCO$^{+}$ flux ratios in the nuclear
$\lesssim$500 pc regions of these two ULIRGs (Table~\ref{tab:result}) 
is unclear from our PPV analysis.

\subsubsection{Summary of the geometry of the elevated 
HCN-to-HCO$^{+}$ flux ratios} 

Table~\ref{tab:result} summarizes our classification of the geometry
of spaxels with elevated HCN-to-HCO$^{+}$ flux ratios. 
Among the 14 ULIRGs showing elevated ($>$1) velocity-integrated
HCN-to-HCO$^{+}$ flux ratios in the nuclear $\lesssim$500 pc regions, 
signatures of elevated flux ratios originated from spatially resolved 
outflow activity are seen in seven ULIRGs 
(IRAS~16090$-$0139, 01298$-$0744, 00456$-$2904, 14348$-$1447,
00188$-$0856, 10378$+$1108, and 10190$+$1322;  
labeled with (i) in Table~\ref{tab:result}, column 2).
This is a significant fraction of ULIRGs, 
as previously suggested by \citet{nis24} based on a limited ULIRG sample.
The elevated HCN-to-HCO$^{+}$ flux ratios of nine ULIRGs
(IRAS~16090$-$0139, 00456$-$2904, 14348$-$1447, 00188$-$0856,
01166$-$0844, 22206$-$2715, 22491$-$1808, 09039$+$0503, and
11095$-$0238; labeled with (ii) or (iii) in Table~\ref{tab:result},
column 2) can be explained by spatially unresolved outflows and/or AGN
effects. 
Our spectrally and spatially resolved PPV study provides valuable
insights into the origin of the elevated velocity-integrated
HCN-to-HCO$^{+}$ flux ratios in nearby ULIRG nuclei 
($\lesssim$500 pc).

%%%%%%%%%% Table 2 (result) %%%%%%%%%
\begin{deluxetable*}{l|l|c|ccc}[!hbt]
\tabletypesize{\scriptsize}
\rotate
\tablecaption{Summary of the Results of the Spectrally and Spatially 
Resolved HCN-to-HCO$^{+}$ Flux Ratios \label{tab:result}}  
\tablewidth{0pt}
\tablehead{
\colhead{Object} & \colhead{Classification} & 
\colhead{HCN-to-HCO$^{+}$ flux ratio} & 
\colhead{AGN} & \multicolumn{2}{c}{Outflow} \\  
\colhead{} & \colhead{} & \colhead{($\lesssim$500 pc)} &
\colhead{IR/(sub)mm} & \colhead{OH abs.} & \colhead{CO or OH emi.} \\   
\colhead{(1)} & \colhead{(2)} & \colhead{(3)} & \colhead{(4)} &
\colhead{(5)} & \colhead{(6)} 
}
\startdata
IRAS~16090$-$0139 & (i) Spherical shell (J21,J32,J43) $+$ (iii)
Filled & 1.3$\pm$0.1 (J21), 1.2$\pm$0.1 (J32), 1.4$\pm$0.1
(J43) & Y & Y$^{a}$ (\tablenotemark{A}) & Y$^{d}$  \\       
IRAS~01298$-$0744 & (i) Spherical shell (J32) & 1.3$\pm$0.2
(J32) & Y & \nodata & Y$^{e}$  \\    
IRAS~00456$-$2904 & (i) Spherical shell (J21) $+$ (iii) Filled &
1.5$\pm$0.1 (J21), 1.8$\pm$0.2 (J32), 1.6$\pm$0.2 (J43) & Y & \nodata &
\nodata \\   
IRAS~14348$-$1447 SW & (i) Spherical shell(?) (J32) $+$ (iii)
Filled & 1.5$\pm$0.1 (J32) & Y & Y$^{b,c}$ & Y$^{d,f}$  \\  
IRAS~00188$-$0856 & (i) Spherical shell(?) (J21) $+$ (iii) Filled
& 1.8$\pm$0.1 (J21), 1.8$\pm$0.1 (J32), 1.9$\pm$0.4 (J43) &
Y & Y$^{a}$ & N$^{d}$ \\      
IRAS~10378$+$1108 & (i) Spherical shell(?) (J32) 
& 1.1$\pm$0.2 (J21), 1.1$\pm$0.1 (J32) & Y & Y$^{a}$ & \nodata \\     
IRAS~10190$+$1322 & (i) Spherical shell (J32) &
1.1$\pm$0.2 (J32) & Y & \nodata & Y$^{d}$  \\  \hline
IRAS~01166$-$0844 & (ii) Spectrally distinct $+$ (iii) Filled &
1.8$\pm$0.2 (J21), 1.3$\pm$0.3 (J32), 1.5$\pm$0.2 (J43) & Y & \nodata &
\nodata \\     
IRAS~22206$-$2715 & (ii) Spectrally distinct $+$ (iii) Filled &
1.5$\pm$0.2 (J21), 1.3$\pm$0.1 (J32), 1.3$\pm$0.2 (J43) & Y
& \nodata & \nodata \\  
IRAS~22491$-$1808 & (iii) Filled & 1.5$\pm$0.1 (J21), 1.4$\pm$0.1 (J32),
1.5$\pm$0.1 (J43) & Y & N$^{b}$ (\tablenotemark{A}) & Y$^{d,f,g}$  \\ 
IRAS~09039$+$0503 & (iii) Filled & 1.7$\pm$0.4 (J32, SW), 1.1$\pm$0.3
(J32, NE) & N & \nodata & \nodata \\    
IRAS~11095$-$0238 NE & (iii) Filled & 1.9$\pm$0.4 (J32) & Y
& N$^{a}$ & Y$^{d}$  \\  \hline   
IRAS~00091$-$0738 & Central absorption dip &
1.3$\pm$0.2 (J21), 1.9$\pm$0.7 (J32), 1.4$\pm$0.4 (J43) & Y & \nodata &
N$^{d}$ \\   
IRAS~12112$+$0305 NE & Central absorption dip &
1.7$\pm$0.3 (J43) & Y & Y$^{b}$ & Y$^{d,f}$  \\   
IRAS~01569$-$2939 & Central absorption dip & 1.0$\pm$0.2 (J21),
0.9$\pm$0.2 (J32), 0.9$\pm$0.1 (J43) & Y & \nodata & \nodata \\ 
IRAS~01004$-$2237 & \nodata & 1.0$\pm$0.2 (J32) & Y & Y$^{a}$ & \nodata \\      
IRAS~13509$+$0442 & \nodata & 1.0$\pm$0.3 (J32) & N & \nodata & \nodata \\    
IRAS~23234$+$0946 & \nodata  & 1.0$\pm$0.2 (J32) & N & \nodata & \nodata \\  
\enddata

\tablenotetext{A}{Inflow signature is seen, based on the 
detection of redshifted far-infrared 60--120 $\mu$m OH absorption
features, relative to the systemic velocity \citep{spo13,vei13}.} 

\tablecomments{
Col.(1): Object name.  
Col.(2): Classification of the spectrally and spatially resolved
investigation of the elevated HCN-to-HCO$^{+}$ flux ratios.
For the (i) ``spherical shell'' geometry, clearly detected line
information is added: J=2--1 (J21), J=3--2 (J32), or J=4--3 (J43).  
The plus (+) mark means that signatures of both features are seen.
Col.(3): Velocity-integrated HCN-to-HCO$^{+}$ flux ratio measured in
the nuclear ($\lesssim$500 pc) spectra, taken from \citet{ima19} or
\citet{ima23b}.  
Col.(4): Detection (``Y'') or non-detection (``N'') of optically 
elusive, but infrared- and/or (sub)millimeter-identified luminous buried
AGN signatures, adopted from columns 10 and 11 in Table~\ref{tab:object}.
Col.(5): Presence (``Y'') of molecular outflow signatures based
on the detection of blueshifted OH absorption lines at far-infrared
60--120 $\mu$m, with respect to the ULIRG systemic velocity.  
``N'': No clear signature. ``$\cdots$'': no data.
$^{a}$: \citet{spo13};  
$^{b}$: \citet{vei13};  
$^{c}$: \citet{gon17}.  
Col.(6): Presence (``Y'') of molecular outflow signatures based
on the detection of CO J-transition broad emission-line wings in the millimeter
wavelength or VLBI detection of the blueshifted 18 cm OH emission line.  
``N'': No clear signature. ``$\cdots$'': no data.
$^{d}$: \citet{lam22};  
$^{e}$: \citet{wu24}.  
$^{f}$: \citet{per18};  
$^{g}$: \citet{flu19};  
}

\end{deluxetable*}
%%%%%%%%%%%%%%%%%%%%%%%%%%%%%%%%%%%

\subsection{ULIRGs with non-elevated ($\lesssim$1.0) velocity-integrated 
HCN-to-HCO$^{+}$ flux ratios}

Four ULIRGs (IRAS~01569$-$2939, 01004$-$2237, 13509$+$0442, and
23234$+$0946) show non-elevated ($\lesssim$1.0) velocity-integrated 
HCN-to-HCO$^{+}$ flux ratios in the nuclear $\lesssim$500 pc regions 
(Table~\ref{tab:result}).
For IRAS~01569$-$2939, a spectrally distinct (and spatially compact)
geometry is seen for all the $>$3$\sigma$-detected spaxels 
(Figure~\ref{fig:3D5}, left) because of
the presence of significant central dips in the beam-sized and
nuclear area-integrated HCN and HCO$^{+}$ emission-line spectra
\citep{ima19,ima23b}.  
This makes it difficult to discuss the origin of the spectrally
distinct (and spatially compact) geometry of the elevated 
HCN-to-HCO$^{+}$ flux ratios in the PPV plots 
(Figure~\ref{fig:3D5}, middle and right).  
For IRAS~01004$-$2237, 13509$+$0442, and 23234$+$0946, 
the number of spaxels with elevated HCN-to-HCO$^{+}$ flux ratios 
(top 10\% and $>$1.5) is too small to discuss their geometry in the
PPV plots (Figure~\ref{fig:3D9}).  
Our classification in the PPV plots of these four ULIRG nuclei 
with non-elevated ($\lesssim$1.0) velocity-integrated HCN-to-HCO$^{+}$
flux ratios is also summarized in Table~\ref{tab:result}.  

\section{Summary}

We have conducted spectrally and spatially resolved investigations of
the (sub)millimeter HCN-to-HCO$^{+}$ flux ratios in 18 
nearby ($z <$ 0.15) ULIRG nuclei 
with bright HCN and HCO$^{+}$ dense molecular line
emission at J=2--1, J=3--2, and/or J=4--3, observed with
$\lesssim$0$\farcs$2 ($\lesssim$500 pc) resolution. 
Of the 18 observed ULIRGs, elevated ($>$1.0) velocity-integrated
HCN-to-HCO$^{+}$ flux ratios were found in 14 ULIRG nuclei
($\lesssim$500 pc), while the remaining four ULIRG nuclei display
non-elevated ($\lesssim$1.0) velocity-integrated HCN-to-HCO$^{+}$ flux
ratios.  
We created position-position-velocity (PPV) plots of the
HCN-to-HCO$^{+}$ flux ratios for 
(a) all spaxels with $>$3$\sigma$ detections for both HCN and HCO$^{+}$,  
(b) spaxels with the top 10\% flux ratios (relatively high) 
(with $>$3$\sigma$ detections), and 
(c) spaxels with flux ratios $>$1.5 (absolutely high) 
(with $>$3$\sigma$ detections).
We investigated how the geometry of spaxels with elevated
HCN-to-HCO$^{+}$ flux ratios differs from that of all the
$>$3$\sigma$-detected spaxels in the PPV plots. 
We found the following main results.

\begin{enumerate}

\item 
In the PPV plots, seven ULIRG nuclei with elevated
velocity-integrated HCN-to-HCO$^{+}$ flux ratios 
(IRAS~16090$-$0139, 01298$-$0744, 00456$-$2904, 14348$-$1447, 
00188$-$0856, 10378$+$1108, and 10190$+$1322)
show signatures of (i) 
``spherical shell'' geometry for the spaxels with elevated
HCN-to-HCO$^{+}$ flux ratios, 
either in a relative (top 10\%) and/or an absolute sense ($>$1.5).
In particular, for IRAS~16090$-$0139, this ``spherical shell''
geometry is clearly seen in all the J=2--1, J=3--2, and J=4--3 lines.
Spatially resolved outflows are the natural explanation for this
geometry.
Indeed, molecular-outflow signatures have been found through the
detection of blueshifted far-infrared 60--120 $\mu$m OH absorption 
lines, millimeter CO broad emission-line wings, and/or 
VLBI $\lesssim$0$\farcs$012-resolution 18 cm OH emission lines,
whenever such observational data are available (except for IRAS~00456$-$2904).
If the elevated HCN-to-HCO$^{+}$ flux ratios in this ``spherical shell''
geometry originate from spatially resolved outflow activity, 
they tend to trace modest-velocity outflow components compared to the
maximum outflow velocity. 
In fact, HCN abundance enhancement, relative to HCO$^{+}$, is
predicted in modest-velocity shocks with a time scale of
10$^{3-5}$ years in several chemical models. 
The high fraction ($\sim$40\%; 7 our of 18) of ULIRG nuclei with
signatures of spatially resolved outflow-origin elevation of the
HCN-to-HCO$^{+}$ flux ratios conforms to the trend of \citet{nis24}
suggested based on a smaller number of ULIRG sample.

\item 
Two ULIRG nuclei with elevated velocity-integrated HCN-to-HCO$^{+}$
flux ratios (IRAS~01166$-$0844 and 22206$-$2715) show signatures of
(ii) ``spectrally distinct (and spatially compact)'' geometry of
the spaxels with elevated HCN-to-HCO$^{+}$ flux ratios,
despite the fact that the geometry is (iii) ``filled'' for all the 
$>$3$\sigma$-detected spaxels.
The elevation can be due to spatially unresolved outflows and/or AGN
effects.

\item 
Three ULIRG nuclei with elevated velocity-integrated HCN-to-HCO$^{+}$
flux ratios (IRAS~22491$-$1808, 09039$+$0503, and 11095$-$0238)
display (iii) ``filled'' (spectrally filled and spatially compact)
geometry for the spaxels with elevated HCN-to-HCO$^{+}$ flux ratios.
We interpret that AGN effects can explain the elevation.
For IRAS~22491$-$1808 and 11095$-$0238, 2.6 mm CO J=2--1 broad 
emission-line wings, possibly caused by molecular outflow activity,
are observed, but the blueshifted and redshifted emission-wing components
are not clearly separated spatially. 
The (iii) ``filled'' geometry of spaxels with elevated
HCN-to-HCO$^{+}$ flux ratios for these two ULIRGs can also be reproduced
by spatially confined molecular outflows with not yet clearly separated
blueshifted and redshifted velocity components.  
The signatures of (iii) ``filled'' geometry are also seen in some 
J-transitions lines in six ULIRGs discussed above and already classified
as (i) or (ii) (IRAS 16090$-$0139, 00456$-$2904, 14348$-$1447,
00188$-$0856, 01166$-$0844, and 22206$-$2715).
The same scenario, namely, AGN effects and/or the
spatially confined outflows, may also apply to these six ULIRG nuclei
and IRAS~09039$+$0503 (in total nine nuclei), as the origin of the
elevated HCN-to-HCO$^{+}$ flux ratios.      
 
\item 
For two ULIRG nuclei with elevated velocity-integrated HCN-to-HCO$^{+}$
flux ratios (IRAS~00091$-$0738 and 12112$+$0305), the physical 
origin of the elevation is difficult to discuss 
because strong absorption dips, due to self-absorption, in the
beam-sized and/or nuclear area-integrated spectra hamper reliable
investigation of the velocity structure of $>$3$\sigma$-detected
spaxels in emission near the systemic velocity. 

\item 
For four ULIRG nuclei that show non-elevated ($\lesssim$1.0) 
velocity-integrated HCN-to-HCO$^{+}$ flux ratios (IRAS~01569$-$2939, 
01004$-$2237, 13509$+$0442, and 23234$+$0946), the presence of
significant absorption dips near the systemic velocity hampers
investigation of the true velocity structure of spaxels with
elevated HCN-to-HCO$^{+}$ flux ratios (IRAS~01569$-$2939), or the 
number of spaxels with elevated HCN-to-HCO$^{+}$ flux ratios 
(with $>$3$\sigma$ detections for both molecular lines) is too small
to investigate the geometry (the remaining three ULIRGs).   

\end{enumerate}

The geometry of the elevated HCN-to-HCO$^{+}$ flux ratios in the PPV
plots was found to fall into various types, including those 
expected to originate from (a) spatially extended outflows and (b) AGN
and/or spatially unresolved outflows.     
We argue that spectrally and spatially resolved investigation of dense 
molecular line emission with elevated HCN-to-HCO$^{+}$ flux ratios can 
provide an important clue to the origin of the elevated 
velocity-integrated HCN-to-HCO$^{+}$ flux ratios ($>$1.0) observed in
the nuclear $\lesssim$500 pc regions of many nearby ULIRGs.

%% If you wish to include an acknowledgments section in your paper,
%% separate it off from the body of the text using the \acknowledgments
%% command.

\begin{acknowledgments}
We thank the anonymous referee for his/her valuable
comments, which helped improve the clarity of this manuscript.
This paper made use of the following ALMA data:
ADS/JAO.ALMA\#2017.1.00057.S. and \#2019.1.00027.S.
ALMA is a partnership of ESO (representing its member states), NSF (USA) 
and NINS (Japan), together with NRC (Canada), NSC and ASIAA
(Taiwan), and KASI (Republic of Korea), in cooperation with the Republic
of Chile. The Joint ALMA Observatory is operated by ESO, AUI/NRAO, and
NAOJ. 
M.I. is supported by JP21K03632 and JP25K07359.
Y.N. gratefully acknowledges support from JSPS KAKENHI Grant 
Numbers JP23K13140 and JP23K20035.  
Data analysis was in part carried out on the open use data analysis
computer system at the Astronomy Data Center, ADC, of the National
Astronomical Observatory of Japan. 
This research has made use of NASA's Astrophysics Data System and the
NASA/IPAC Extragalactic Database (NED) which is operated by the Jet
Propulsion Laboratory, California Institute of Technology, under
contract with the National Aeronautics and Space Administration. 
\end{acknowledgments}

%% To help institutions obtain information on the effectiveness of their 
%% telescopes the AAS Journals has created a group of keywords for telescope 
%% facilities.
%
%% Following the acknowledgments section, use the following syntax and the
%% \facility{} or \facilities{} macros to list the keywords of facilities used 
%% in the research for the paper.  Each keyword is check against the master 
%% list during copy editing.  Individual instruments can be provided in 
%% parentheses, after the keyword, but they are not verified.

\vspace{5mm}
\facilities{ALMA}
\software{CASA \citep{CASA22},
Astropy \citep{Astropy},
NumPy \citep{NumPy},
Matplotlib \citep{Matplotlib}}
%RADEX \citep{RADEX},
%pyradex (\url{https://github.com/keflavich/pyradex}),
%IPython \citep{IPython},
%Jupyter Notebook \citep{JupyterNotebook},
%SciPy \citep{SciPy},
%Pandas \citep{Pandas},
%lmfit \citep{lmfit},
%emcee \citep{emcee},
%corner \citep{corner}}

%% Similar to \facility{}, there is the optional \software command to allow 
%% authors a place to specify which programs were used during the creation of 
%% the manusscript. Authors should list each code and include either a
%% citation or url to the code inside ()s when available.

%\software{astropy \citep{2013A&A...558A..33A},  
%          Cloudy \citep{2013RMxAA..49..137F}, 
%          SExtractor \citep{1996A&AS..117..393B}
%          }

%% Appendix material should be preceded with a single \appendix command.
%% There should be a \section command for each appendix. Mark appendix
%% subsections with the same markup you use in the main body of the paper.

%% Each Appendix (indicated with \section) will be lettered A, B, C, etc.
%% The equation counter will reset when it encounters the \appendix
%% command and will number appendix equations (A1), (A2), etc. The
%% Figure and Table counter will not reset.

%\section{Appendix information}

%\appendix

%% This command is needed to show the entire author+affilation list when
%% the collaboration and author truncation commands are used.  It has to
%% go at the end of the manuscript.
%\allauthors

%% Include this line if you are using the \added, \replaced, \deleted
%% commands to see a summary list of all changes at the end of the article.
%\listofchanges

\end{document}